\documentclass[fleqn,usenatbib]{mnras}

\usepackage{lscape}
\usepackage{subfigure}
\usepackage{rotating}
\usepackage{graphicx}	
\usepackage{amsmath}	
\usepackage{amssymb}	
\usepackage{multicol}        
\usepackage{bm}		
\usepackage{pdflscape}	
\usepackage{longtable}
\usepackage{threeparttable}
\usepackage{multirow}

\usepackage{mathptmx}
\usepackage{txfonts}

\usepackage[T1]{fontenc}

\DeclareRobustCommand{\VAN}[3]{#2}
\let\VANthebibliography\thebibliography
\def\thebibliography{\DeclareRobustCommand{\VAN}[3]{##3}\VANthebibliography}

\usepackage{graphicx}	
\usepackage{amsmath}	
\usepackage{amssymb}	

\title[ATOMS X]{ATOMS: ALMA Three-millimeter Observations of Massive Star-forming regions – X. Chemical differentiation among the massive cores in G9.62+0.19}

\author[Y. P. PENG et al.]{
Yaping Peng,$^{1,2}$\thanks{E-mail: pyp893@163.com}
Tie Liu,$^{3}$\thanks{E-mail: liutie@shao.ac.cn}
Sheng-Li Qin,$^{2}$
Tapas Baug,$^{4}$
Hong-Li Liu,$^{2}$
Ke Wang,$^{5}$
Guido Garay,$^{6}$
Chao Zhang,$^{7}$
\newauthor
Long-Fei Chen,$^{8}$
Chang Won Lee,$^{9,10}$
Mika Juvela,$^{11}$
Dalei Li,$^{12}$
Ken’ichi Tatematsu,$^{13}$
Xun-Chuan Liu,$^{3}$
\newauthor
Jeong-Eun Lee,$^{14}$
Gan Luo,$^{15}$
Lokesh Dewangan,$^{16}$
Yue-Fang Wu,$^{17}$
Li Zhang,$^{2}$
Leonardo Bronfman,$^{6}$
\newauthor
Jixing Ge,$^{18}$
Mengyao Tang,$^{2,19}$
Yong Zhang,$^{20,21,22}$
Feng-Wei Xu,$^{17}$
Yao Wang, $^{23}$ 
and Bing Zhou $^{1,2}$
\\
Affiliations are listed at the end of the paper }

\date{Accepted XXX. Received YYY; in original form ZZZ}

\pubyear{2022}

\begin{document}
\label{firstpage}
\pagerange{\pageref{firstpage}--\pageref{lastpage}}
\maketitle

\begin{abstract}
 Investigating the physical and chemical structures of massive star-forming regions is critical for understanding the formation and the early evolution of massive stars. We performed a detailed line survey toward six dense cores named as MM1, MM4, MM6, MM7, MM8, and MM11 in G9.62+0.19 star-forming region resolved in ALMA band 3 observations. Toward these cores, about 172 transitions have been identified and attributed to 16 species including organic Oxygen-, Nitrogen-, Sulfur-bearing molecules and their isotopologues. Four dense cores MM7, MM8, MM4, and MM11 are line rich sources. Modeling of these spectral lines reveals the rotational temperature in a range of 72$-$115~K, 100$-$163~K, 102$-$204~K, and 84$-$123~K for the MM7, MM8, MM4, and MM11, respectively. The molecular column densities are 1.6 $\times$ 10$^{15}$ $-$ 9.2 $\times$ 10$^{17}$~cm$^{-2}$ toward the four cores. The cores MM8 and MM4 show chemical difference between Oxygen- and Nitrogen-bearing species, i.e., MM4 is rich in oxygen-bearing molecules while nitrogen-bearing molecules especially vibrationally excited HC$_{3}$N lines are mainly observed in MM8. The distinct initial temperature at accretion phase may lead to this N/O differentiation. Through analyzing column densities and spatial distributions of O-bearing Complex Organic Molecules (COMs), we found that C$_{2}$H$_{5}$OH and CH$_{3}$OCH$_{3}$ might have a common precursor, CH$_{3}$OH. CH$_{3}$OCHO and CH$_{3}$OCH$_{3}$ are likely chemically linked. In addition, the observed variation in HC$_{3}$N and HC$_{5}$N emission may indicate that their different formation mechanism at hot and cold regions. 
\end{abstract}




\section{Introduction}
  High-mass stars play a major role in the evolution of the galaxies. They affect the formation and evolution of planets, stars, and galaxies \citep{Kennicutt98, Bally05}. But the formation process of high-mass star remains much less understood because of their comparatively larger distances and shorter lifetimes than low-mass star. In addition, high-mass stars are usually born in complicated cluster environments. The process of star formation will inevitably be influenced by nearby protostars \citep{Krumholz14}, which complicates the physical and chemical structures of high-mass star formation. During the earliest phases high-mass stars remain embedded in their natal molecular clouds. Observationally, their embedded phases can generally be subdivided into four different evolutionary stages: massive dense cores (including starless cores), high-mass protostellar objects (HMPOs), hot molecular cores (HMCs), and Hyper/ultra compact H{\sc ii} regions (H/UC H{\sc ii}) (e.g., \citealt{Kurtz2000,Hoare07,Beltran09,Tan13,Liu2021}). Investigating the physical and chemical properties of high-mass star forming cores at different evolutionary stages is important for understanding the formation and early evolution of massive stars. 
  
  G9.62+0.19 is a well known high-mass star forming region at a distance of 5.2 kpc \citep{Sanna09}, which hosts several massive cores at different evolutionary phases (e.g., \citealt{Testi98, Hofner94, Hofner01,Liu17}). Multi-wavelength VLA observations identified nine radio continuum sources (denoted as A-I) in this region \citep{Garay93, Testi2000}. The extended H{\sc ii} region "A", cometary-shaped H{\sc ii} region "B" and compact H{\sc ii} region "C" are three evolved H{\sc ii} regions \citep{Kurtz94}. The JCMT/SCUBA 450 $\mu$m observations found that new generations of high-mass young stellar objects ("G9.62 clump") are forming in the region which is located to the east of the evolved H{\sc ii} regions \citep{Liu11, Liu17}. Twelve dense cores (MM1-MM12) have been identified in this clump through 1.3 mm observations with ALMA \citep{Liu17}. Recent ALMA observations at 1 mm wavelength (band 7) with higher resolution further resolved "G9.62 clump" into 23 dense cores \citep{Dallolio19}. Among of them, MM4, MM7, MM8, and MM11 are associated with radio sources "E", "G", "F", and "D", respectively. Five cores (MM4/E, MM6, MM7/G, MM8/F and MM11/D) are considered as high-mass protostars. Based on observations of CH$_{3}$OH~v$_{t}$ =1 line and the existence of the outflow, \citet{Liu17} further classified their evolutionary sequence as following: MM6 (HMPO), MM7 (early HMC), MM8 (HMC), MM4 (late HMC or HC H{\sc ii}), and MM11 (UC H{\sc ii}). Presence of collimated SiO (5-4) and CO (2-1) outflows and absence of CH$_{3}$OH~v$_{t}$ =1 emission in MM6 are indicative of it to be at HMPO stage. MM7 and MM8 are associated with centimeter continuum emission \citep{Testi2000}, outflows, and strong CH$_{3}$OH~v$_{t}$ =1 line emission, and are thus at comparatively more evolved stage than MM6. CH$_{3}$OCHO~v$_{t}$=1 was detected toward the MM8 while not for MM7, indicating that MM8 is hotter and older than MM7. MM4 has centimeter continuum counterpart \citep{Testi2000}, while no outflows implying MM4 to be at a more evolved phase. Weaker 
  CH$_{3}$OH~v$_{t}$ =1 emission toward MM11 than MM4 suggest that MM11 should be at a more evolved phase, since the abundance of CH$_{3}$OH should be close to maximum at hot core stage and decrease in the UC H{\sc ii} phase \citep{Gerner14}. The other seven cores (MM1, MM2, MM3, MM5, MM9, MM10, and MM12) are considered to be starless core as they do not show any signature of outflows. They only show the signature of rare interstellar molecules, especially complex organic molecules (COMs). \citet{Su05} reported that many molecular lines including Nitrogen-bearing molecules and organic molecules have been detected toward the source F(MM8) and E(MM4) in G9.62+0.19 by Submillimeter Array (SMA) observations (345 GHz). In general, G9.62+0.19 contains different evolutionary types of dense cores, and harbours rich chemistry indicating that it is an ideal object to explore the distinct chemical properties of cores at different evolutionary stages. 
 
  Interstellar molecules especially the COMs are useful diagnosis to study the interstellar environments and evolutionary stages of various astronomical sources since they usually exist in special physical environments. COMs have been detected in warm regions including hot cores in high-mass star-forming regions such as SgrB2(N) (e.g., \citealt{Belloche13,Belloche16,Belloche17,Bonfand19}), G31.41+0.31 (e.g., \citealt{Rivilla17b,Mininni20,Colzi21}), AFGL 4176 \citep{Bogelund19}, NGC 6334I (e.g., \citealt{Zernickel12,Bogelund18,Ligterink20}), W51 e2 \citep{Remijan04a,Rivilla17a}, G34.3+0.2 \citep{Lykke15}, Orion KL (e.g., \citealt{Crockett14,Cernicharo16,Tercero18}), G331.512-0.103 \citep{Hervias2019,Duronea2019} and hot corinos in low-mass star-forming regions such as NGC 1333-IRAS 2A and -IRAS 4A \citep{Taquet15}, IRAS 16293-2422 \citep{Jorgensen12,Richard13}, and in cold environments such as dark clouds (TMC-1 (e.g., \citealt{McGuire18,Agundez21}) and L134N \citep{Dickens2000}), starless cores (e.g., \citealt{Bacmann12,Vastel14,Jimenez16,Molet19}), and cold outer envelopes of protostars (e.g., \citealt{Oberg2010,Bergner17}), Galatic Center molecular clouds such as G-0.02-0.07, G-0.11-0.08, and G+0.693-0.027 \citep{Requena-Torres06,Requena-Torres08,Zeng18,Rodriguez-Almeida2021}, and the disk of an outbursting young star, V883 Ori \citep{Lee2019}. The chemistry in different interstellar environments, and the formation of COMs in these environments are complex and long-standing quest that has been explored by many researches in the past several years (e.g., \citealt{Herbst09,Laas11,Bacmann12,Woods13,Vasyunin13,Ruaud15,Abplanalp16,Skouteris18,Shingledecker19}). Yet, there remain huge challenges for revealing a clearer picture of COMs formation. Fundamentally, the chemistry is the result of interactions of different physical and chemical parameters that are tangled with each other, with the former includes but not limited to the temperature, density and evolutionary stages of the cores, and the later primarily depends on the elemental content, grain mantle compositions and reaction network that represents the chemical contents of the cores. Thus, when the physical properties of the core are well constrained, the observational results, such as the column density and abundance of individual molecule and the correlations between different molecules, could be used to retrieve the intrinsic chemical properties.
 
  Here we preform a detailed analysis of the physical and chemical conditions (e.g., temperature, density, and possible chemical formation routes of COMs) of dense cores in G9.62+0.19 through ALMA band 3 observations. The paper is organized as follows. In Sect. 2 we describe the observations and the data reduction. The results for continuum and line survey are presented in Sect. 3. We model the spectral lines and calculate the parameters under the LTE assumption in Sect. 4. A general discussion about properties of each core and chemical differentiation among these cores is presented in Sect. 5. In Sect.6 we give our summary, with a possibility of future investigations of massive star formation at different evolutionary stages.
  
\section{Observations}
 G9.62+0.19 was observed with ALMA in band 3 (Project ID: 2019.1.00685.S; PI: Tie Liu), as part of the ATOMS survey, which was conducted towards 146 IRAS clumps \citep{Liu20a,Liu20b,Liu2021}. The Atacama
 Compact 7-m Array (ACA) observations of G9.62+0.19 were conducted on 1st and 13th October, 2019 with two executions, and the 12-m array observations were conducted on 31th October, 2019. The phase center in both ACA and 12-m array observations is R.A.(J2000)=$18^\mathrm{h}06^\mathrm{m}14^\mathrm{s}.99$ and Decl.(J2000)=$-20^{\circ}31^{'}35^{\prime\prime}.4$. The spectral windows (SPWs) 1$-$6 at the lower sideband have a bandwidth of 58.59 MHz, which cover dense gas tracers such as J=1-0 transition of HCO$^{\rm +}$, H$^{\rm 13}$CO$^{\rm +}$, HCN, and H$^{\rm 13}$CN, shock tracer SiO J=2-1 and photodissociation region tracer CCH J=1-0. The other two wide SPWs 7$-$8 at the upper sideband covering frequency range of 97.530$-$101.341 GHz, each with a bandwidth of 1875 MHz, which are used for continuum emission and line survey. The spectral resolution is $\sim$0.2 km s$^{-1}$, $\sim$0.1 km s$^{-1}$, and $\sim$1.5 km s$^{-1}$ for SPWs 1$-$4, SPWs 5$-$6, and SPWs 7$-$8, respectively \citep{Liu2021}. 

  The calibrated UV data and images were processed using the Common Astronomy Software Applications (CASA). All images are primary beam corrected. The calibration uncertainty on the flux density is $\sim$ 10\%. Continuum image was constructed from line-free channels. The synthesized beam size and 1$\sigma$ rms noise level for the continuum image from the 12-m array are 1.56$\arcsec\times1.38\arcsec$ ($\sim$ 8112 $\times$ 7176 au at a distance of 5.2~kpc, P.A.=87.62$^{\circ}$) and 0.4 mJy beam$^{-1}$, respectively. For spectral images, their synthesized beam size is approximately 1.9$\arcsec\times1.7\arcsec$ ($\sim$ 9880 $\times$ 8840 au) at the lower band (SPWs 1$-$6), and $\sim$1.6$\arcsec\times1.4\arcsec$ ($\sim$ 8320 $\times$ 7280 au) at the upper band (SPWs 7$-$8). The 1$\sigma$ rms noise is about 8 mJy beam$^{-1}$ per channel, 10~mJy beam$^{-1}$ per channel, and 3 mJy beam$^{-1}$ per channel for SPWs 1$-$4, 5$-$6, and 7$-$8, respectively. The 12-m+ACA combined continuum image of G9.62+0.19 has a beam size of $2.3\arcsec \times 1.7\arcsec$ (P.A.=85.14$^{\circ}$), and a sensitivity of $\sim$0.5~mJy~beam$^{-1}$. For spectral lines, combined SiO (2-1), HCO$^{+}$ (1-0) and CS (2-1) are utilized to probe the kinematics of this source. The synthesized beams for SiO, HCO$^{+}$, and CS are 2.7$\arcsec\times2.1\arcsec$, 2.5$\arcsec\times1.9\arcsec$, and 2.3$\arcsec\times1.7\arcsec$, respectively, and the sensitivity levels are $\sim$13, 60, and 48~mJy beam$^{-1}$ per channel, respectively.

\section{Results}
\subsection{ALMA 3 mm continuum emission}
\begin{figure}
   \centering
   \includegraphics[angle=0,width=9cm]{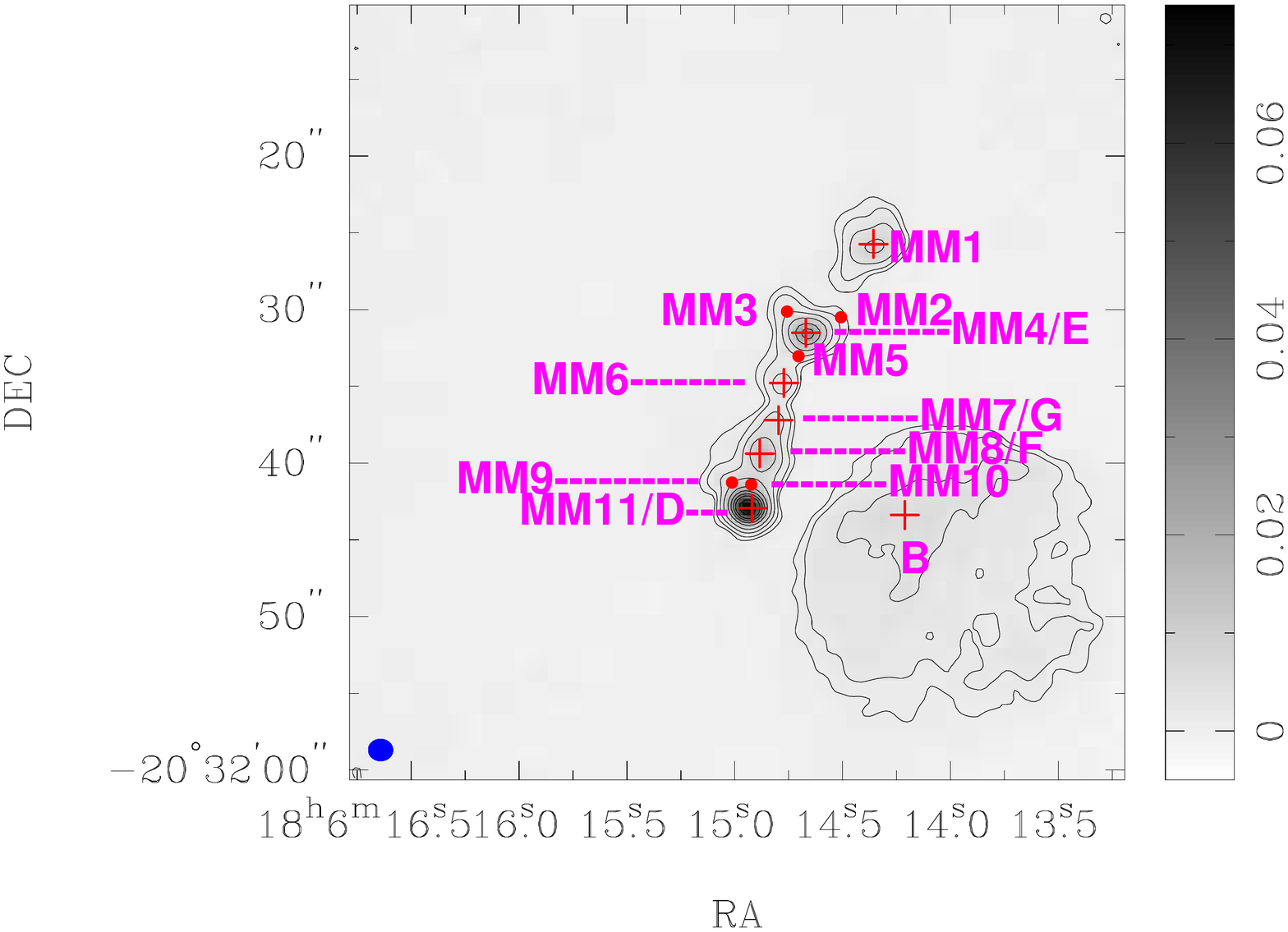}
   \includegraphics[angle=0,width=9cm]{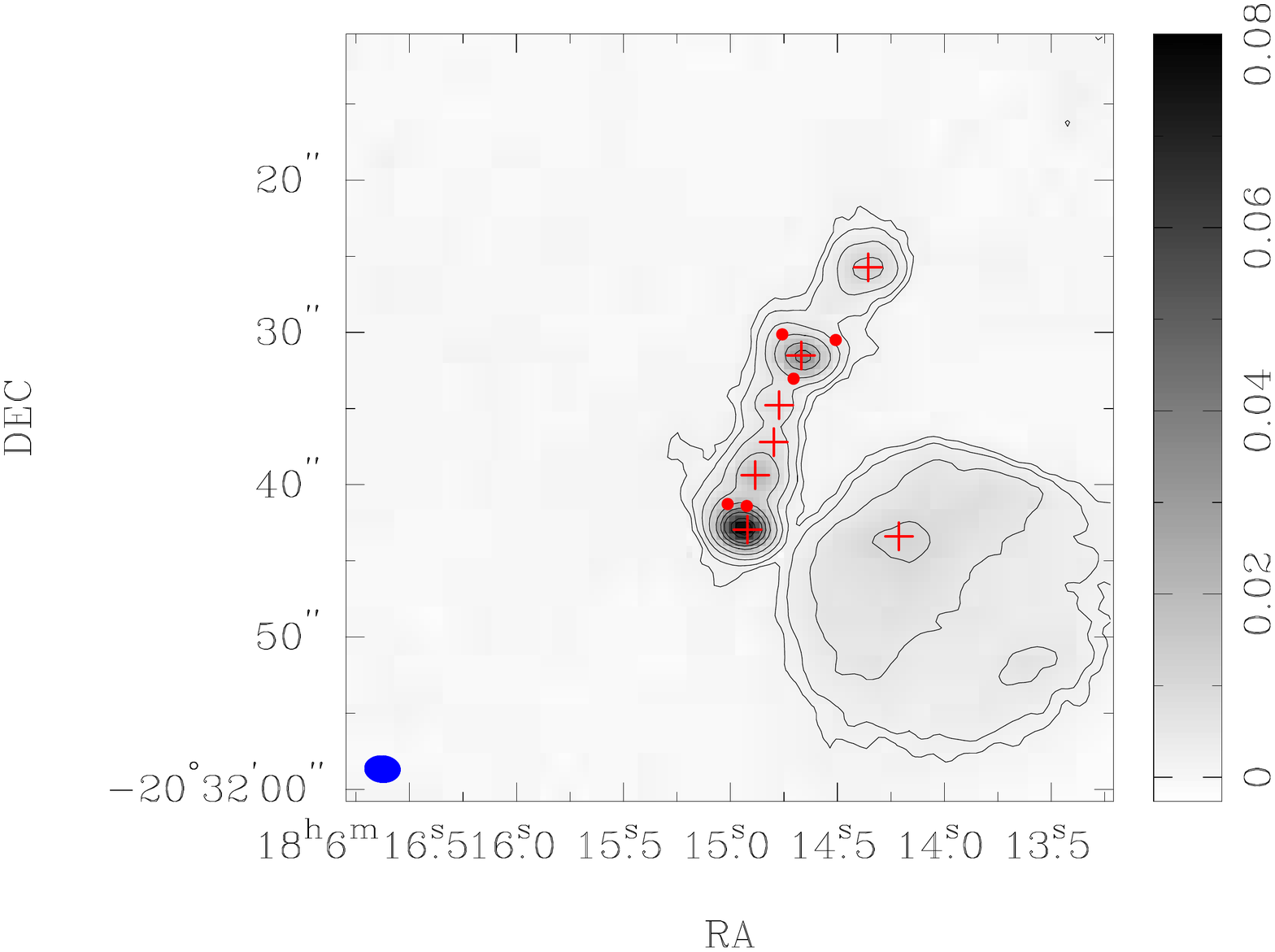}
      \caption{Upper panel: the 3 mm continuum emission of G9.62+0.19 from 12-m array observations. Lower panel: the 3 mm continuum emission of G9.62+0.19 from 12-m+ACA combined observations. The synthesized beam size is shown in the bottom left corner. The contour levels are (3, 5, 10, 20, 40, 60, 80, 100, 140, 180) $\times$ 0.4~mJy~beam$^{-1}$, and (3, 5, 10, 20, 40, 60, 80, 120, 140) $\times$ 0.5~mJy~beam$^{-1}$ for 12-m and 12-m+ACA combined data, respectively. The cross symbols mark the 3 mm continuum peaks of MM1, MM4, MM6, MM7, MM8, MM11, and H{\sc ii} region "B". The filled circles indicate the 1.3 mm continuum peaks (MM2, MM3, MM5, MM9, and MM10) \citep{Liu17}. The unit of the gray scale bar on the right is in Jy/beam. 
              }
      \label{fig:continuum-3mm}
   \end{figure}

\begin{figure}
   \hspace{-2cm}
   \includegraphics[angle=0,width=14cm]{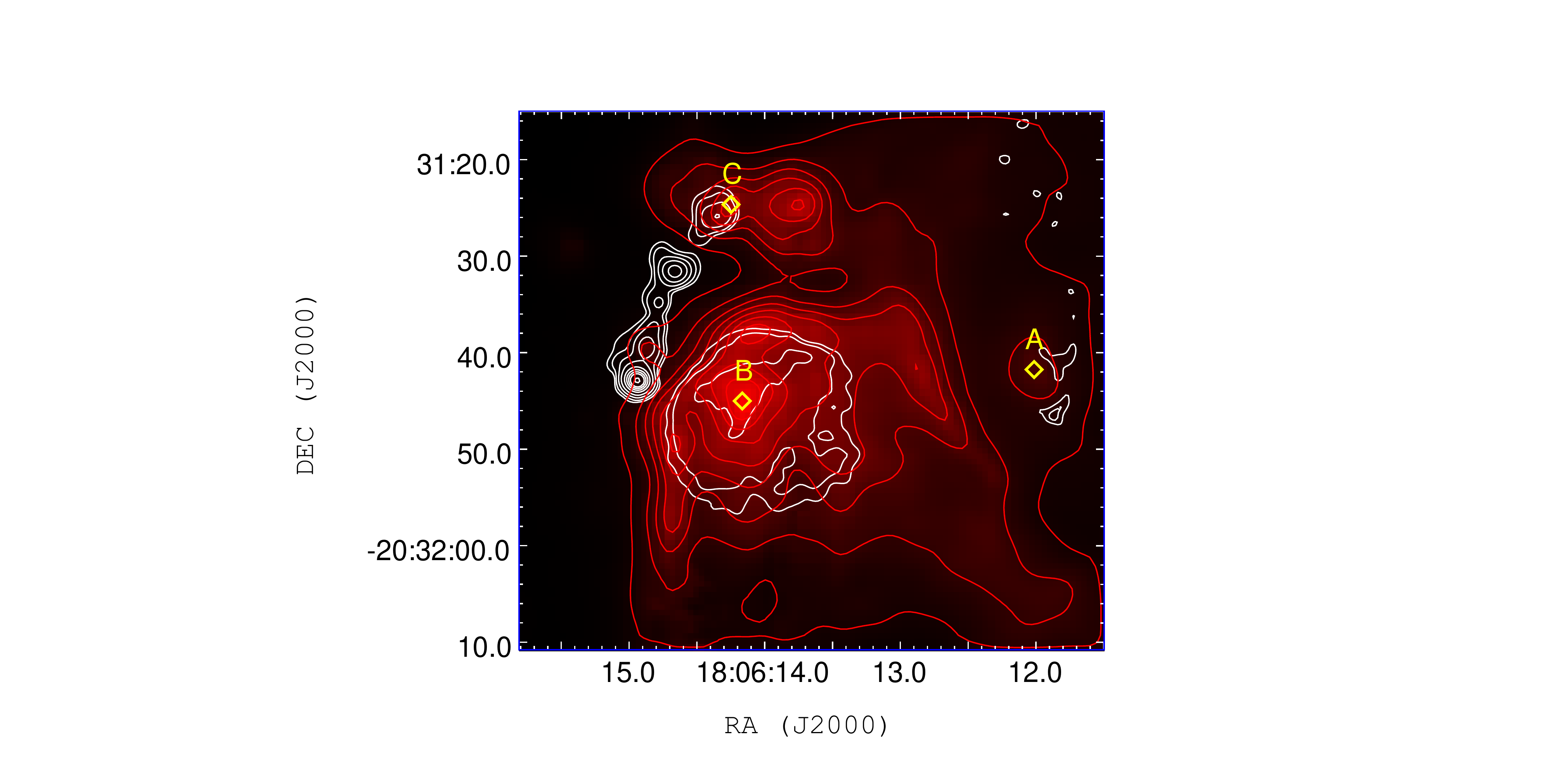}
      \caption{ Three millimeter continuum emission (white contours) overlaid on 8 $\mu$m emission (red contours) with Spitzer/IRAC observation. The red contours are (10\%, 20\%, 30\%, 40\%, 50\%, 60\%, 70\%, 80\%, 90\%, and 100\%) $\times$ 2648~MJy/sr. Three evolved H{\sc ii} regions A, B, and C are marked with yellow diamond shapes. 
              }
      \label{fig:3mm+IR}
   \end{figure}

  Figure \ref{fig:continuum-3mm} shows the 3 mm continuum emission of G9.62+0.19 observed by the 12-m array alone and 12-m+ACA combined data. Six dense cores (MM1, MM4, MM6, MM7, MM8, and MM11) that have been previously observed by 1.3 mm continuum emission with higher angular resolution (0.94$\arcsec\times0.71\arcsec$) \citep{Liu17} in G9.62 clump are also resolved in our 3 mm data. The positions of MM2, MM3, MM5, MM9, and MM10 are also marked in Figure 1. However, these cores are not resolved in 3 mm map possibly because of the relatively lower spatial resolution of our observations. The expanding cometary-like H{\sc ii} region "B" is observed to the west of the G9.62 clump (see Figure \ref{fig:3mm+IR}). The positions, sizes, peak flux density ($I_{\rm peak}$) and total flux density ($S_{\nu}$) of six dense cores obtained by two dimensional Gaussian fitting are listed in Table \ref{tab:continuum}. We found that the size, $I_{\rm peak}$ and $S_{\nu}$ of each core observed from combined data are larger than those from 12-m observations alone, since 12-m array observations are affected by missing flux, then extended emission is not sampled completely. The observations of 12-m array missed $\sim$ 5\% of extended emissions compared to the observations of 12-m + ACA combined data. More information on continuum emission also can be found in \citet{Liu20a}.
  
 \begin{table*}
 \caption{Parameters of Continuum Sources}             
 \label{tab:continuum}     
 \centering                          
 \begin{tabular}{c c c c c c c c c c c c c c c c c c c c c}        
 \hline\hline                 
 Name &  R.A.(J2000) & Decl.(J2000) & \multicolumn{3}{c}{12-m array} & & \multicolumn{3}{c}{12-m and ACA combined data} \\
 \cline{4-6} \cline{8-10} \\
 & & & size & $I_{\rm peak}$ & $S_{\nu}$ & & size & $I_{\rm peak}$ & $S_{\nu}$ \\    
      &  &  &  & (mJy beam$^{-1}$) & (mJy) & & & (mJy beam$^{-1}$) & (mJy) \\
 \hline                        
   MM1  & $18^\mathrm{h}06^\mathrm{m}14^\mathrm{s}.357$ & $-20^{\circ}31^{'}25^{\prime\prime}.73$ &  3.01$^{\prime\prime}\times2.09^{\prime\prime}$ & 8.2$\pm$0.6 & 32.4$\pm$2.3 & & 3.73$^{\prime\prime}\times2.54^{\prime\prime}$ & 11.1$\pm$0.8 & 38.3$\pm$2.1 \\
   MM4  &$18^\mathrm{h}06^\mathrm{m}14^\mathrm{s}.67$ & $-20^{\circ}31^{'}31^{\prime\prime}.52$ & 1.32$^{\prime\prime}\times1.06^{\prime\prime}$ & 25.2$\pm$1.9 & 41.4$\pm$2.1  &  & 1.97$^{\prime\prime}\times1.79^{\prime\prime}$ & 30.1$\pm$2.0 & 56.9$\pm$5.3 \\
   MM6  & $18^\mathrm{h}06^\mathrm{m}14^\mathrm{s}.772$ & $-20^{\circ}31^{'}34^{\prime\prime}.78$ & 2.05$^{\prime\prime}\times0.88^{\prime\prime}$ & 5.3$\pm$0.4 & 10.8$\pm$0.7 & & 2.43$^{\prime\prime}\times1.95^{\prime\prime}$ & 8.2$\pm$0.2 & 18.5$\pm$1.2 \\
   MM7$^{a}$& $18^\mathrm{h}06^\mathrm{m}14^\mathrm{s}.798$ & $-20^{\circ}31^{'}37^{\prime\prime}.21$ & $\cdots$ & $\cdots$ & $\cdots$ & & $\cdots$ & $\cdots$ & $\cdots$ \\ 
   MM8  & $18^\mathrm{h}06^\mathrm{m}14^\mathrm{s}.884$ & $-20^{\circ}31^{'}39^{\prime\prime}.37$ &  2.67$^{\prime\prime}\times1.23^{\prime\prime}$ & 13.1$\pm$1.0 & 36.3$\pm$4.1 & & 3.05$^{\prime\prime}\times1.58^{\prime\prime}$ & 19.2$\pm$0.8 & 46.9$\pm$3.2\\
   MM11 &$18^\mathrm{h}06^\mathrm{m}14^\mathrm{s}.92$ & $-20^{\circ}31^{'}42^{\prime\prime}.95$ & 1.05$^{\prime\prime}\times0.75^{\prime\prime}$ & 71.4$\pm$3.7 & 99.1$\pm$4.3 & & 1.40$^{\prime\prime}\times1.07^{\prime\prime}$ & 77.5$\pm$3.3 & 109.7$\pm$4.6 \\                
\hline   
\multicolumn{10}{l}{$^a$ The 2-D Gaussian fitting toward the MM7 suffered failed convergence due to this core is not well resolved by 3~mm continuum emission.}\\
\end{tabular}
\end{table*}

\subsection{Molecular lines identification}
  We identified spectral line transitions using the eXtended CASA Line Analysis Software Suite (XCLASS\footnote{https://xclass.astro.uni-koeln.de}; \citealt{Moller17}). The molecular spectroscopic parameters in XCLASS are obtained from Cologne Database for Molecular Spectroscopy (CDMS\footnote{http://cdms.de}; \citealt{Muller01,Muller05}) and Jet Propulsion Laboratory (JPL\footnote{http://spec.jpl.nasa.gov}; \citealt{Pickett98}) molecular databases. The software includes myXCLASS program, which is used to model data by solving the radiative transfer equation. Under the assumption of local thermodynamical equilibrium (LTE), XCLASS takes the beam dilution, dust attenuation, the line opacity, and line blending into account \citep{Moller17}. A few constraints are set for proper identification of the molecular species. The rest frequency must be consistent with the value from laboratory measurement; for a certain species, all detected transitions from the same source must have a similar LSR velocity; and transitions for a specific species should have similar spatial distribution owing to that they are excited by similar conditions \citep{Peng17}. Furthermore, all lines above the 3$\sigma$ "detection level" are available in our identification and follow-up analyses. The spectrum from 12-m array observations are used for line identification and following data modeling. In the Appendix A, we make a comparison between 12-m array and combined data, suggesting there is no difference when using them to study the hot core chemistry.  
 
\subsubsection{Molecular lines from different cores}
\begin{figure*}
   \centering
   \includegraphics[width=\textwidth]{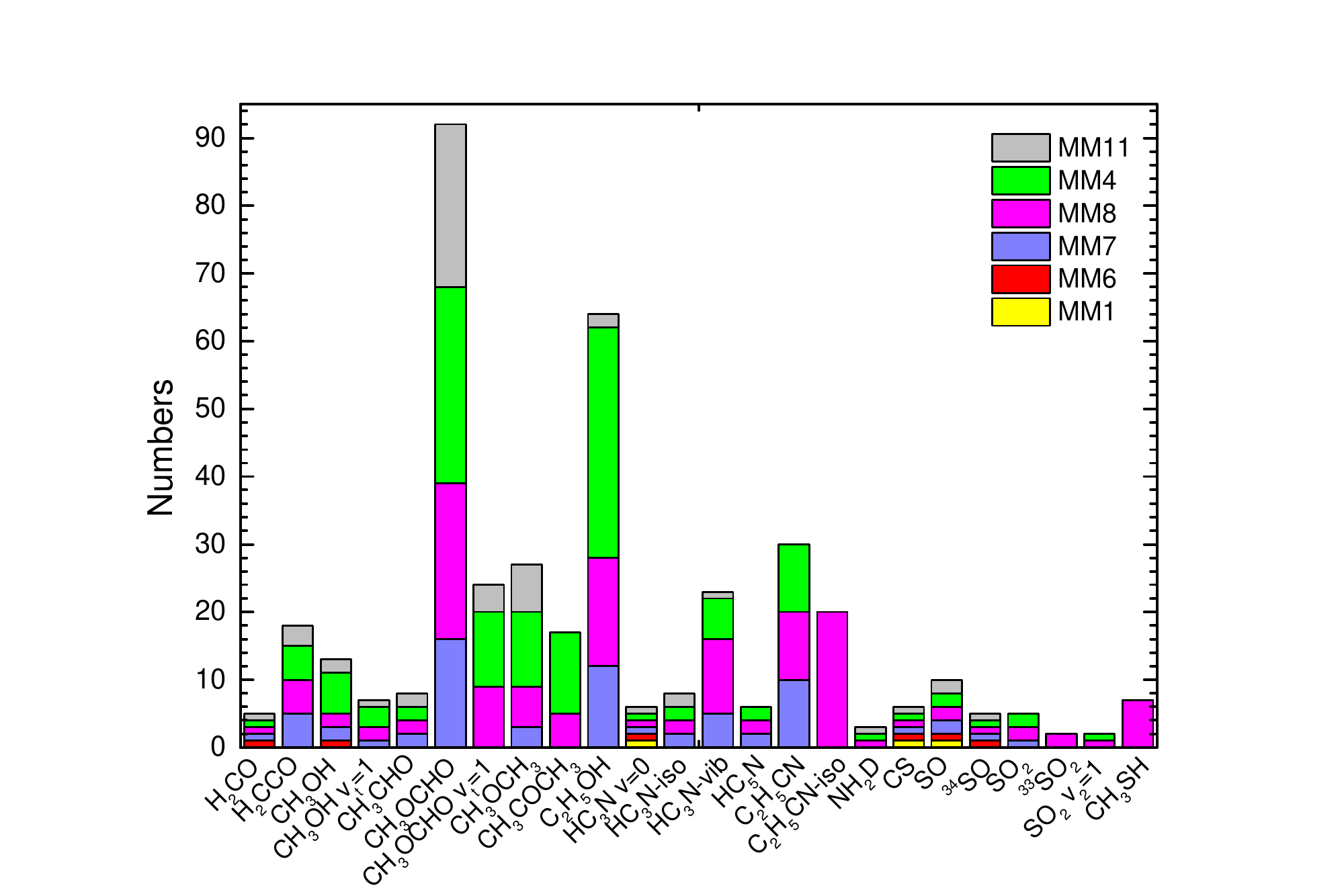}
      \caption{Molecules including O-bearing (H$_{2}$CO, H$_{2}$CCO, CH$_{3}$OH, CH$_{3}$OH~v$_{t}$=1, CH$_{3}$CHO, CH$_{3}$OCHO, CH$_{3}$OCHO~v$_{t}$=1, CH$_{3}$OCH$_{3}$, CH$_{3}$COCH$_{3}$, and C$_{2}$H$_{5}$OH), N-bearing (HC$_{3}$N, HC$_{5}$N, C$_{2}$H$_{5}$CN, and NH$_{2}$D), and S-bearing (CS, SO, $^{34}$SO, SO$_{2}$, $^{33}$SO$_{2}$, SO$_{2}$~v$_{2}$=1, and CH$_{3}$SH) molecules that have been identified toward the six cores. The numbers on the Y-axis indicate the detected molecular transition number of each core. X-axis presents the molecular name, in which the label "HC$_{3}$N-iso" indicates the HC$_{3}$N isotopologues HC$^{13}$CCN and HCC$^{13}$CN; "HC$_{3}$N-vib" means HC$_{3}$N in vibrationally excited states v$_{7}$=1, v$_{7}$=2, v$_{7}$=3, v$_{6}$=1, as well as HC$^{13}$CCN~v$_{7}$=1 and HCC$^{13}$CN~v$_{7}$=1; C$_{2}$H$_{5}$CN-iso contains CH$_{3}$CH$_{2}^{13}$CN, CH$_{3}^{13}$CH$_{2}$CN, and $^{13}$CH$_{3}$CH$_{2}$CN. 
              }
         \label{fig:species}
   \end{figure*}

\begin{figure*}
   \centering
   \includegraphics[width=0.9\hsize]{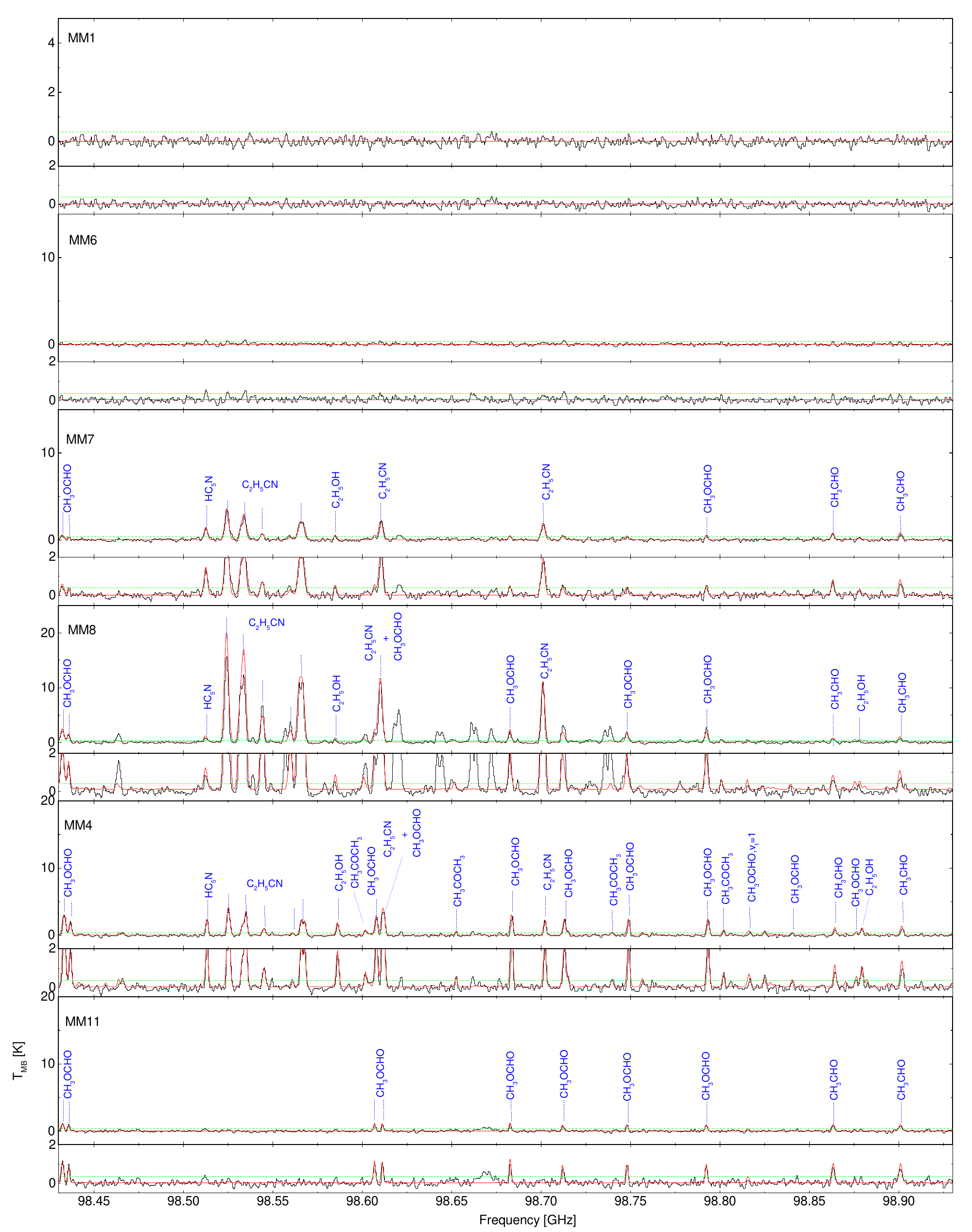}
      \caption{Beam-averaged spectra toward the MM1, MM6, MM7, MM8, MM4, and MM11. This segment covers frequency range from 98.43 to 98.93 GHz, the rest of other frequencies are showed in Figure~\ref{fig:spec_2}. The frequency scale is in terms of rest frequency. The black curves are the observed spectra, and the red curves indicate the simulated LTE spectra. The horizontal green dashed line indicates the 3 $\sigma$ noise level in each core. The small panel below each spectrum shows the transitions that have been zoomed in. 
              }
         \label{fig:survey_1}
   \end{figure*}

 We identified 16 species toward 6 dense cores of G9.62+0.19. Since each of the dense core is at different evolutionary stage, the emission of molecules have diverse characteristics in species and intensity. Figure \ref{fig:species} presents the species and transition numbers in each core. Figure \ref{fig:survey_1} and Figure \ref{fig:spec_2} display the full-band beam-averaged spectra at the six continuum peaks. Transitions with intensity higher than 3$\sigma$ are listed in Table \ref{tab:Transitions}. The rest frequencies, quantum numbers, the line strength (S$\mu^{2}$), and the upper-level energy ($E_{\rm u}$) of each transitions are given in Columns 1 to 4 in the Table. Estimated 1$\sigma$ RMS is 0.13~K, 0.12~K, 0.13~K, 0.14~K, 0.12~K, and 0.12~K for MM1, MM6, MM7, MM8, MM4, and MM11, respectively. The noise level $\sigma$ in each spectrum is computed in line-free channels.
 
 Below we discuss the line identification toward each core:
 
 \emph{MM1:} Simple molecular transitions CS J=2-1 and SO 3(2)-2(1), and a dense gas tracer HC$_{3}$N J=11-10 are detected in this region. The ionized gas tracer H$_{40\alpha}$ line is also detected in this core.
 
 \emph{MM4:} This is a line-rich source, in which a number of molecules have been detected. Line identification results showed that O-bearing COMs (e.g., CH$_{3}$OH~v=0, CH$_{3}$OH~v$_{t}$=1, CH$_{3}$OCHO, CH$_{3}$OCHO~v$_{t}$=1, CH$_{3}$OCH$_{3}$, CH$_{3}$COCH$_{3}$, and C$_{2}$H$_{5}$OH) are detected. The line strength of O-bearing COMs is higher toward the MM4 than the other cores. Additionally, a few lines of N-bearing species (e.g., HC$_{3}$N v$_{7}$=1, HC$_{3}$N v$_{7}$=2, HC$_{5}$N, and C$_{2}$H$_{5}$CN) are detected toward MM4. 
 
  \emph{MM6:} MM6 contains a few species such as H$_{2}$CO, CH$_{3}$OH with low $E_{\rm u}$ of 22 K, HC$_{3}$N~v=0, CS, SO, and $^{34}$SO, all of which cover only one transition. Other COMs are not detected. 
  
  \emph{MM7:} O-bearing species containing H$_{2}$CO, H$_{2}$CCO, CH$_{3}$OH, CH$_{3}$OH~v$_{t}$=1, CH$_{3}$CHO, CH$_{3}$OCHO, CH$_{3}$OCH$_{3}$, and C$_{2}$H$_{5}$OH, N-bearing molecules including HC$_{3}$N~v=0 and its $^{13}$C isotopologues, HC$_{3}$N in vibrational states (v$_{7}$=1, v$_{7}$=2), HC$_{5}$N, C$_{2}$H$_{5}$CN, along with S-bearing species (CS, SO, $^{34}$SO, SO$_{2}$) are identified in this core. CH$_{3}$OCHO~v$_{t}$=1 is not excited toward this core. 
  
  \emph{MM8:} MM8 is the richest in molecular species. Sixteen species are identified in this core including simple species such as H$_{2}$CO, CS, SO, and SO$_{2}$, N-, O-, and S-bearing COMs (e.g., CH$_{3}$OH, CH$_{3}$CHO, CH$_{3}$OCHO, CH$_{3}$OCH$_{3}$, CH$_{3}$COCH$_{3}$ C$_{2}$H$_{5}$OH, C$_{2}$H$_{5}$CN, and CH$_{3}$SH), with a few in torsionally excited state (CH$_{3}$OH~v$_{t}$=1 and CH$_{3}$OCHO~v$_{t}$=1). Cyanopolyynes species HC$_{3}$N in its vibrationally excited states v$_{7}$=1, v$_{7}$=2, v$_{7}$=3, and v$_{6}$=1 are observed in this core. We note that three transitions of HC$_{3}$N~v$_{7}$=3 at 100.88044, 101.02780, and 101.16989 GHz are blended with SO$_{2}$ at 100.87810 GHz, H$_{2}$CCO at 101.02447 GHz, and CH$_{3}$SH at 101.16830 GHz, respectively.
  
  \emph{MM11:} H$_{40\alpha}$, H$_{50\beta}$ and a few O-bearing COMs species such as CH$_{3}$OH, CH$_{3}$CHO, CH$_{3}$OCHO, CH$_{3}$OCH$_{3}$ and C$_{2}$H$_{5}$OH are detected toward this core. Cyanoacetylene and its $^{13}$C isotopologues in ground state and in vibrational state v$_{7}$=1 are also detected, while cyanoacetylene in higher vibrationally excited states and C$_{2}$H$_{5}$CN are not detected.  
  
  It is worth to mention that MM3 and MM9 have larger volume density (6.1$\times$10$^{6}$ cm$^{-3}$ and 15.9$\times$10$^{6}$ cm$^{-3}$) than other four starless cores (1.9$\times$10$^{6}$, 2.2$\times$10$^{6}$, 2.8$\times$10$^{6}$, and 0.6$\times$10$^{6}$ cm$^{-3}$ for MM2, MM5, MM10, and MM12, respectively; see more details in Table 1 of \citealt{Liu17}). Although MM3 and MM9 are not resolved in our 3~mm continuum emission, we extracted the spectra for the position of those cores which are shown in Figure~\ref{fig:B1-1}. It shows that species such as CS, SO, H$_{2}$CO, CH$_{3}$OH~v=0, HC$_{3}$N~v=0, HC$^{13}$CCN~v=0, HCC$^{13}$CN~v=0, HC$_{3}$N~v$_{7}$=1, and HC$_{5}$N are detected toward MM3; CS, SO, $^{34}$SO, H$_{2}$CO, CH$_{3}$OH~v=0, HC$_{3}$N~v=0, HC$^{13}$CCN~v=0, HCC$^{13}$CN~v=0, and CH$_{3}$CHO are detected at MM9.

\subsection{Spatial distributions}   
\begin{figure*}
\centering    
\subfigure[H$_{2}$CO; -6 to 22 km/s] {
 \label{fig:a}     
\includegraphics[width=0.6\columnwidth]{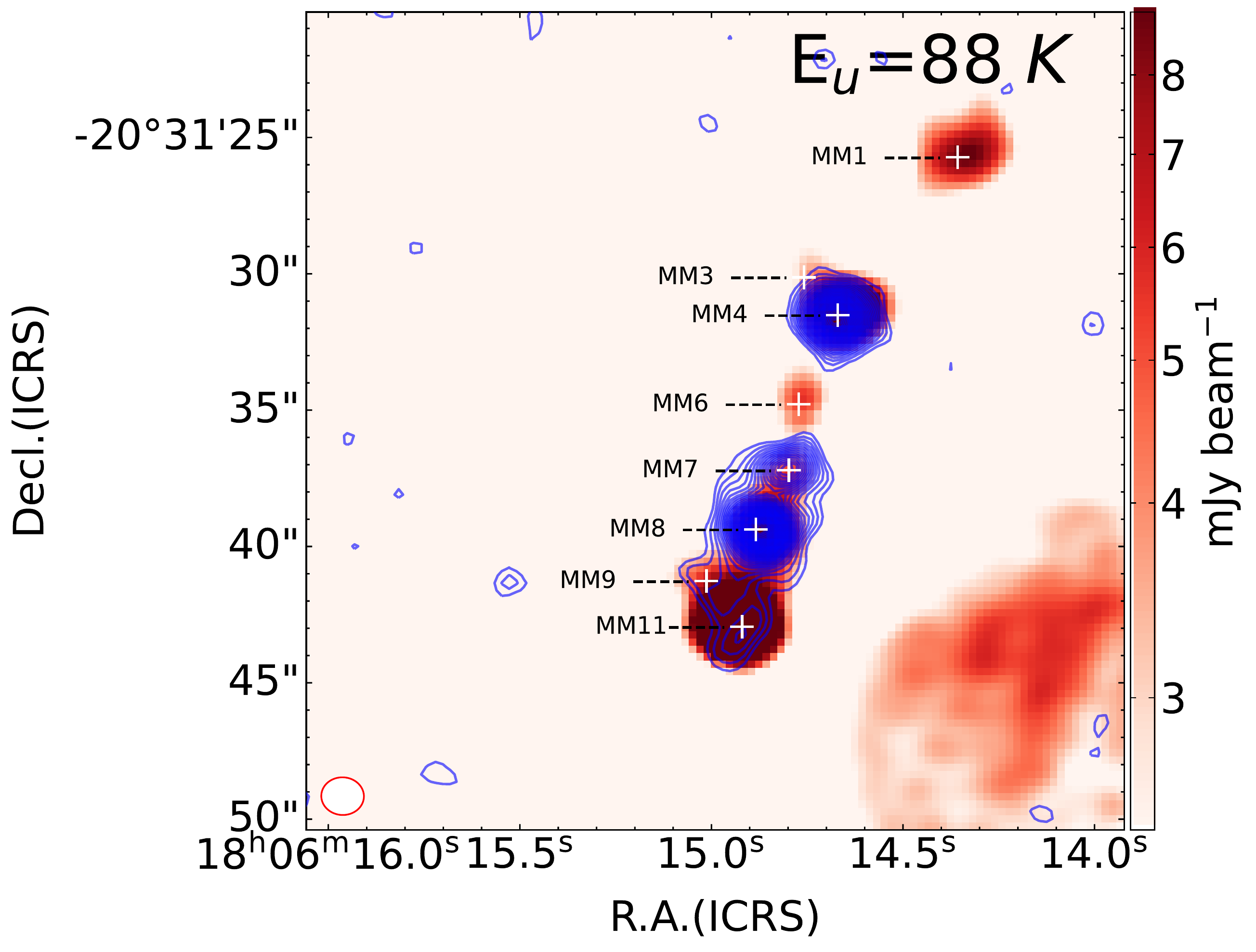}  
}     
\subfigure[H$_{2}$CCO; -1 to 12 km/s] { 
\label{fig:b}     
\includegraphics[width=0.6\columnwidth]{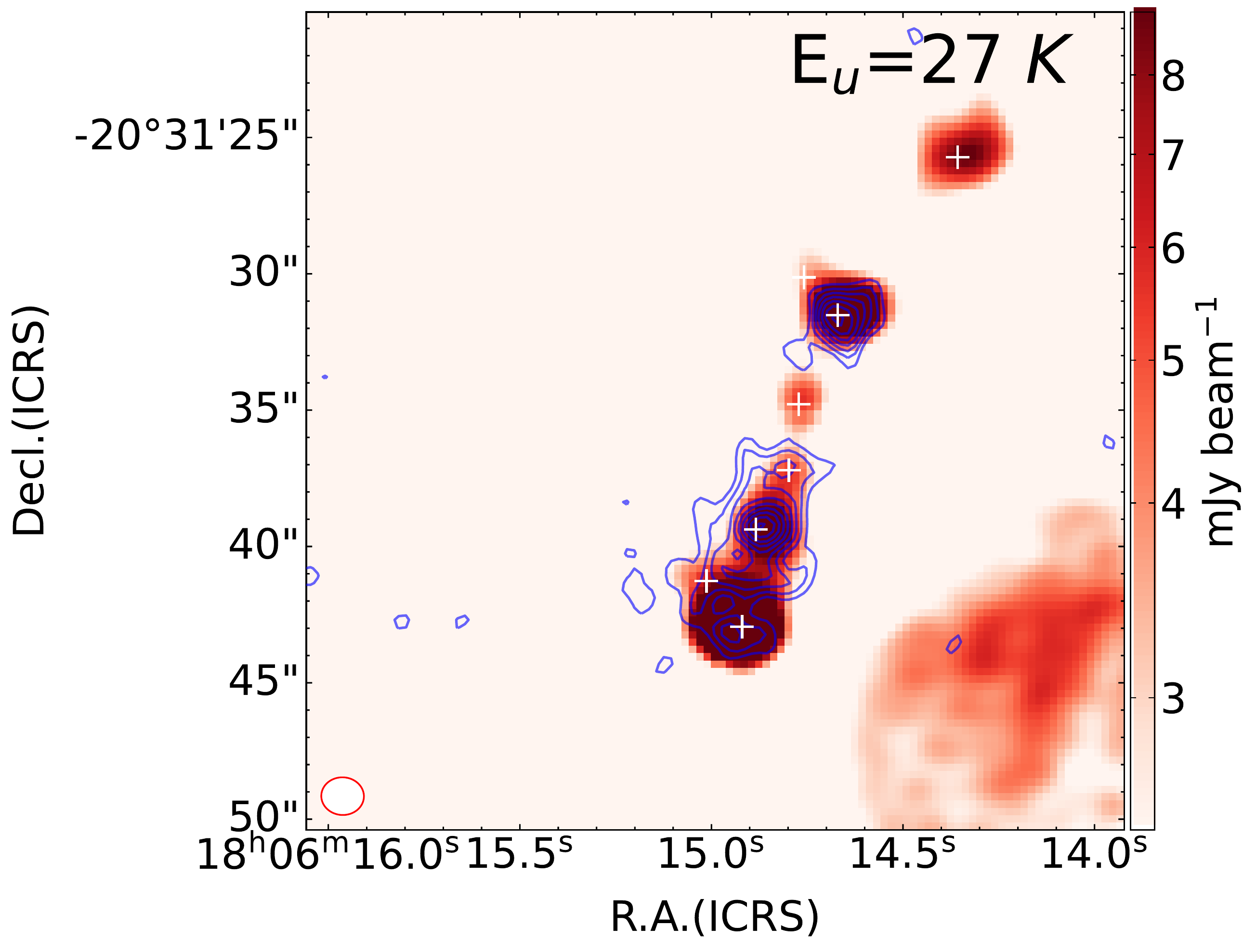}     
}    
\subfigure[CH$_{3}$OH~v$_{t}$=0; -2 to 11 km/s] { 
\label{fig:c}     
\includegraphics[width=0.6\columnwidth]{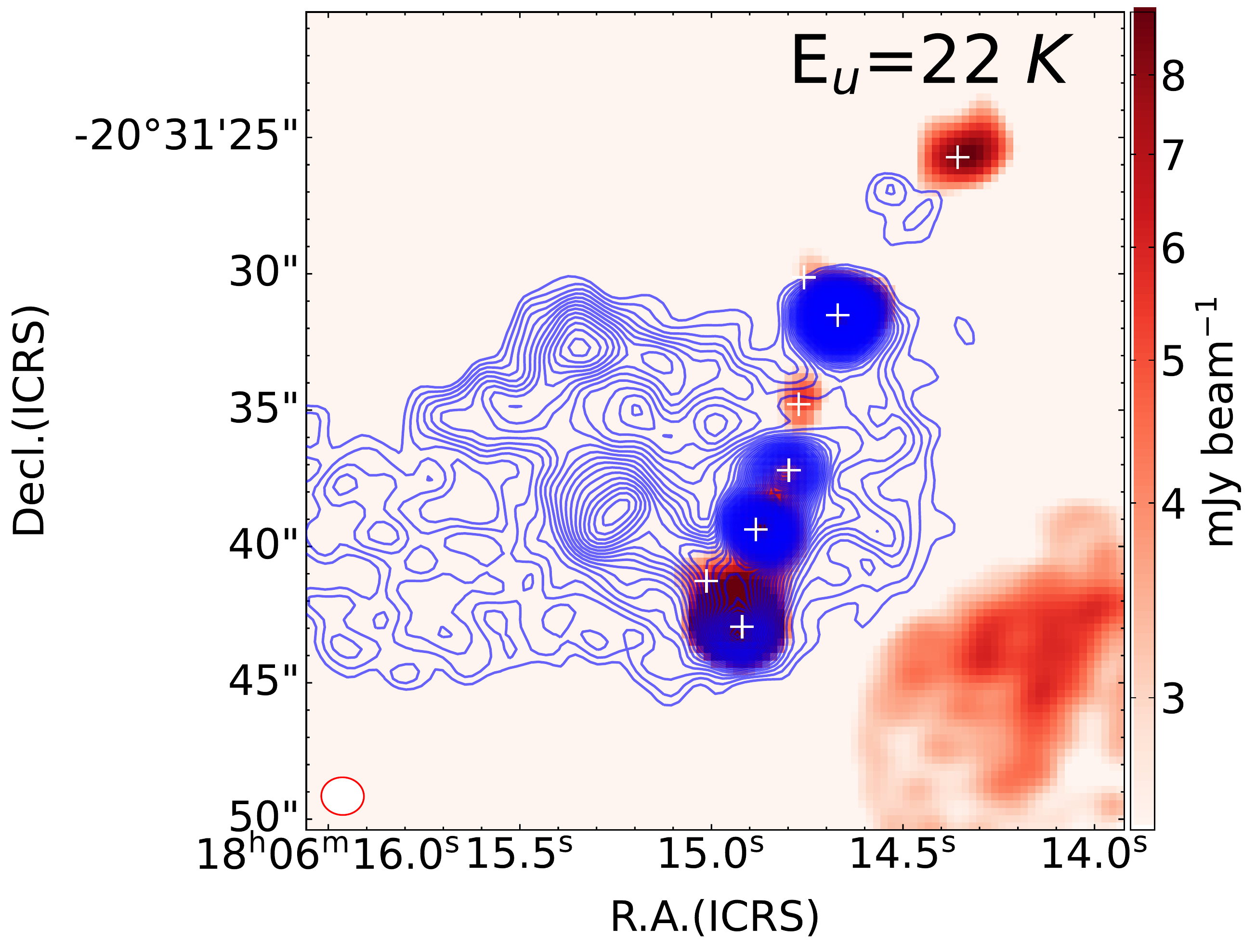}     
}   
\subfigure[CH$_{3}$OH~v$_{t}$=1; 0 to 11 km/s] { 
\label{fig:d}     
\includegraphics[width=0.6\columnwidth]{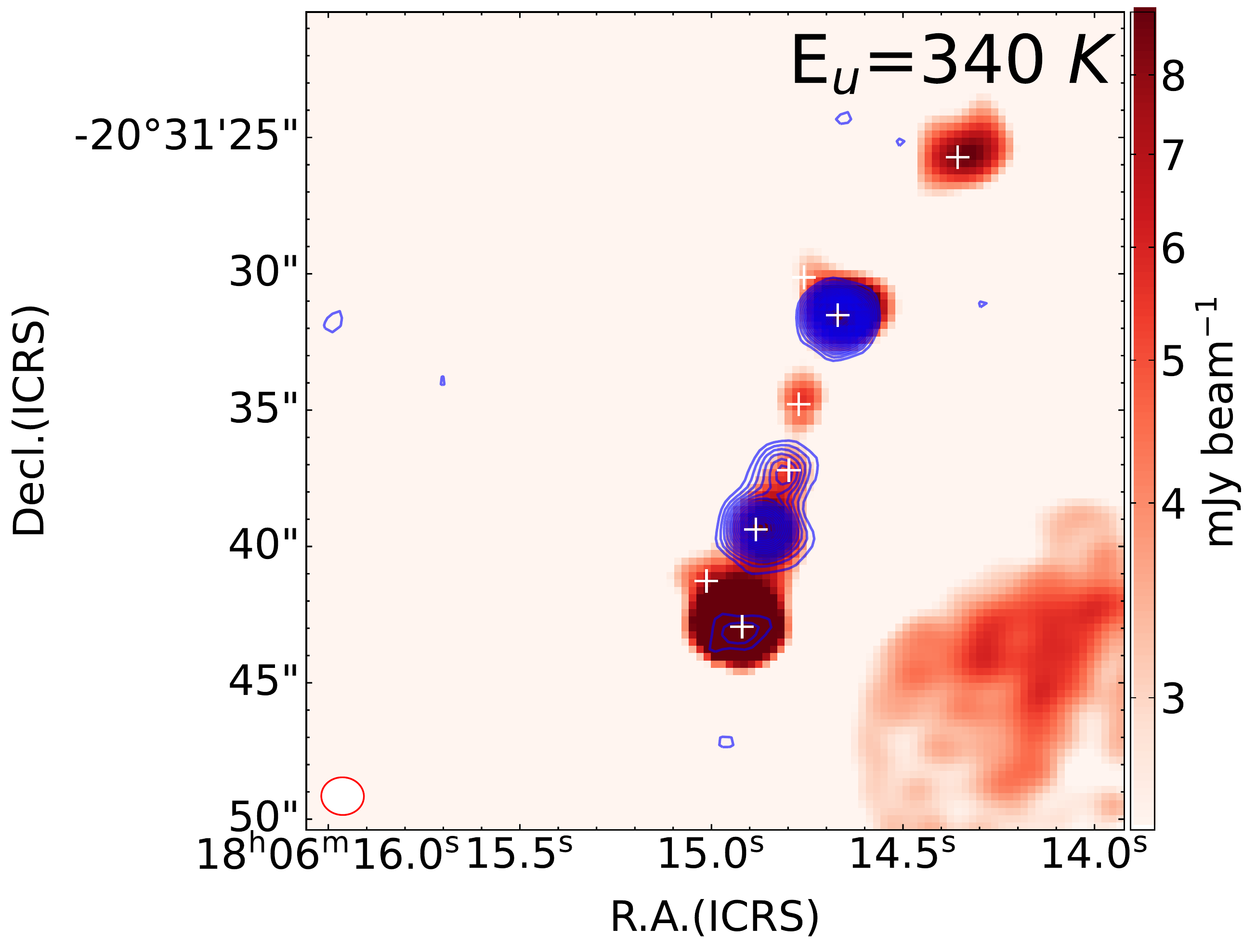}     
}   
\subfigure[CH$_{3}$CHO; -2 to 12 km/s] { 
\label{fig:e}     
\includegraphics[width=0.6\columnwidth]{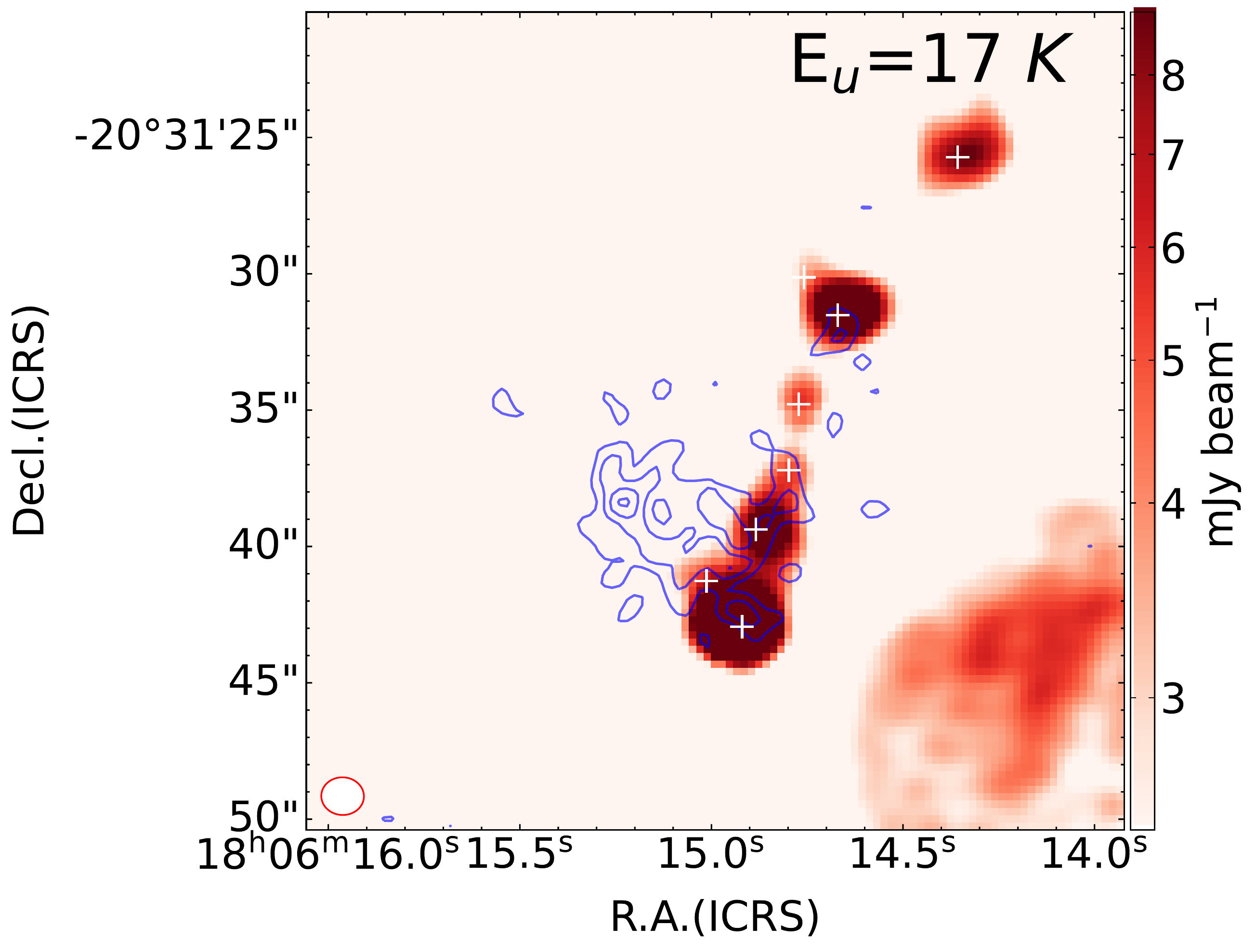}    
}   
\subfigure[CH$_{3}$OCHO; -7 to 12 km/s] { 
\label{fig:f}     
\includegraphics[width=0.6\columnwidth]{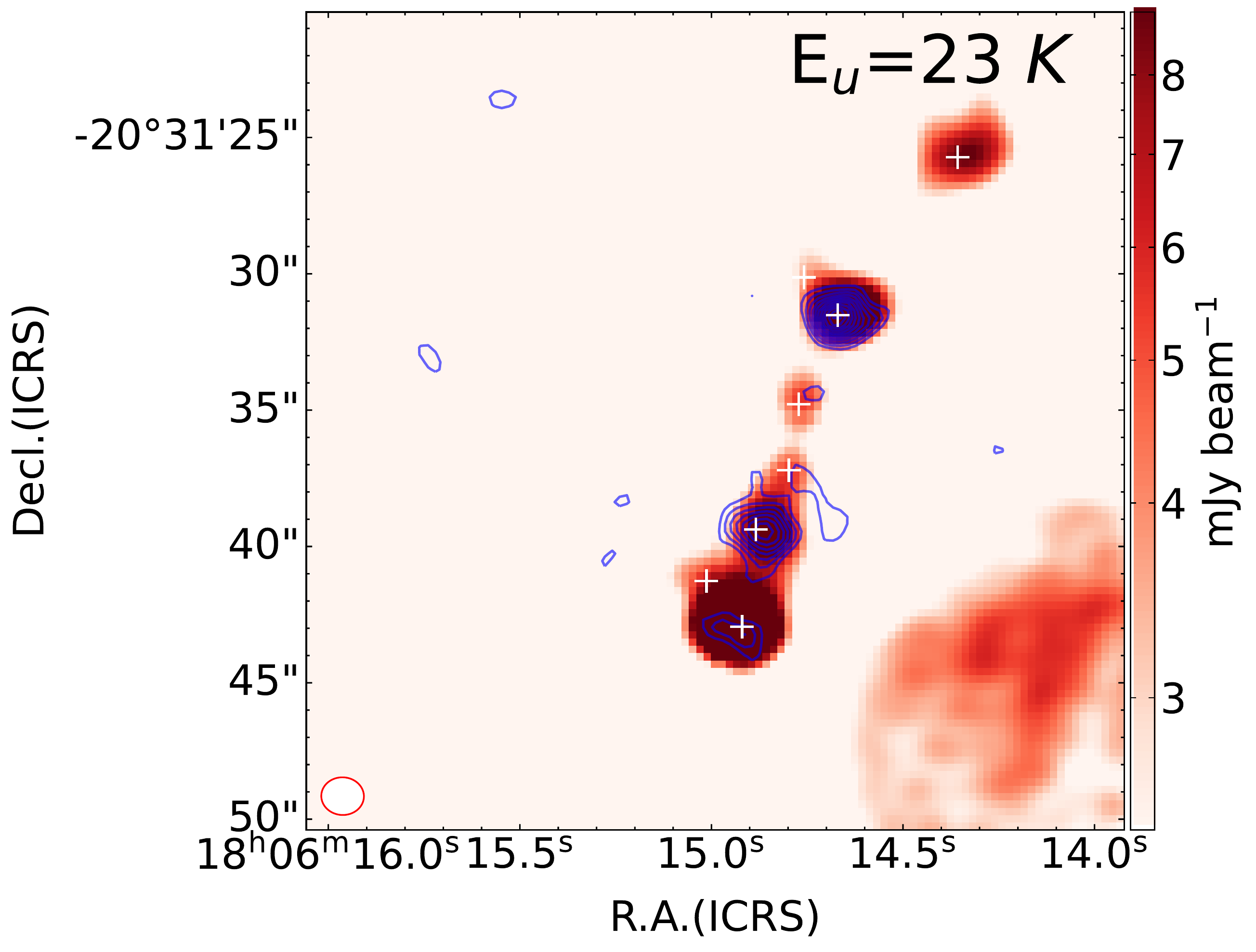} 
}   
\subfigure[CH$_{3}$OCHO~v$_{t}$=1; 0 to 12 km/s] { 
\label{fig:g}     
\includegraphics[width=0.6\columnwidth]{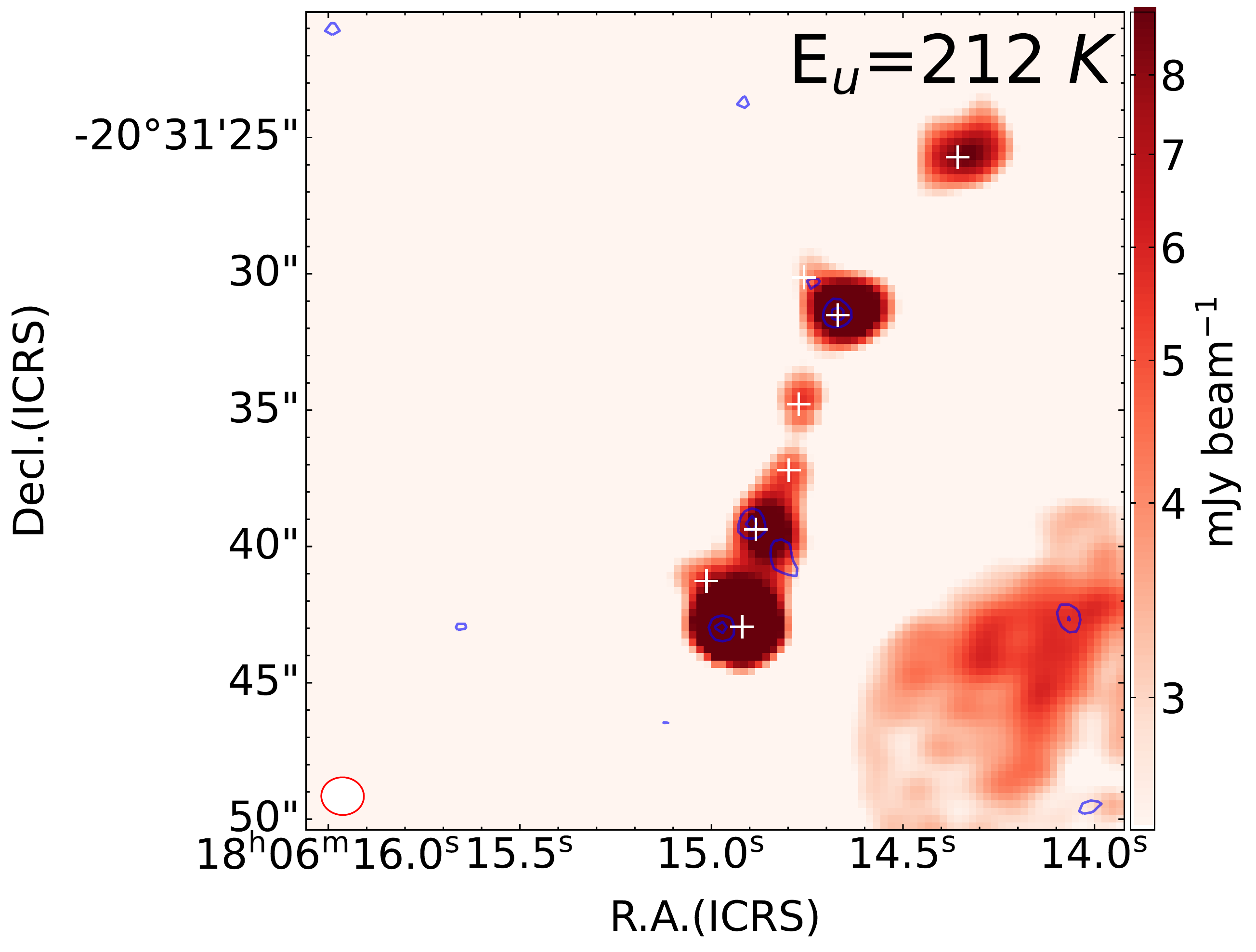}     
}   
\subfigure[C$_{2}$H$_{5}$OH; -1 to 11 km/s] { 
\label{fig:h}     
\includegraphics[width=0.6\columnwidth]{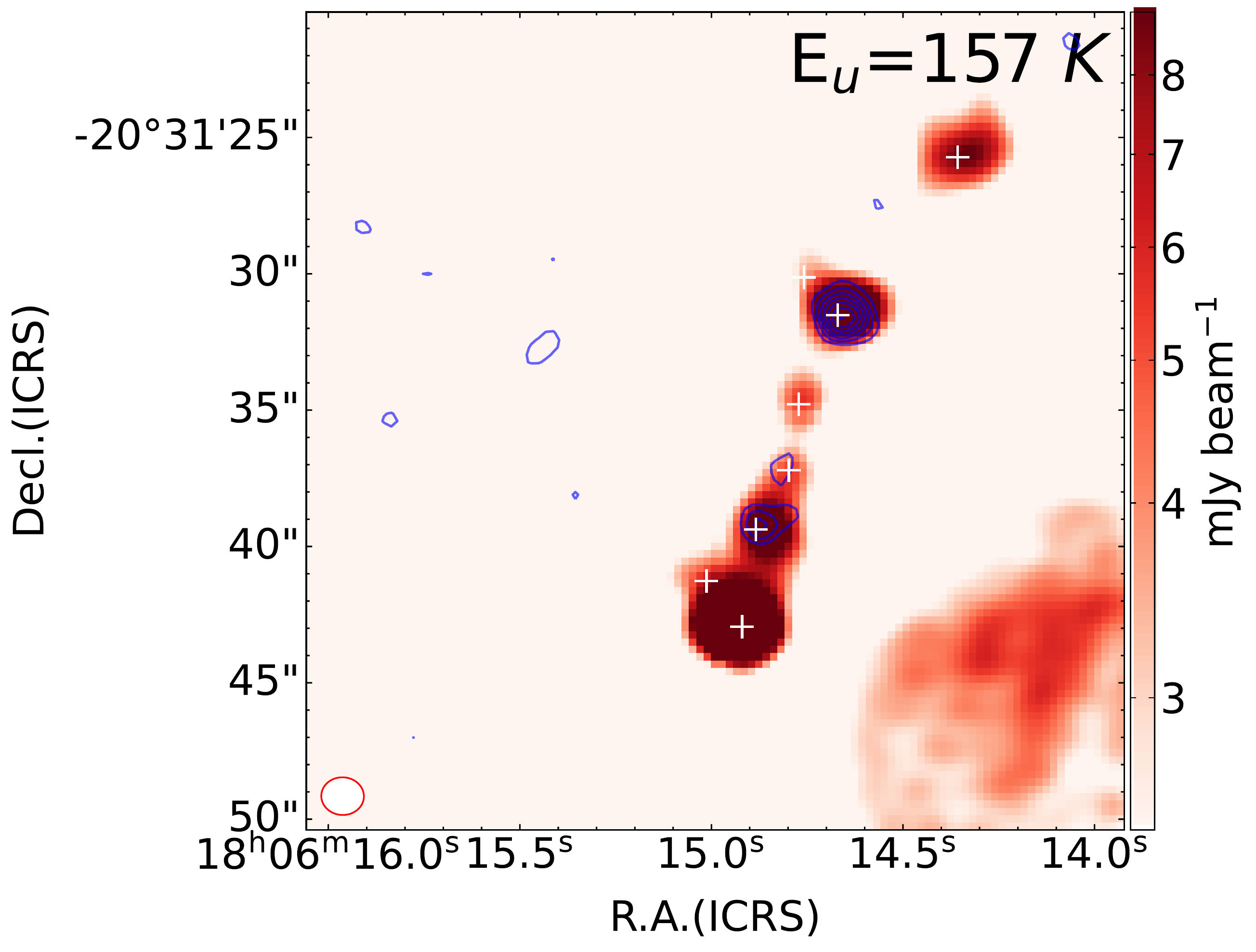}     
}   
\subfigure[CH$_{3}$OCH$_{3}$; -2 to 13 km/s] { 
\label{fig:i}     
\includegraphics[width=0.6\columnwidth]{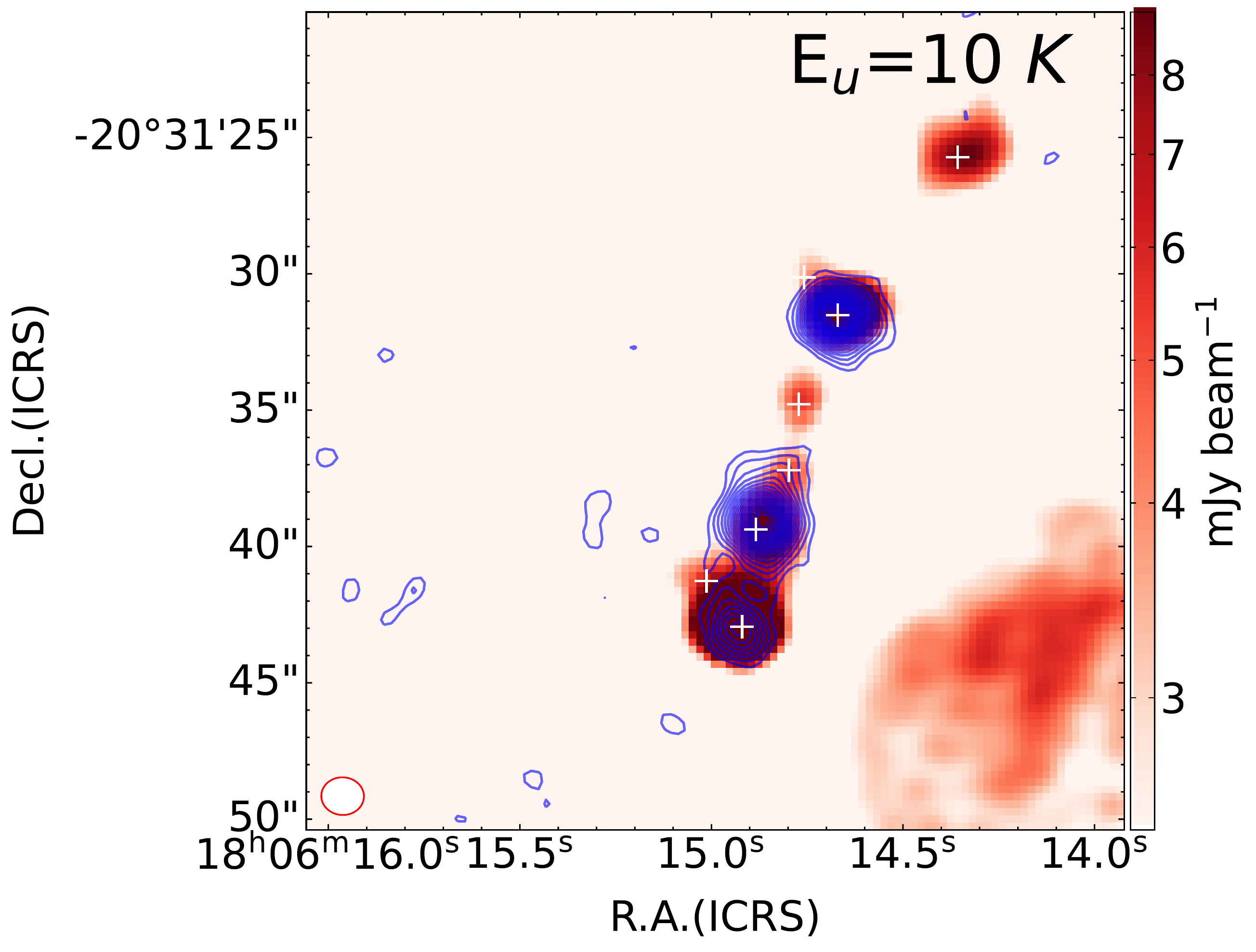}     
}   
\caption{Spatial distributions of O-bearing molecules observed by 12-m array. The orange background shows the 12-m continuum emission at 3 mm. The synthetic beam for continuum is indicated in the bottom left corner by the open ellipse. Continuum peaks MM1, MM3, MM4, MM6, MM7, MM8, MM9, and MM11 are denoted by the white cross symbols. The blue contour levels show the line emission starting from 5$\sigma$ to the maximum by 2$\sigma$, where $\sigma$ is the rms noise. The 1$\sigma$ of H$_{2}$CO, H$_{2}$CCO, CH$_{3}$OH~v$_{t}$=0, CH$_{3}$OH~v$_{t}$=1, CH$_{3}$CHO, CH$_{3}$OCHO, CH$_{3}$OCHO~v$_{t}$=1, C$_{2}$H$_{5}$OH, and CH$_{3}$OCH$_{3}$ are 1.17, 0.81, 0.81, 0.77, 0.81, 0.95, 0.77, 0.77, and 0.85~K~km/s, respectively. The value of upper-level energy $E_{u}$ for each species is shown in the right-upper corner of each panel.}
\label{fig:fig-O}     
\end{figure*}

\begin{figure*}
\centering    
\subfigure[HC$_{3}$N~v=0; -8 to 24 km/s] {
 \label{fig:a1}     
\includegraphics[width=0.6\columnwidth]{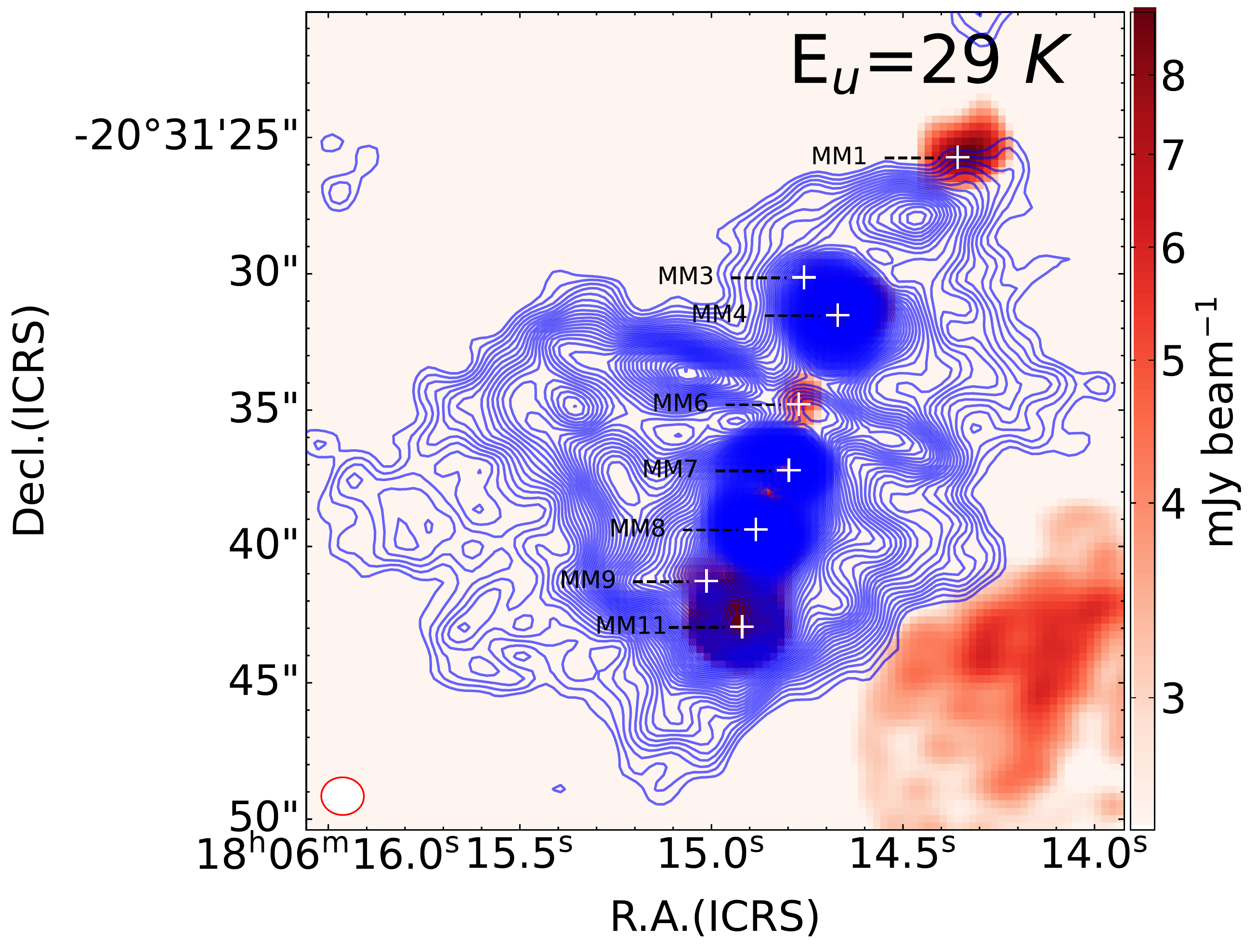}  
}     
\subfigure[HC$^{13}$CCN~v=0; -8 to 22 km/s] { 
\label{fig:b1}     
\includegraphics[width=0.6\columnwidth]{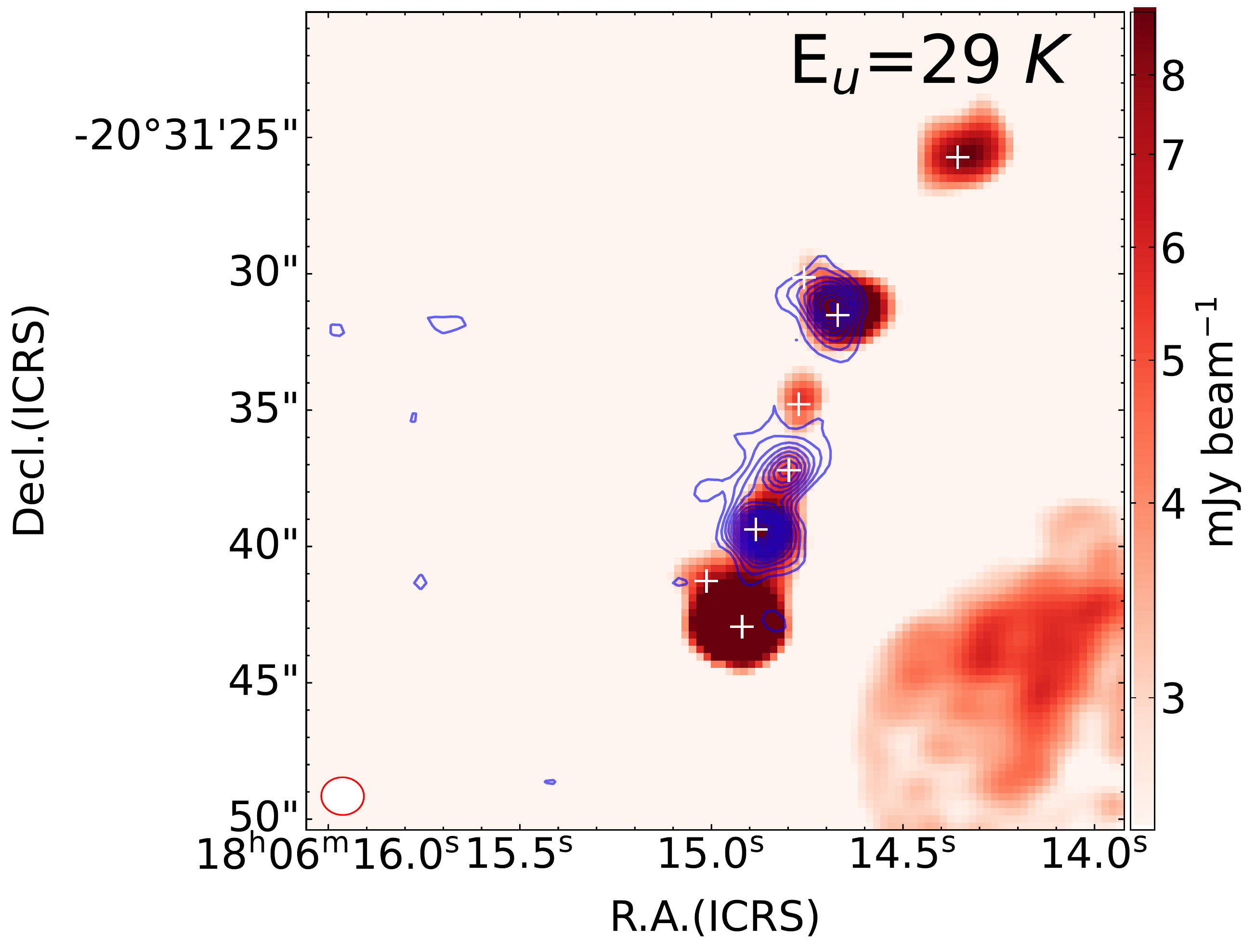}     
}    
\subfigure[HC$_{3}$N~v$_{7}$=1; -3 to 22 km/s] { 
\label{fig:c1}     
\includegraphics[width=0.6\columnwidth]{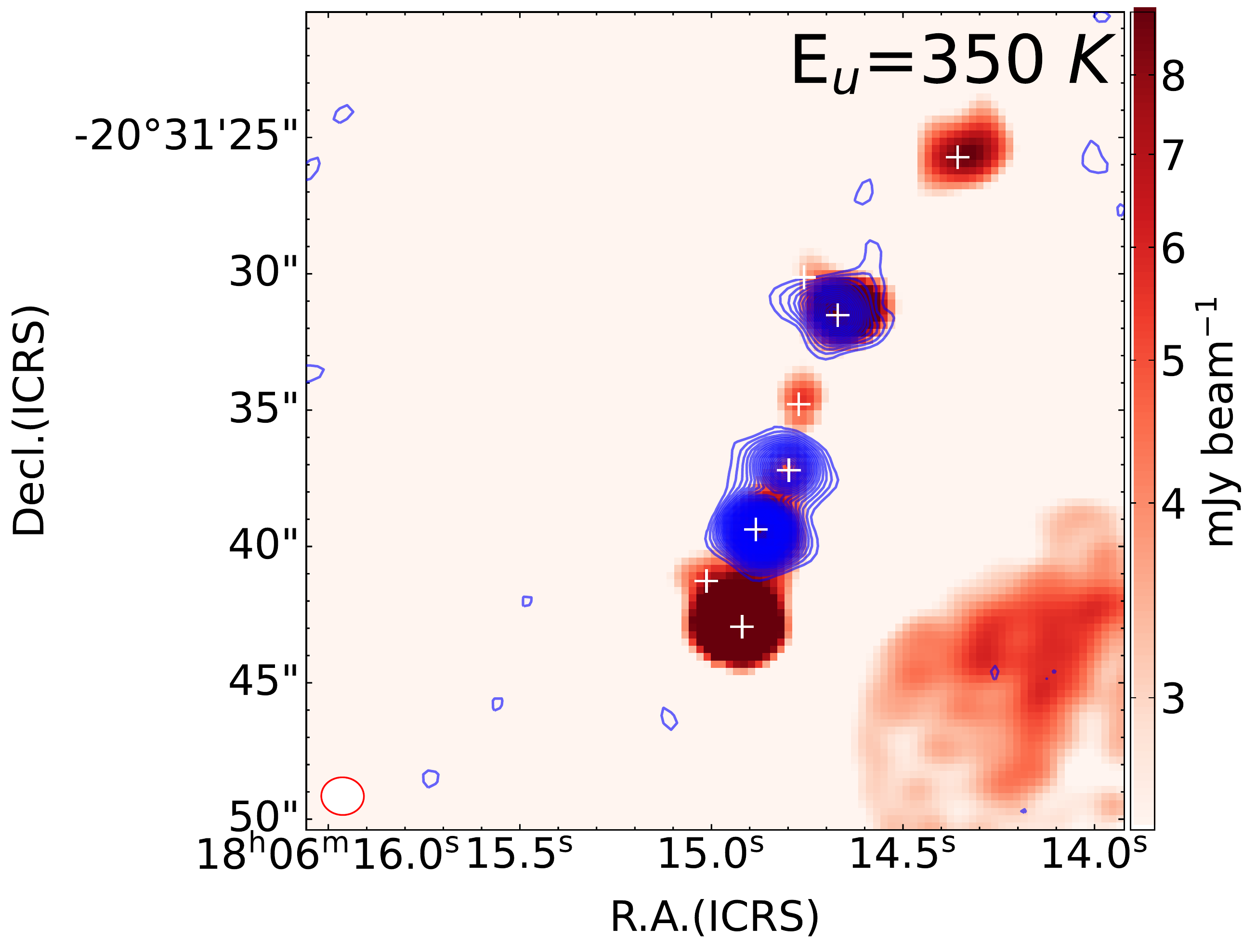}     
}   
\subfigure[HC$_{3}$N~v$_{6}$=1; -2 to 15 km/s] { 
\label{fig:d1}     
\includegraphics[width=0.6\columnwidth]{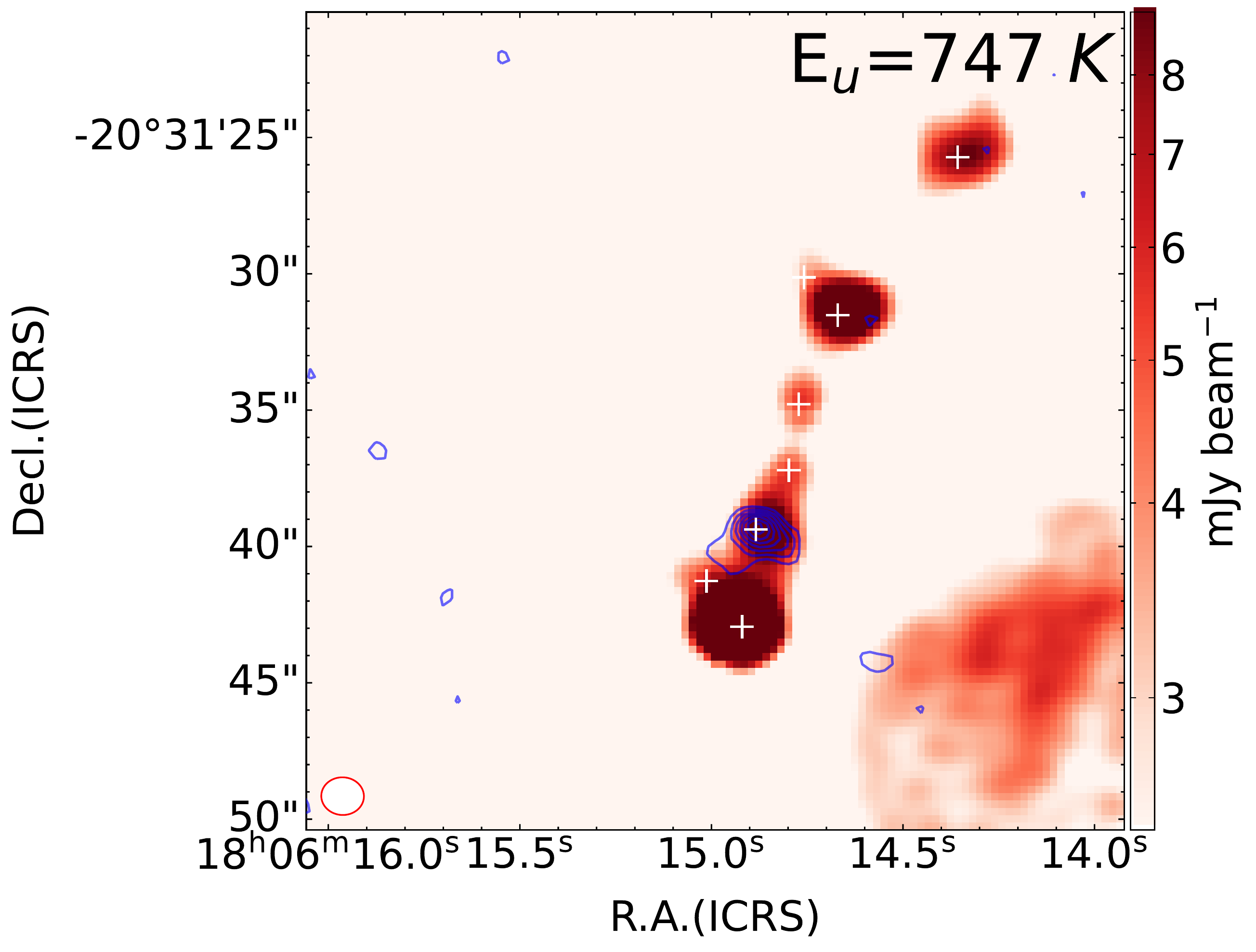}     
}   
\subfigure[HC$_{5}$N; -1 to 12 km/s] { 
\label{fig:e1}     
\includegraphics[width=0.6\columnwidth]{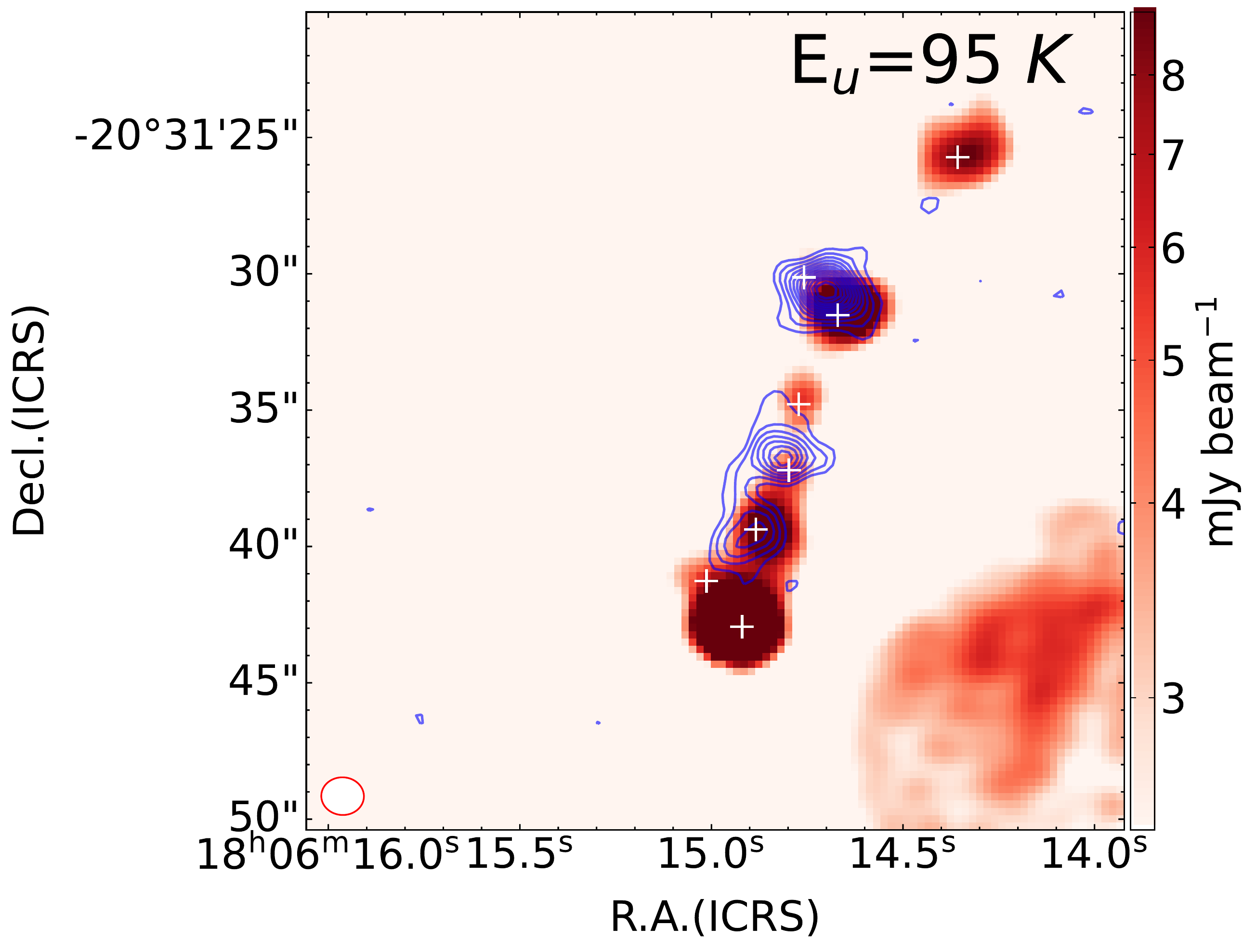}    
}   
\subfigure[C$_{2}$H$_{5}$CN; -4 to 17 km/s] { 
\label{fig:f1}     
\includegraphics[width=0.6\columnwidth]{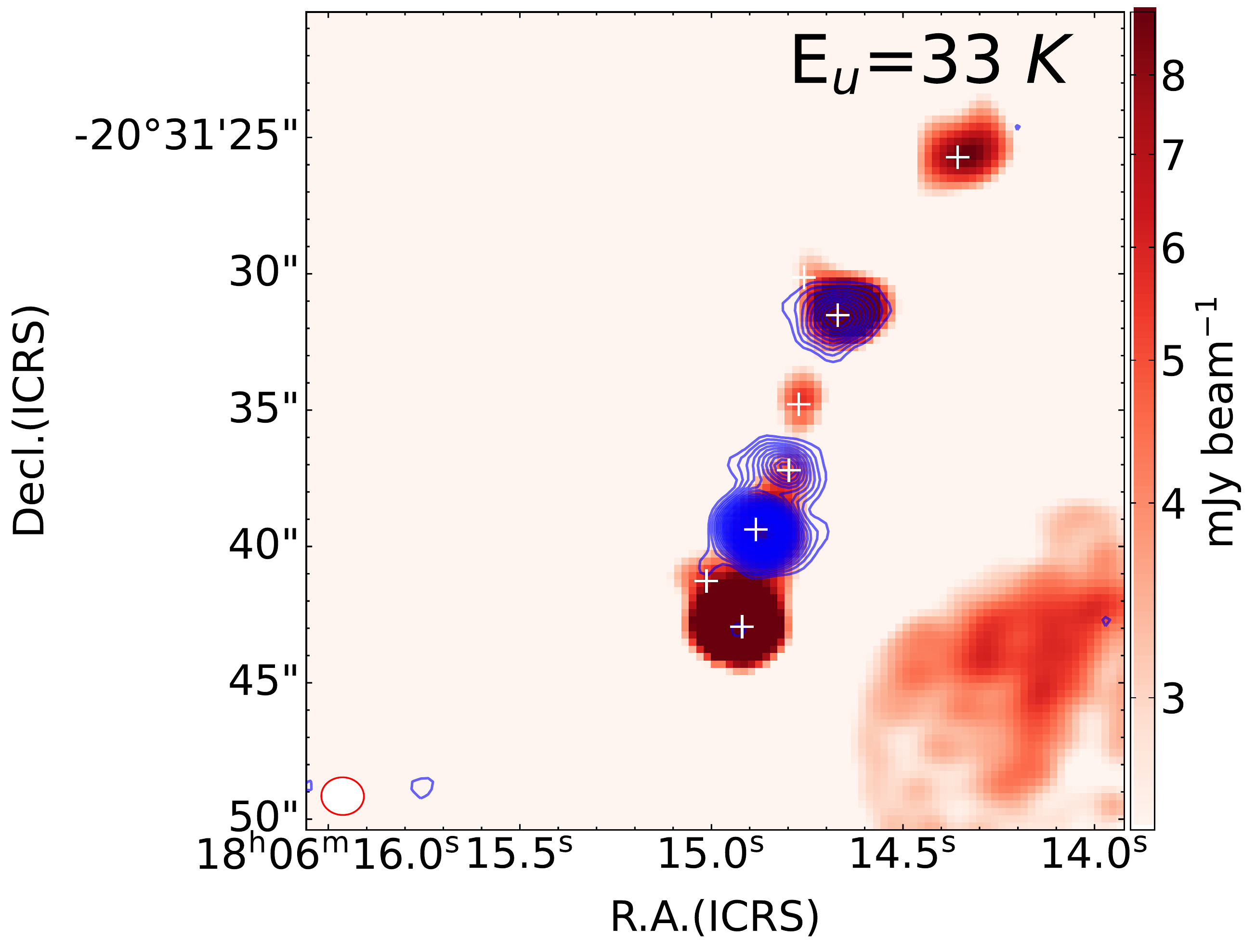} 
}   
\caption{ Spatial distributions of N-bearing molecules. See caption of Figure \ref{fig:fig-O} for more details. The 1$\sigma$ noise level of HC$_{3}$N~v=0, HC$^{13}$CCN~v=0, HC$_{3}$N~v$_{7}$=1, HC$_{3}$N~v$_{6}$=1, HC$_{5}$N, and  C$_{2}$H$_{5}$CN are 1.22, 1.17, 1.08, 0.92, 0.81, and 0.99~K~km/s, respectively. The value of upper-level energy $E_{u}$ for each species is shown in the right-upper corner of each panel.}    
\label{fig:fig-N}     
\end{figure*}

\begin{figure*} 
\centering    
\subfigure[CS~v=0; -22 to 48 km/s] {
 \label{fig:a2}     
\includegraphics[width=0.6\columnwidth]{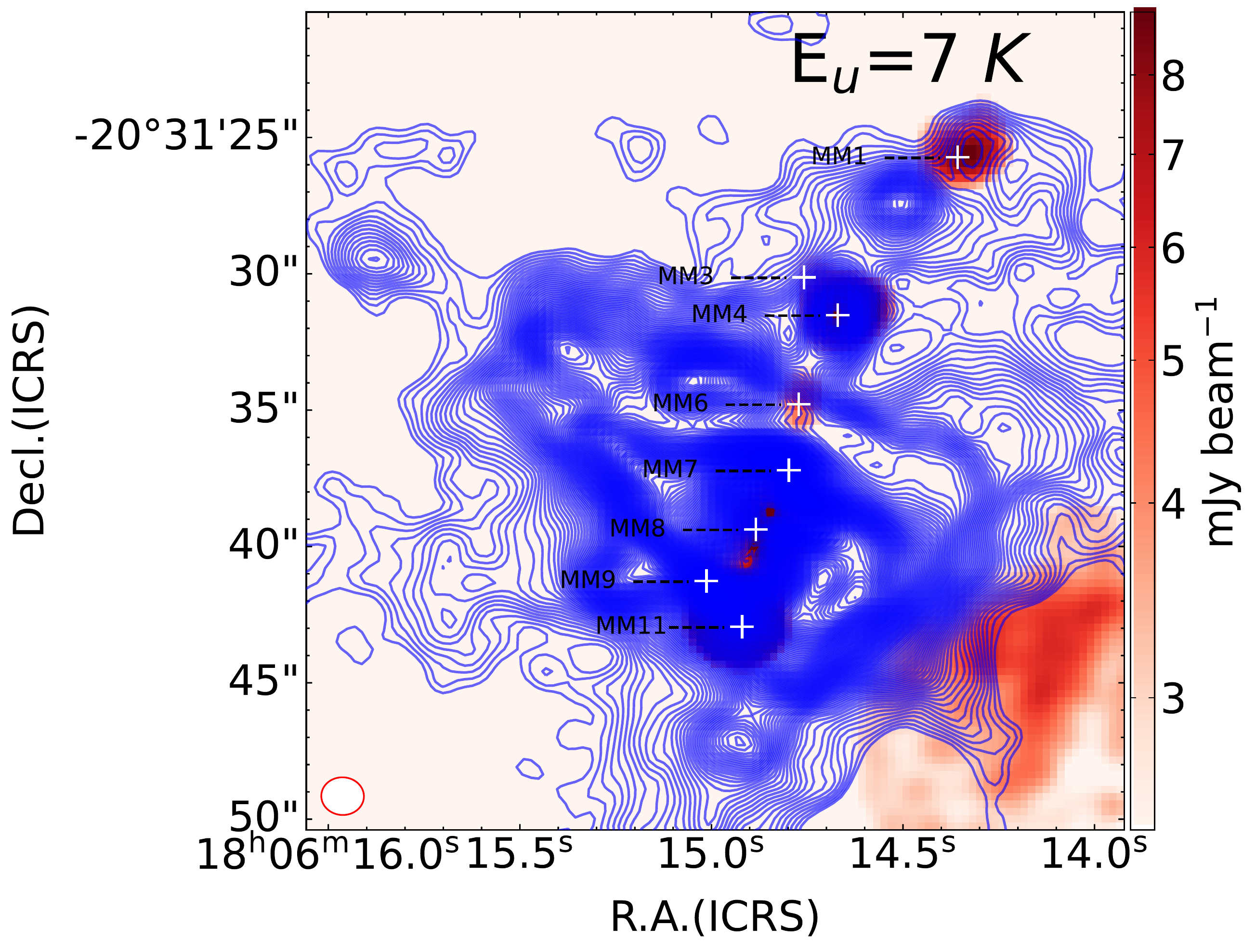}  
}     
\subfigure[SO (E$_{u}$=9 k); -38 to 50 km/s] { 
\label{fig:b2}     
\includegraphics[width=0.6\columnwidth]{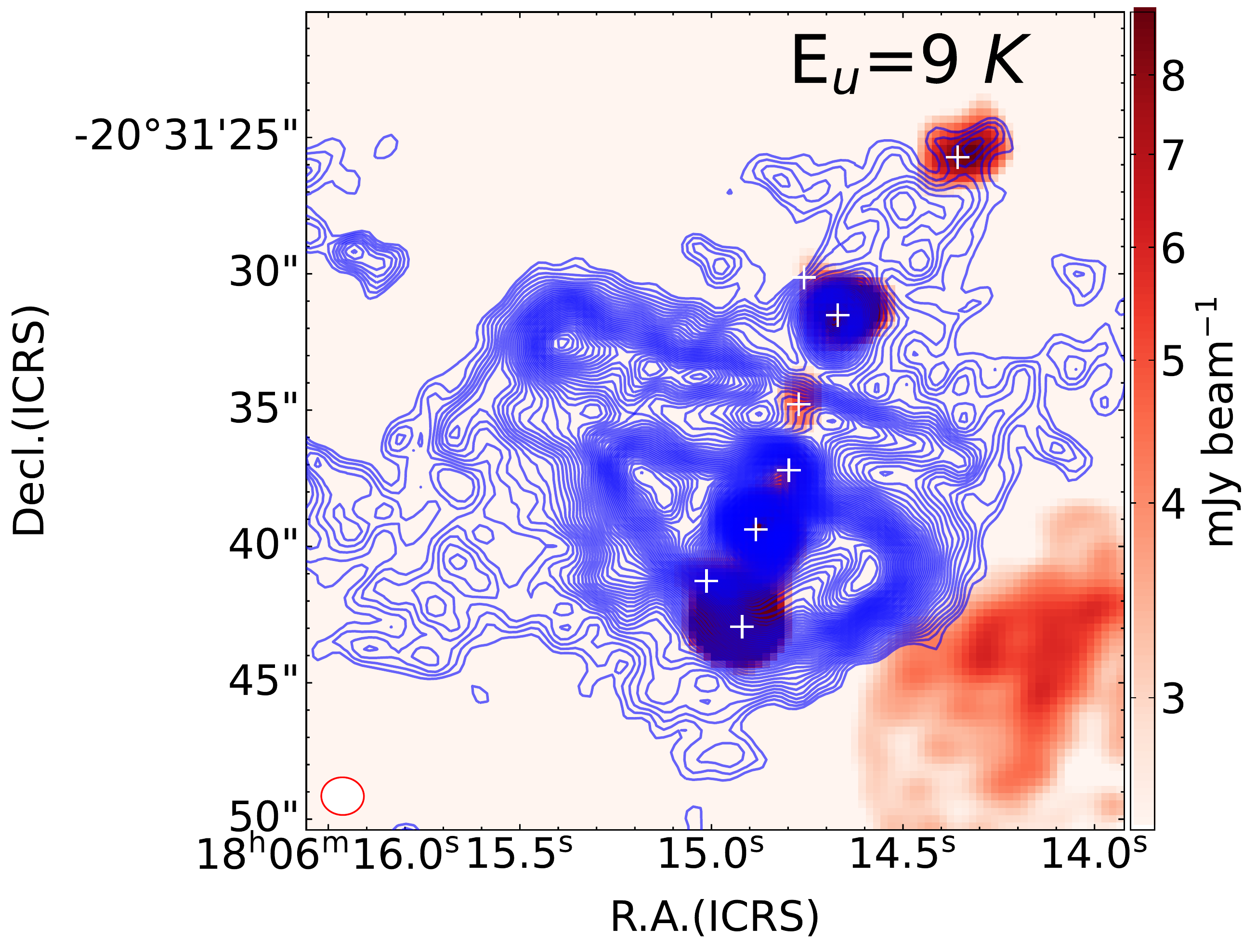}     
}    
\subfigure[SO (E$_{u}$=39 k); -5 to 12 km/s] { 
\label{fig:c2}     
\includegraphics[width=0.6\columnwidth]{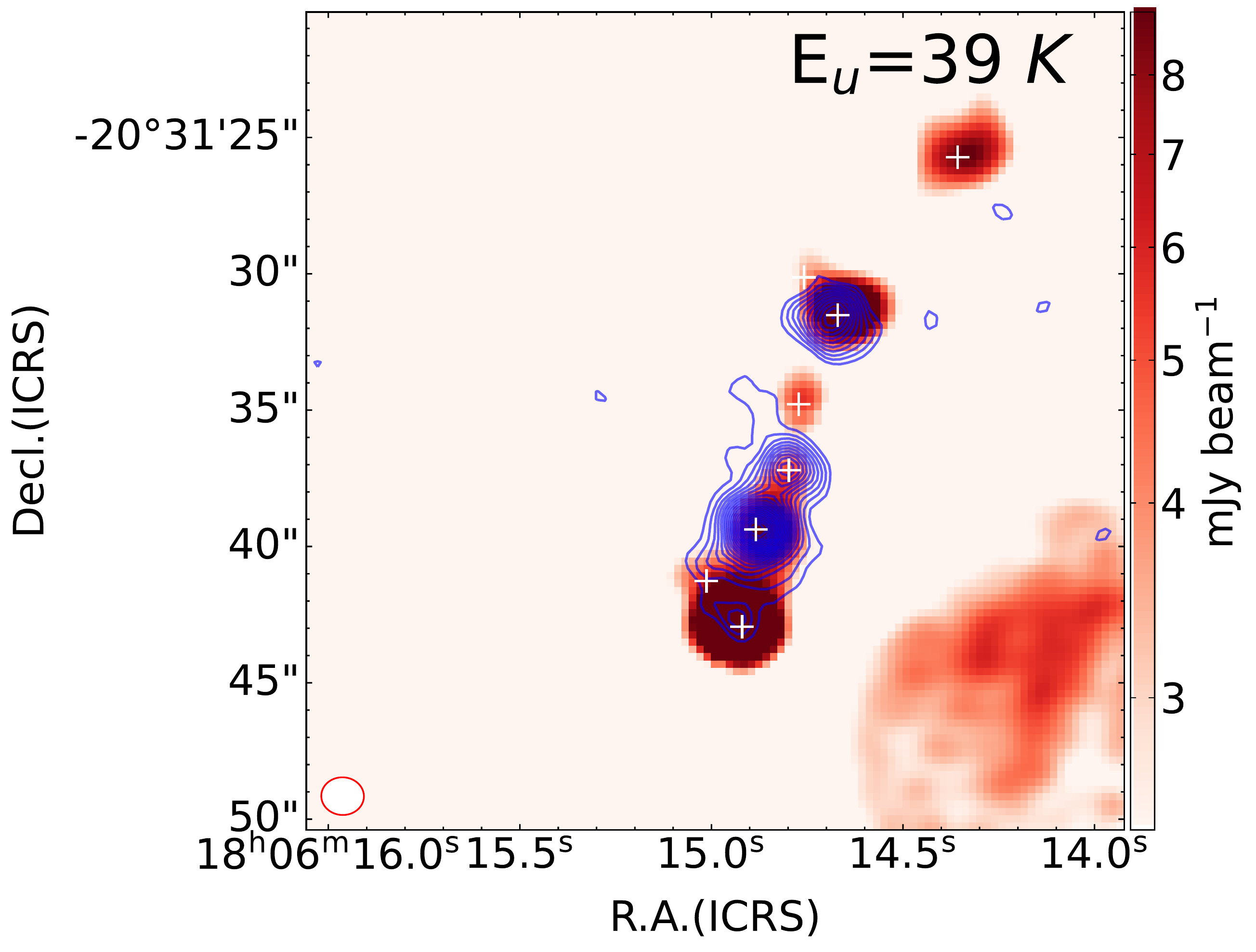}     
}   
\subfigure[$^{34}$SO; -5 to 12 km/s] { 
\label{fig:d2}     
\includegraphics[width=0.6\columnwidth]{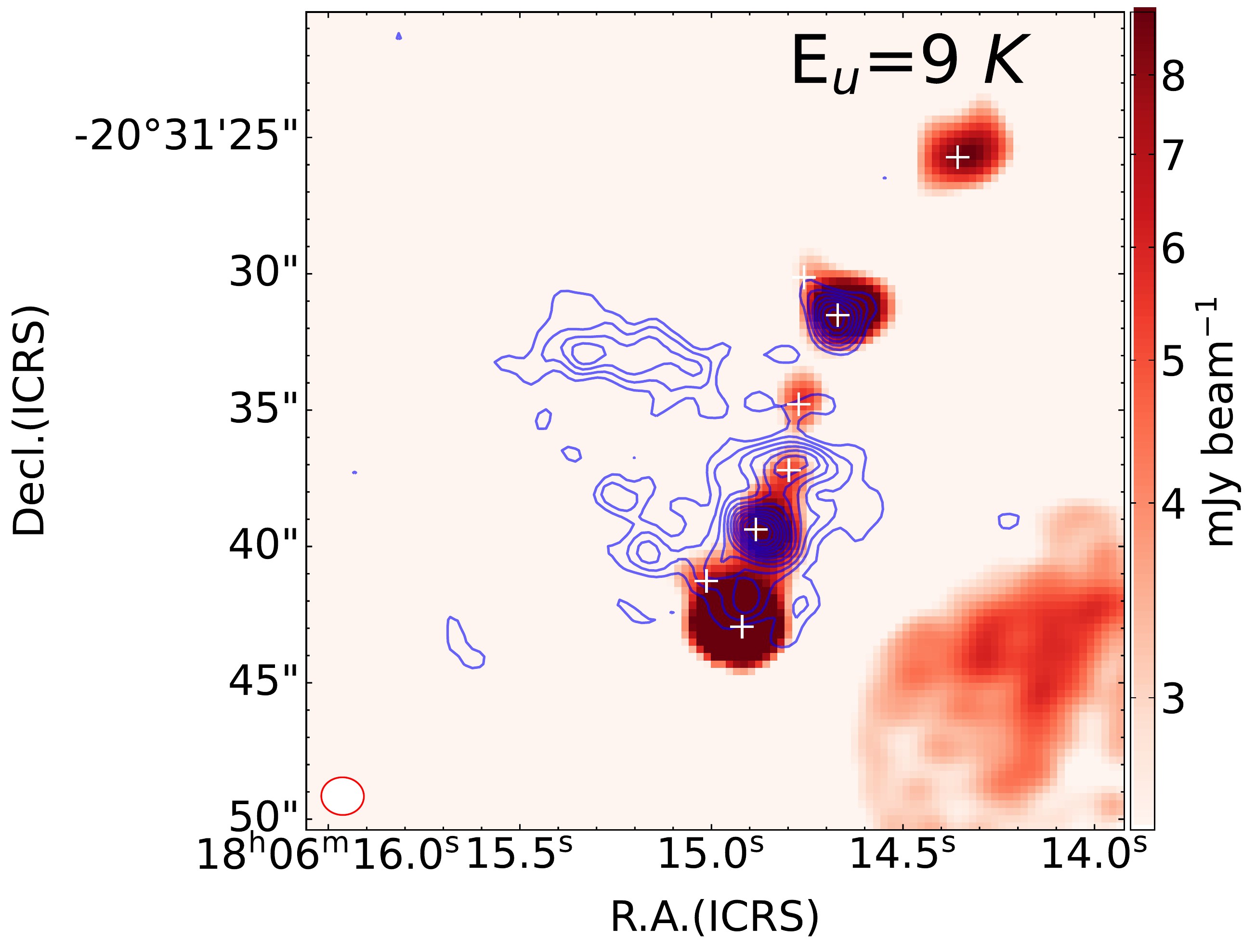}     
}   
\subfigure[SO$_{2}$; -5 to 14 km/s] { 
\label{fig:e2}     
\includegraphics[width=0.6\columnwidth]{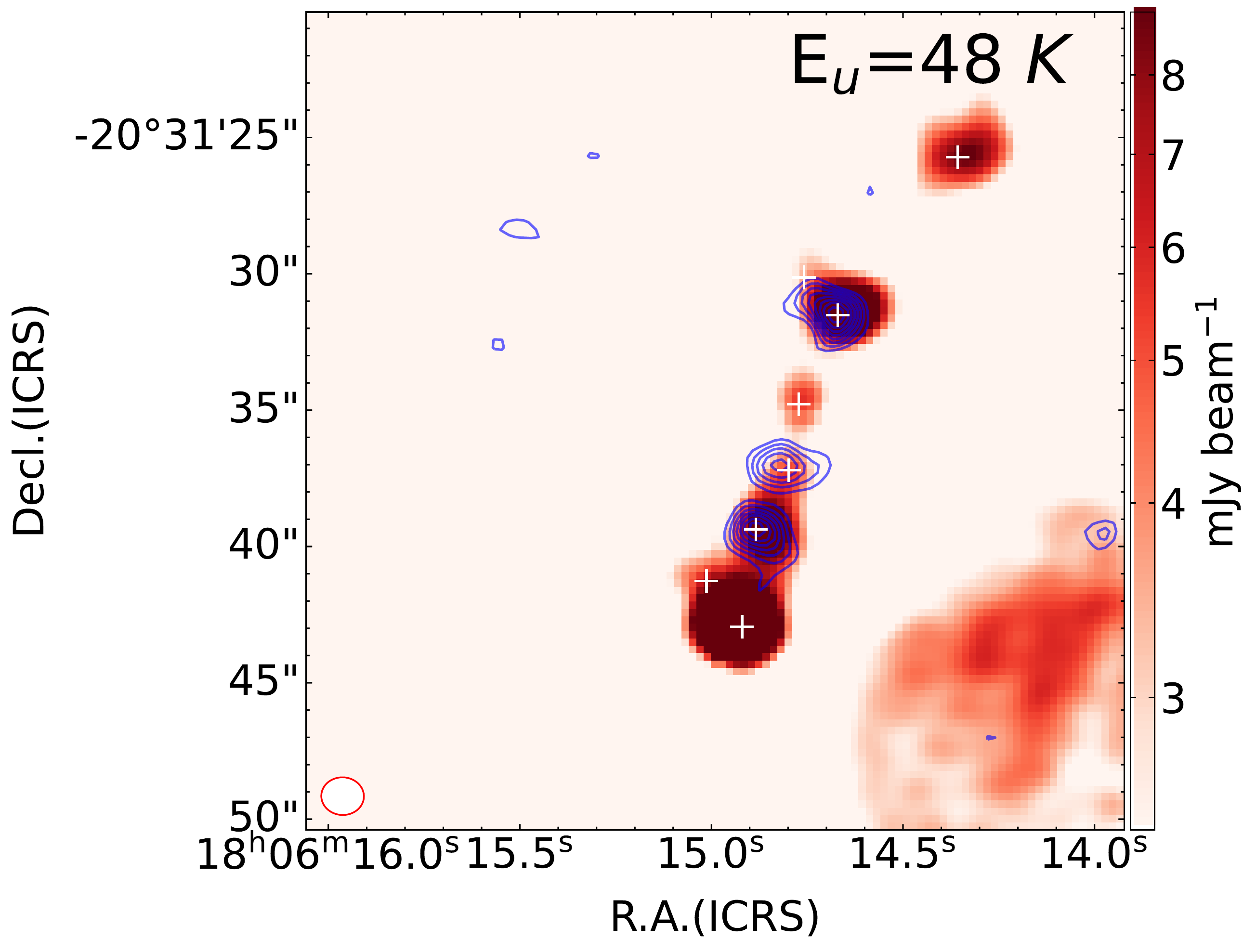}    
}   
\subfigure[SO$_{2}$~v$_{2}$=1; 0 to 13 km/s] { 
\label{fig:f2}     
\includegraphics[width=0.6\columnwidth]{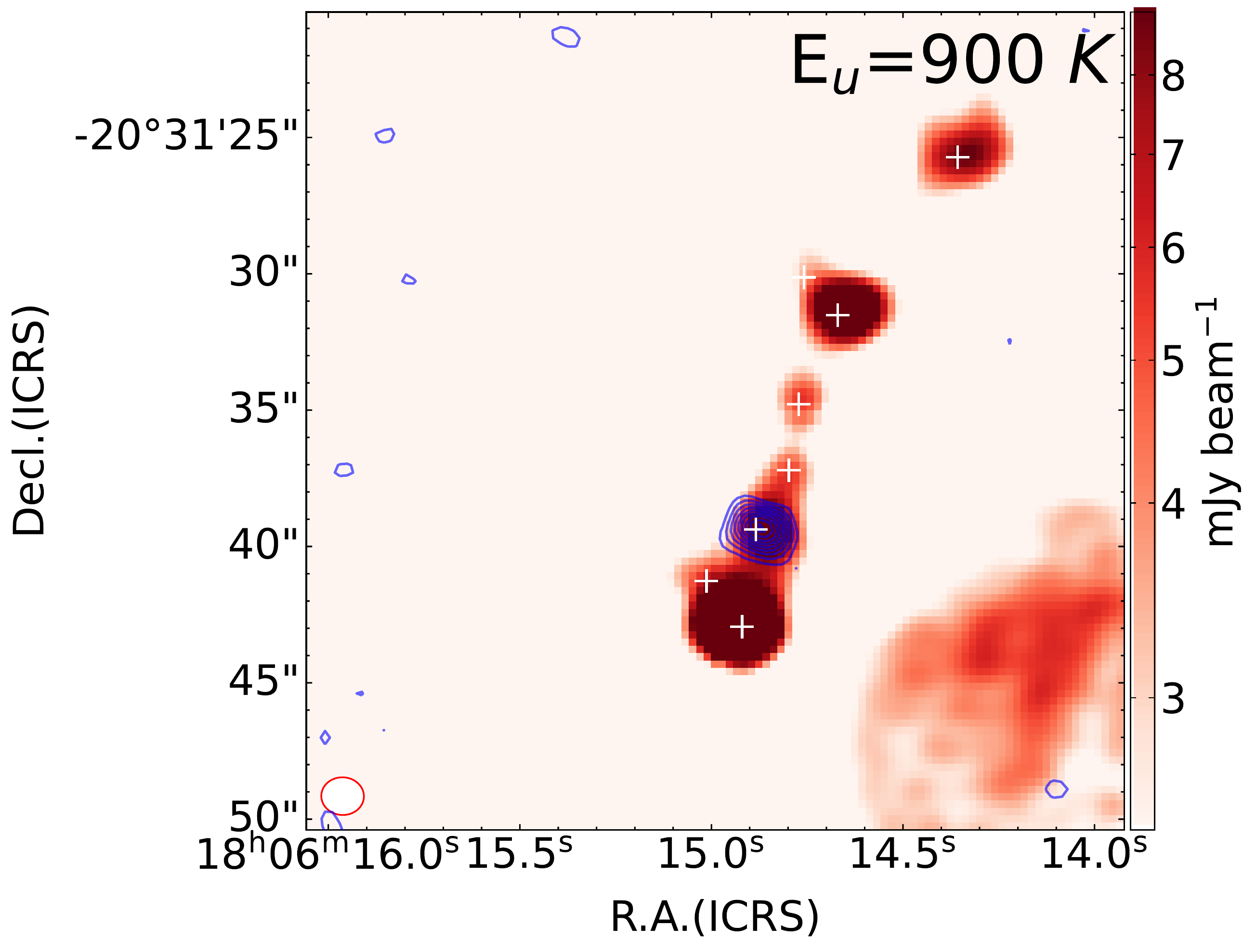} 
}   
\subfigure[CH$_{3}$SH; 0 to 14 km/s] { 
\label{fig:g2}     
\includegraphics[width=0.6\columnwidth]{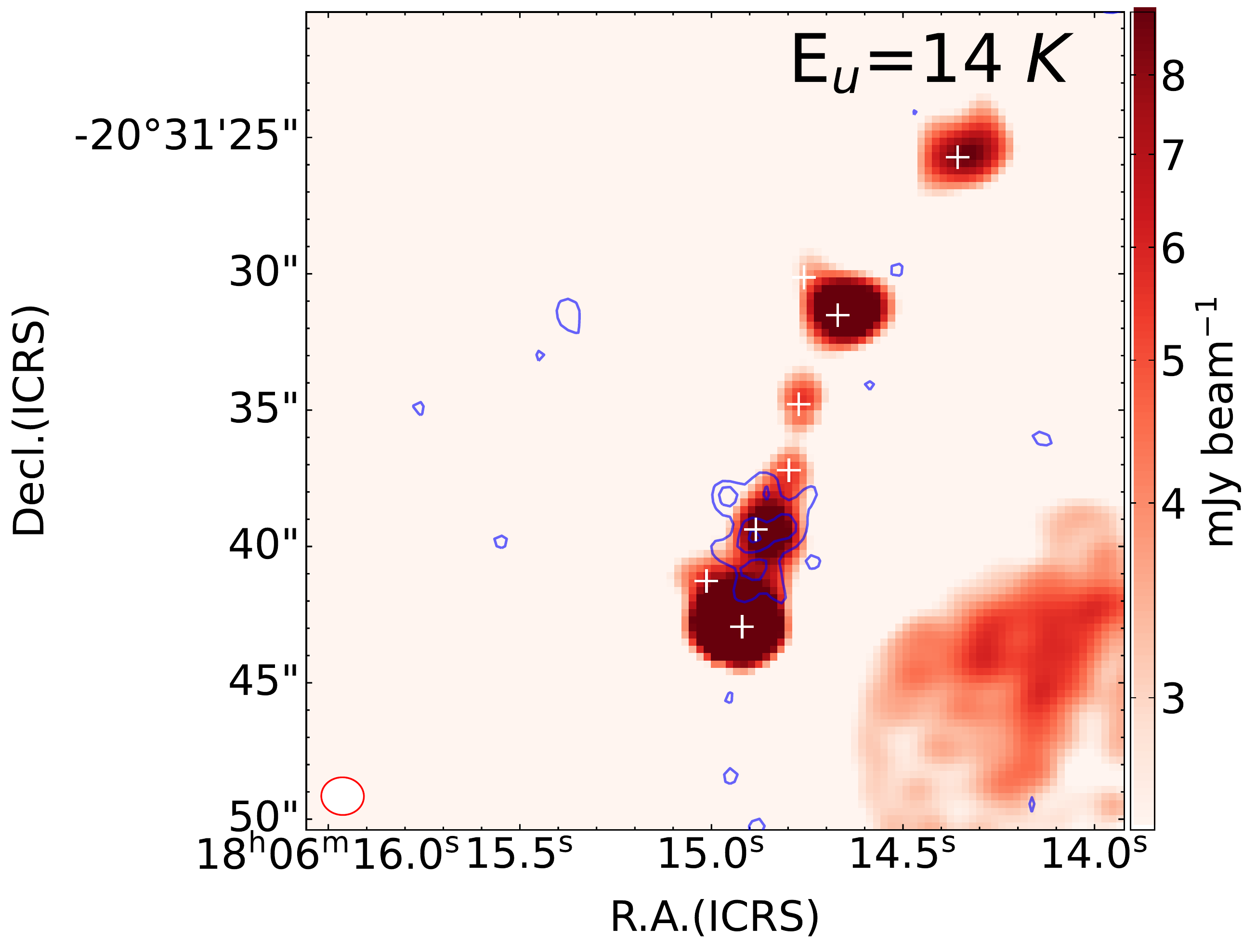} 
}   
\caption{ Spatial distributions of S-bearing molecules. See caption of Figure \ref{fig:fig-O} for more details. The 1$\sigma$ noise level of CS~v=0, SO ($E_{u}$=9 k), SO ($E_{u}$=39 k), $^{34}$SO, SO$_{2}$, SO$_{2}$~v$_{2}$=1, and CH$_{3}$SH are 1.77, 1.51, 0.92, 0.88, 0.95, 0.81, and 0.85~K~km/s, respectively. The value of upper-level energy $E_{u}$ for each species is shown in the right-upper corner of each panel.}    
\label{fig:fig-S}     
\end{figure*}

The integrated intensity maps of main transitions of Oxygen-, Nitrogen- and Sulphur-bearing species are shown in Figure \ref{fig:fig-O}$-$\ref{fig:fig-S}. Analyzed the spatial distribution, we note that on the whole, transitions with low upper-level energy are expected to trace more extended and diffuse gas structures than with high upper-level energy. Three main results for emissions of O-, N-, and S-bearing species are as follows: 
\begin{enumerate} 
\item Emission of CH$_{3}$OH~v=0 covers extended parts, while emission of its torsional state CH$_{3}$OH~v$_{t}$=1 with higher upper-level energy only distributes over the four dense cores MM4, MM7, MM8, and MM11. The spatial distributions of CH$_{3}$OCHO~v$_{t}$=1 and C$_{2}$H$_{5}$OH concentrate on MM4 and MM8, the other O-bearing molecules primarily originate from MM4, MM7, MM8, and MM11 (see Figure \ref{fig:fig-O}). Meanwhile, the emission from MM4 is stronger than that from MM8 for O-bearing species except for CH$_{3}$CHO. The gas distributions of CH$_{3}$CHO extended to the eastern region, which is likely to be affected by an outflow. 
\item Figure \ref{fig:fig-N}a shows the spatial distribution of HC$_{3}$N~v=0, covering a large scale and extending to MM1 and MM11. In addition, an outflow originating from MM6 along the northeast-southwest direction is also traced by HC$_{3}$N~v=0. Emission of its isotopologues HC$^{13}$CCN~v=0 peaks at MM4, MM7, and MM8, since isotopic molecules trace denser and more optically thin gas compare to its main species. HC$_{3}$N in vibrationally excited state v$_{7}$=1 peaks at MM4, MM7, and MM8, and higher energy v$_{6}$=1 (747~K) only peaks toward the MM8, implying MM8 to have very warm environment. Nitrogen-bearing molecule C$_{2}$H$_{5}$CN has similar emission with those of HC$^{13}$CCN~v=0 and HC$_{3}$N~v$_{7}$=1. Unlike the above Oxygen-bearing species that mainly peak at the position of MM4, the strongest emission of these N-bearing species is mostly located at MM8 with the exception of HC$_{5}$N for which the strongest emission is associated with MM4. Moreover, there is offset between its main emission peak and continuum emission peak MM4. A similar scenario is also seen for MM7 (see Figure \ref{fig:fig-N}e).  
\item Similar to N-bearing species, S-bearing molecules mainly come from the MM4, MM7, and MM8. Observation shows that SO with low $E_{\rm u}$=9 K and CS trace very extended regions. Furthermore, the collimated northeast-southwest outflow that comes from MM6 can be seen in CS and SO ($E_{\rm u}$ = 9 K) emission. The gas emission of SO$_{2}$~v$_{2}$=1 with E$_{\rm u}$ of 900~K is very compact and peaks at MM8. Emission of CH$_{3}$SH in low energy (E$_{\rm u}$ = 14~K) also comes from MM8, but it is not extremely compact.  
\end{enumerate}

\section{LTE Analyses}
 In this section, we investigate the physical and chemical properties of six cores with sufficient number of molecular lines to derive their physical parameters (rotational temperature, column density, and abundance) under LTE condition. All LTE modeled spectra are shown in Figure~\ref{fig:survey_1} and Figure~\ref{fig:spec_2}.
 
\subsection{LTE calculation}
 We model the observed spectra using the XCLASS suite. The software contains an interface for the model optimizer package MAGIX (Modeling and Analysis Generic Interface for eXternal numerical codes; \citealt{Moller13}), which helps to optimize the fit, finds the best solutions of parameters, and provides corresponding error estimates by using different optimization algorithms or algorithm chain. In this work, three algorithm including Genetic, Levenberg–Marquardt, and Markov chain Monte Carlo were adopted. The main modeling parameters of XCLASS for each molecule are the source size, rotational temperature, column density, full width at half maximum of the observed line, and the velocity offset of the line with respect to the systemic velocity of the object. For the compact components, the source sizes is obtained by two-dimensional Gaussian fits to the line images. For extended components, the beam filling factor is always $\sim$~1, we adopt very large source size. While for some species, the source size is comparable to the beam size, thus the source size is set as free parameters. The velocity offsets and line widths are derived from Gaussian fits to the spectral lines. The rest of the parameters are set as free parameters, mainly the rotational temperature and column density. Then MAGIX is used to minimizes the $\chi^{2}$ in the given parameter space.

\subsection{Rotational Temperatures and Column Densities}
 The molecular transitions that are unaffected by blending effect and cover a large upper-level energy $E_{u}$ range, are most suitable for constraining the physical parameters. In our line survey SPW 7$-$8, molecular species with more than three detected transitions along with wide $E_{u}$ range are used to evaluate the rotational temperature $T_{\rm rot}$ and source-averaged total column density $N_{\rm T}$. Table \ref{tab:parameterstable} summarizes the best fitting results. All lines above 3$\sigma$ noise level in the spectra are analyzed. For each detected line, the optical depth ($\tau^{\rm line}$) computed with XCLASS is listed in Table \ref{tab:Transitions}. The CS (2-1) transition in MM7 and MM8 has the highest optical depths of  1.32 and 1.17, respectively. The optical depth of the other molecular transitions are lower than 1. The fitted rotational temperature and column density are discussed below.

\emph{Ketene (H$_{2}$CCO):}  Five unblended transitions of H$_{2}$CCO spanning $E_{\rm u}$ of 15$-$133~K are detected in MM7, MM8, and MM4. Toward the MM11, only three transitions are well detected with high S/N ($>3$). The rotational temperatures of 72$\pm$19~K, 102$\pm$10~K, 102$\pm$30~K and 84$\pm$23~K, and column densities of (1.6$\pm$0.7) $\times$ 10$^{15}$~cm$^{-2}$, (5.5$\pm$0.1) $\times$ 10$^{15}$~cm$^{-2}$, (5.0$\pm$1.7) $\times$ 10$^{15}$~cm$^{-2}$, and (1.6$\pm$0.7) $\times$ 10$^{15}$~cm$^{-2}$ toward the MM7, MM8, MM4, and MM11, respectively, are derived by fitting those uncontaminated transitions (see Table~\ref{tab:parameterstable}).  

\emph{Methanol (CH$_{3}$OH):} Methanol is detected in ground state toward MM6, MM7, MM8, MM4, and MM11, and in torsionally excited state toward MM7, MM8, MM4, and MM11. The number of transitions over three are identified only toward MM4, allowing the determination of parameters. Transitions of CH$_{3}$OH~v=0 cover a $E_{\rm u}$ range from 22 to 889~K and  CH$_{3}$OH~v$_{t}$=1 cover $E_{\rm u}$ range from 340 to 902~K. The excitation temperature and column density are 159$\pm$8~K and (7.5$\pm$0.3) $\times$ 10$^{17}$~cm$^{-2}$ for ground state, 204$\pm$20~K and (9.2$\pm$0.7) $\times$ 10$^{17}$~cm$^{-2}$ for torsional state.

\emph{Methyl Formate (CH$_{3}$OCHO):} Methyl Formate in ground state is identified toward MM7, MM8, MM4, and MM11. Excluding lines with intensity lower than 3$\sigma$, the largest number of uncontaminated transitions (15$-$54~K) are detected toward MM4. The rotational temperatures of 108$\pm$35~K, 152$\pm$25~K, 146$\pm$10~K, and 123$\pm$23~K are obtained toward the MM7, MM8, MM4, and MM11, respectively. The largest column density of (1.3$\pm$0.1) $\times$ 10$^{17}$~cm$^{-2}$ is obtained toward MM4. Its first excited torsional states v$_{t}$=1 (210$-$225~K) are identified only toward MM8 and MM4. The fitting of CH$_{3}$OCHO~v$_{t}$=1 transitions give higher rotational temperatures and column densities (MM8: 163$\pm$35~K, (9.1$\pm$0.6) $\times$ 10$^{16}$~cm$^{-2}$; MM4: 150$\pm$33~K, (1.0$\pm$0.4) $\times$ 10$^{17}$~cm$^{-2}$) than those derived from CH$_{3}$OCHO~v=0. 

\emph{Dimethyl ether (CH$_{3}$OCH$_{3}$):} The largest number of transitions (10 transitions with $E_{u}$ of 10$-$196~K) of CH$_{3}$OCH$_{3}$ are detected toward MM4, which have $T_{\rm rot}$ of 106$\pm$13 K and $N_{\rm T}$ of (6.6$\pm$0.6) $\times$ 10$^{16}$~cm$^{-2}$. Six unblended transitions with $E_{\rm u}$ of 10$-$101 K, are identified toward MM8, resulting $T_{\rm rot}$ of 111$\pm$30~K and $N_{\rm T}$ of (4.8$\pm$0.3) $\times$ 10$^{16}$~cm$^{-2}$. Seven transitions spanning 10$-$101~K give the values of $T_{\rm rot}$ of 95$\pm$32 K and $N_{\rm T}$ of (2.9$\pm$0.9) $\times$ 10$^{16}$~cm$^{-2}$ toward MM11. For MM7, three transitions with $E_{\rm u}$ of 10 K are observed. Therefore, its temperatures and densities cannot be calculated due to lack of sufficient $E_{u}$ range.

\emph{Acetone (CH$_{3}$COCH$_{3}$):} Acetone is excited only toward MM8 and MM4, but all lines in our frequency range have relatively weak intensities. By modeling 5 transitions ($E_{u}$ of 14$-$81~K) toward the MM8 and 12 transitions ($E_{u}$ of 14$-$111~K) toward MM4, we obtain the rotational temperatures and column densities are 100$\pm$38~K, (1.0$\pm$0.5)$\times$ 10$^{16}$~cm$^{-2}$ for MM8, and 113$\pm$31~K, (1.1$\pm$0.2) $\times$ 10$^{16}$~cm$^{-2}$ for MM4.

\emph{Ethanol (C$_{2}$H$_{5}$OH):} Ethanol is the isomer of dimethyl ether. Thirty transitions of gauche-Ethanol and 4 transitions of trans-Ethanol toard MM4 cover a wide $E_{u}$ range (35$-$444 K). A smaller number of lines are detected toward MM7 and MM8 (MM7: 12; MM8: 16). Many spectral lines are below 3$\sigma$ in these two cores. The rotational temperature is 149$\pm$32 K toward MM8, slightly lower value of 131$\pm$4 K toward MM4, and 99$\pm$28 K toward MM7. Among the three cores, the highest column density of (4.6$\pm$0.3) $\times$ 10$^{16}$~cm$^{-2}$ is estimated for MM4. Only two lines are slightly stronger than 3$\sigma$ toward MM11. 

\emph{Ethyl Cyanide (C$_{2}$H$_{5}$CN):} Ethyl Cyanide is detected toward MM7, MM8, and MM4. Ten lines of C$_{2}$H$_{5}$CN with $E_{\rm u}$ ranging from 33 to 118 K are used for model fitting. The best fit parameters are $T_{\rm rot}$=115$\pm$26 K and $N_{\rm T}$=(4.4$\pm$0.9)$\times10^{15}$ cm$^{-2}$ toward MM7, 140$\pm$21 K and (3.4$\pm$0.7)$\times10^{16}$ cm$^{-2}$ toward MM8, 114$\pm$14 K and (4.3$\pm$0.3)$\times10^{15}$ cm$^{-2}$ toward MM4. Additionally, within the observed bandwidth we detected 20 unblended lines of C$_{2}$H$_{5}$CN isotopologues (10 CH$_{3}$CH$_{2}^{13}$CN lines, 9 CH$_{3}^{13}$CH$_{2}$CN and 1 $^{13}$CH$_{3}$CH$_{2}$CN line) toward MM8. The column densities of  (1.4$\pm$0.5)$\times10^{15}$ cm$^{-2}$ have been estimated by fixing the rotational temperature similar to that found from main isotopologue C$_{2}$H$_{5}$CN. Therefore, $^{12}$C/$^{13}$C ratio of 24$\pm$8 is derived (see Section 4.4 for more details).

\emph{Methyl Mercaptan (CH$_{3}$SH):} Methyl Mercaptan with $E_{\rm u}$ covering from 17 K to 53 K are identified toward MM8. $T_{\rm rot}$ and $N_{\rm T}$ are 156$\pm$21 K and (3.7$\pm$0.9)$\times10^{16}$ cm$^{-2}$, respectively. 

\emph{Parameters from molecules with one or two identified transitions:} As mentioned above, some molecules (e.g., methanol, acetaldehyde, dimethyl ether, etc) are observed with only one or two detected transitions. In this case, we fix the rotational temperature, then get the column densities for those species. Assuming the gas temperature of each core is equal to the average value of rotational temperatures derived from spectral fits, we use 99~K, 134~K, 137~K, and 100~K as the gas temperatures toward MM7, MM8, MM4, and MM11. The derived column densities are listed in Table \ref{tab:parameterstable}. For CH$_{3}$CHO, four cores have $N_{\rm T}$ ranging from $3.4\times10^{15}$ cm$^{-2}$ to $5.3\times10^{15}$ cm$^{-2}$. The column densities of CH$_{3}$OH toward MM7, MM8, and MM11 are $3.0\times10^{17}$ cm$^{-2}$, $4.0\times10^{17}$ cm$^{-2}$, and $2.0\times10^{17}$ cm$^{-2}$, respectively. The value of $1.8\times10^{16}$ cm$^{-2}$ for CH$_{3}$OCH$_{3}$ at MM7 , and $9.0\times10^{15}$ cm$^{-2}$ for C$_{2}$H$_{5}$OH at MM11 have been estimated. Two lines of HC$_{5}$N and two lines of $^{13}$C isotopic of HC$_{3}$N v=0 (HC$^{13}$CCN and HCC$^{13}$CN) are identified toward MM7, MM8, and MM4, with molecular column densities of $\sim10^{14}$ cm$^{-2}$. 

\begin{landscape}
 \begin{table}
  \caption{Model fitting results of detected molecules.}
  \label{tab:parameterstable}
  \begin{tabular}{cccccccccccccccccccccccccc}
    \hline
    Molecular name &\multicolumn{3}{c}{MM7} & & \multicolumn{3}{c}{MM8} & & \multicolumn{3}{c}{MM4} & & \multicolumn{3}{c}{MM11}\\
\cline{2-4} \cline{6-8} \cline{10-12} \cline{14-16} \\
& $T_{\rm rot}$& $N_{\rm T}$ & $\chi$ & &  $T_{\rm rot}$& $N_{\rm T}$ & $\chi$& & $T_{\rm rot}$& $N_{\rm T}$ & $\chi$ & &  $T_{\rm rot}$& $N_{\rm T}$ & $\chi$\\
&($\rm K$)&$10^{16}$cm$^{-2}$)& ($10^{-8}$)&&($\rm K$)&($10^{16}$cm$^{-2}$) & ($10^{-8}$) & & ($\rm K$)&$10^{16}$cm$^{-2}$)& ($10^{-8}$)&&($\rm K$)&($10^{16}$cm$^{-2}$) & ($10^{-8}$) \\
    \hline
    H$_{2}$CCO & 72$\pm$19 & 0.16$\pm$0.07 & 0.7$\pm$0.3 & & 102$\pm$10 & 0.55$\pm$0.01 & 1.5$\pm$0.1 & & 102$\pm$30 & 0.50$\pm$0.17 & 1.5$\pm$0.6 & & 84$\pm$23 & 0.16$\pm$0.07 & 0.6$\pm$0.3\\ 
 CH$_{3}$OH &99 & 30  & 130.0$\pm$11.3 & & 134 & 40 & 105.0$\pm$8.3 & & 159$\pm$8 & 75.0$\pm$3.2 & 227.0$\pm$29.2 & & 100 & 20 & 71.4$\pm$5.1 \\
 CH$_{3}$OH v$_{t}$=1 & $\cdots$ &$\cdots$ & $\cdots$ &  & $\cdots$ &$\cdots$ & $\cdots$ & & 204$\pm$20 & 92.3$\pm$6.5 & 280.0$\pm$39.2 & & $\cdots$ &$\cdots$ & $\cdots$ \\
 CH$_{3}$CHO & 99 & 0.34 & 1.5$\pm$0.1 & & 134 & 0.50 & 1.3$\pm$0.1 & & 137 & 0.53 & 1.6$\pm$0.2 & & 100 & 0.38 & 1.4$\pm$0.1\\
 C$_{2}$H$_{5}$OH & 99$\pm$28 & 1.1$\pm$0.2 & 4.8$\pm$1.0 & & 149$\pm$32 & 2.1$\pm$0.7 & 5.5$\pm$1.9 & & 131$\pm$4 & 4.6$\pm$0.3 & 13.9$\pm$1.9 & & 100 & 0.9 & 3.2$\pm$0.2 \\ 
 CH$_{3}$OCHO & 108$\pm$35 & 1.1$\pm$0.5 & 4.8$\pm$2.2 & & 152$\pm$25 & 5.9$\pm$0.5 & 15.5$\pm$1.8 & & 146$\pm$10 & 13.0$\pm$1.0 & 39.4$\pm$5.7 & & 123$\pm$23 & 3.1$\pm$0.8 & 11.1$\pm$3.0\\
 CH$_{3}$OCHO v$_{t}$=1 & $\cdots$ &$\cdots$ & $\cdots$ & & 163$\pm$35 & 9.1$\pm$0.6 & 23.9$\pm$2.5 & & 150$\pm$33 & 10.5$\pm$4.2 & 31.8$\pm$13.3 & & $\cdots$ &$\cdots$ & $\cdots$\\
 CH$_{3}$OCH$_{3}$ & 99 & 1.8 & 7.8$\pm$0.7 & & 111$\pm$30 & 4.8$\pm$0.3 & 12.6$\pm$1.3 & & 106$\pm$13 & 6.6$\pm$0.6 & 20.0$\pm$3.0 & & 95$\pm$31 & 2.9$\pm$0.9 & 10.4$\pm$3.3 \\
 CH$_{3}$COCH$_{3}$ & $\cdots$ & $\cdots$ & $\cdots$ & & 100$\pm$34 & 1.0$\pm$0.5 & 2.6$\pm$1.3 & & 113$\pm$31 & 1.1$\pm$0.2 & 3.3$\pm$0.7 & & $\cdots$ & $\cdots$ & $\cdots$ \\
 C$_{2}$H$_{5}$CN & 115$\pm$26 & 0.44$\pm$0.09 & 1.9$\pm$0.4 & & 140$\pm$21 & 3.35$\pm$0.69 & 9.0$\pm$2.0 & & 124$\pm$14 & 0.43$\pm$0.03 & 1.3$\pm$0.2 & & $\cdots$ & $\cdots$ & $\cdots$ \\
 CH$_{3}$CH$_{2}^{13}$CN & $\cdots$ & $\cdots$ & $\cdots$ & & 140 & 0.14$\pm$0.05 &  0.37$\pm$0.14 & & $\cdots$ & $\cdots$ & $\cdots$ & & $\cdots$ & $\cdots$ & $\cdots$ \\
 CH$_{3}^{13}$CH$_{2}$CN & $\cdots$ & $\cdots$ & $\cdots$ & & 140 & 0.14$\pm$0.04 &  0.37$\pm$0.14 & & $\cdots$ &$\cdots$ & $\cdots$ & & $\cdots$ & $\cdots$ & $\cdots$ & \\
 CH$_{3}$SH & $\cdots$ & $\cdots$ & $\cdots$ & & 156$\pm$21 & 3.7$\pm$0.9 & 9.7$\pm$2.5 & & $\cdots$ &$\cdots$ & $\cdots$ & & $\cdots$ &$\cdots$ & $\cdots$ \\
 HC$^{13}$CCN v=0 & 99 & 0.019 & 0.082$\pm$0.007 & & 134 & 0.056 & 0.15$\pm$0.01 & & 137 & 0.035 & 0.10$\pm$0.01 & & $\cdots$ & $\cdots$ & $\cdots$ \\
 HCC$^{13}$CN v=0 & 99 & 0.022 & 0.096$\pm$0.008 & & 134 & 0.060 & 0.16$\pm$0.01 & & 137 & 0.035 & 0.10$\pm$0.01 & & $\cdots$ & $\cdots$ & $\cdots$ \\
 HC$_{5}$N & 99 & 0.030 & 0.13$\pm$0.01 & & 134 & 0.026 & 0.070$\pm$0.006 & & 137 & 0.060 & 0.18$\pm$0.02 & & $\cdots$ & $\cdots$ & $\cdots$ \\
 \hline
  \end{tabular}
  \begin{tablenotes}
        \footnotesize
         \item[ ] Notes. For some species, their fitting errors of column densities are not given because detected transitions are below three. The column densities are derived by assuming the fixed $T_{\rm rot}$ of 99~K, 134~K, 137~K, and 100~K toward MM7, MM8, MM4, and MM11, respectively. The "$\cdots$" means no molecular lines were detected. $\chi$ is abundance of specific molecule relative to H$_{2}$. 
      \end{tablenotes}
 \end{table}
\end{landscape}

\subsection{Molecular abundance}
The fractional abundance of a certain molecule relative to H$_{2}$ ($\chi$ = $N_{\rm T}/N_{\rm H_{2}}$) can be estimated from the source-averaged total column density of the certain molecule and that of the molecular hydrogen. With assumption of optically thin emission from dust, the column density of H$_{2}$ can be derived by using the following equation (e.g., \citealt{Frau10,Bonfand19,Gieser21}):
\begin{equation}\label{equ-nh2}
N_{H_{2}} = \frac{S_{\nu}\eta}{\mu m_{H}\Omega\kappa_{\nu}B_{\nu}(T_{d})} ,
\end{equation}
where S$_{\nu}$ is the total integrated flux of the continuum, $\eta$ = 100 is gas-to-dust mass ratio \citep{Lis91,Hasegawa1992}, $\mu$ =2.8 is the mean molecular mass per H$_{2}$ molecule \citep{Kauffmann08}, $m_{\rm H}$ is the mass of the hydrogen atom, $\Omega$ is the solid angle subtended by the source, $\kappa_{\nu}$ is the dust absorption coefficient per unit density, and $B_{\nu}(T_{\rm d})$ is the Planck function at the dust temperature $T_{\rm d}$. 

The dust temperature should equal to the rotation temperature of the core derived from COMs, since gas and dust are expected to be thermally coupled well at density higher than 10$^{6}$~cm$^{-3}$ \citep{Goldsmith2001}. In our case, the H$_{2}$ densities at four cores are larger than 10$^{6}$~cm$^{-3}$ (see Table~\ref{tab:nh2}). We use the dust continuum emission to estimate the column density of H$_{2}$. However, at 3 mm observations, free-free emission from ionized gas could contribute to the continuum emission (detailed analyses are provided in \citealt{Liu20a,Liu2021}), which will overestimate the column density of H$_{2}$. Instead, we use 1.3 mm continuum data from \citet{Liu17}, and smooth it to the same spatial resolution as that of 3 mm continuum data. $\kappa_{\nu}$ of 0.899~cm$^{2}$g$^{-1}$ at 1.3 mm for dust grains with thin ice mantles at a gas density of 10$^{6}$ cm$^{-3}$ is adopted \citep{Ossenkopf1994}. The $N_{\rm H_{2}}$ derived from 1.3 mm continuum emission are: (2.3$\pm$0.2) $\times$ 10$^{23}$ cm$^{-2}$ for MM7,  (3.8$\pm$0.3) $\times$ 10$^{23}$ cm$^{-2}$ for MM8, (3.3$\pm$0.4) $\times$ 10$^{23}$ cm$^{-2}$ for MM4 and (2.8$\pm$0.2) $\times$ 10$^{23}$ cm$^{-2}$ for MM11. The mean optical depth of continuum can be calculated by (e.g., \citealt{Frau10,Gieser21})
\begin{equation}\label{equ-tau}
\tau_{\nu} = -\rm ln \left( 1-\frac{S_{\nu}}{\Omega B_{\nu}(T_{d})}\right) .
\end{equation}
The derived $\tau_{\nu}$ toward four cores are on the order of $10^{-2}$, guaranteeing reliable H$_{2}$ column density. Toward four cores, the deconvolved sizes, dust temperatures, total flux densities, values of optical depth, column densities of H$_{2}$, and volume densities of H$_{2}$ are listed in Table~\ref{tab:nh2}.

\begin{table*}
\caption{\label{tab:nh2}Properties of four dense cores derived by 1.3 mm smooth data.}
\centering
\begin{threeparttable}
\begin{tabular}{lcccccccccc}
\hline\hline
Position & source size $^{a}$ & $T_{\rm d}$ $^{b}$ & $I_{\rm peak}$ & $S_{\nu}$ $^{a}$ & $\tau_{\nu}$ & $N_{H_{2}}$ & $n_{H_{2}}$ \\
& & (K) & mJy beam$^{-1}$ & (mJy) & ($10^{-2}$) & ($10^{23}$cm$^{-2}$) & ($10^{6}$cm$^{-3}$) \\
\hline
MM7 & 2.3$^{\prime\prime}\times1.3^{\prime\prime}$ & 99 & 16.8$\pm$0.9 & 109.7$\pm$0.6 & 0.96 & 2.3$\pm$0.2 & $\sim$1.7\\
MM8 & 2.6$^{\prime\prime}\times1.4^{\prime\prime}$ & 134 & 45.2$\pm$1.8 & 305$\pm$15 & 1.60 & 3.8$\pm$0.3 & $\sim$2.6\\
MM4 & 2.1$^{\prime\prime}\times1.8^{\prime\prime}$ & 137 & 40.4$\pm$1.9 & 266$\pm$14 & 1.38 & 3.3$\pm$0.4 & $\sim$3.6\\
MM11 & 2.8$^{\prime\prime}\times1.8^{\prime\prime}$ & 100 & 26.0$\pm$1.1 & 229$\pm$11 & 1.18 & 2.8$\pm$0.2 & $\sim$1.6\\
\hline
\end{tabular}
\begin{tablenotes}
        \footnotesize
         \item[ ] Notes. $^{a}$ The deconvolved size and total flux density of each core are obtained from 2D Gaussian fitting to 1.3 mm continuum.
         $^{b}$ The dust temperatures for four dense cores are estimated from the average rotational temperature for detected species.
      \end{tablenotes}
\end{threeparttable}
\end{table*}

The fractional abundances of different species at each position are given in Table \ref{tab:parameterstable}. Toward four cores (MM4, MM7, MM8 and MM11), H$_{2}$CCO has the lowest abundance, compared with other O-bearing molecules. The abundances of O-bearing COMs such as CH$_{3}$OH, CH$_{3}$OCHO, C$_{2}$H$_{5}$OH and CH$_{3}$OCH$_{3}$ toward MM4 are higher than that toward MM8, while the abundance of C$_{2}$H$_{5}$CN is largest toward MM8. 

\subsection{Carbon Isotopic Ratio} 
A large number of studies toward molecular clouds in the Galaxy showed that $^{12}$C/$^{13}$C ratio is a function of Galactocentric distance D$_{\rm GC}$ \citep{Langer1990,Langer1993}. Recently, \citet{Yan19} has reported a linear fit by observing H$_{2}$CO and H$_{2}^{13}$CO from 112 sources: 
 \begin{equation}\label{equ-iso}
 ^{12}C/^{13}C  = (5.08\pm1.10) kpc^{-1} \times D_{GC} + (11.86\pm6.60),
  \end{equation}
 At a Galactocentric distances of 3.4 kpc, the $^{12}$C/$^{13}$C ratio is 29$\pm$8. We have estimated the $^{12}$C/$^{13}$C ratio from HC$_{3}$N, C$_{2}$H$_{5}$CN and their $^{13}$C isotopologues, which are shown in Table \ref{tab:iso}. The ratios between HC$_{3}$N~v=0 and its isotopologues are underestimated by a factor of six, which may due to the large optical depth and possible missing-flux of HC$_{3}$N~v=0. Other five values are in the range of 23.9$-$25.6, which is in good agreement with the value derived from equation \ref{equ-iso}.

\begin{table}
\caption{\label{tab:iso}Isotopologue Ratios.}
\centering
\begin{threeparttable}
\begin{tabular}{lccc}
\hline\hline
Ratio & MM8\\
\hline
HC$_{3}$N~v=0/HC$^{13}$CCN~v=0 & $\sim$4.2 \\
HC$_{3}$N~v=0/HCC$^{13}$CN~v=0 & $\sim$4.0     \\
HC$_{3}$N~v$_{7}$=1/HC$^{13}$CCN~v$_{7}$=1 & $\sim$25.6 \\
HC$_{3}$N~v$_{7}$=1/HCC$^{13}$CN~v$_{7}$=1 & $\sim$24.2 \\
CH$_{3}$CH$_{2}$CN/CH$_{3}$CH$_{2}^{13}$CN & 23.9$\pm$9.8 \\
CH$_{3}$CH$_{2}$CN/CH$_{3}^{13}$CH$_{2}$CN & 23.9$\pm$8.4\\
CH$_{3}$CH$_{2}$CN/$^{13}$CH$_{3}$CH$_{2}$CN & 23.9\\
\hline
\end{tabular}
\begin{tablenotes}
        \footnotesize
         \item[ ] Notes. Detected spectral lines of HC$_{3}$N~v=0, HC$_{3}$N~v$_{7}$=1, and their $^{13}$C isotopologues are below three, the column densities of these species are modelled by assuming the rotational temperatures is equal to C$_{2}$H$_{5}$CN.
      \end{tablenotes}
    \end{threeparttable}
\end{table}

\section{Discussions}
 \subsection{Physical parameters of individual molecule}
\begin{figure*}
  \centering
 \includegraphics[width=17cm]{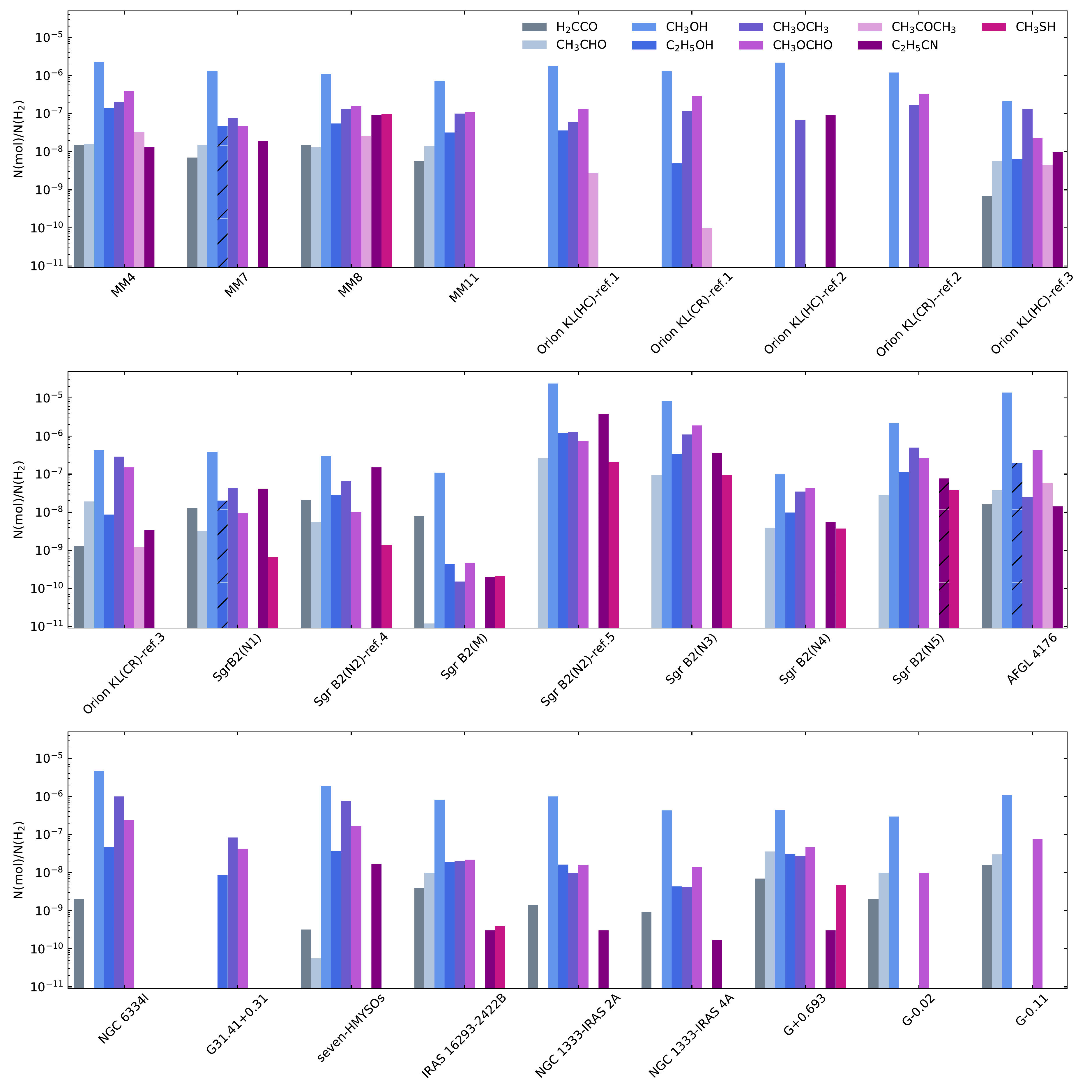}
\caption{Abundances of COMs relative to H$_{2}$ toward four dense cores of G9.62+0.19 and other sources listed in Table~\ref{tab:column density ratio}. The sources labeled as Orion KL (HC)-ref.1 and Orion KL (CR)-ref.1 refer to that the observed results are from Reference 1 in Table~\ref{tab:column density ratio}. Same for Orion KL (HC)-ref.2, Orion KL (CR)-ref.2, Orion KL (HC)-ref.3, Orion KL (CR)-ref.3, Sgr B2(N2)-ref.4 and Sgr B2(N2)-ref.5. 
}
\label{fig:abundances_H2}
\end{figure*}

 In this section we mainly discuss the physical parameters of detected H$_{2}$CCO, HC$_{5}$N and other complex molecules. We also compare molecular abundances with respect to H$_{2}$ and CH$_{3}$OH derived in our work with those derived in other sources including high-mass star forming regions (HMSFRs), low-mass star forming regions (LMSFRs) and Galactic Center molecular clouds (GCMCs). Molecular abundances for individual source and corresponding references can be found in Table~\ref{tab:column density ratio}. Figure~\ref{fig:abundances_H2} shows the molecular abundance with respect to H$_{2}$ toward the sources in Table~\ref{tab:column density ratio}. On average, H$_{2}$CCO, CH$_{3}$CHO, and CH$_{3}$COCH$_{3}$, which have mean abundances of 8.8$\times$10$^{-9}$ $-$ 3.4$\times$10$^{-8}$ (see Table~\ref{tab:column density ratio}), are relatively less abundant than other six COMs in HMSFRs, while CH$_{3}$OH, C$_{2}$H$_{5}$OH, CH$_{3}$OCH$_{3}$, CH$_{3}$OCHO and C$_{2}$H$_{5}$CN are usually rich in these sources. They are 1.2$\times$10$^{-7}$ $-$ 3.4$\times$10$^{-6}$ in molecular abundances. In addition, averaged abundances of C$_{2}$H$_{5}$OH, CH$_{3}$OCH$_{3}$, CH$_{3}$OCHO, C$_{2}$H$_{5}$CN and CH$_{3}$SH from HMSFRs are higher than those from LMSFRs and GCMCs by almost 1-3 orders of magnitude. The following is the detailed discussion for each species:
 
 \emph{Ketene (H$_{2}$CCO)}: When compared the rotational temperatures $T_{\rm rot}$ (72$-$102 K) and abundance $\chi$ ((6.0$-$15.0) $\times$ 10$^{-9}$) of H$_{2}$CCO with those of other O-bearing species, H$_{2}$CCO has the lowest $T_{\rm rot}$ and $\chi$ in the four cores (MM4, MM7, MM8, and MM11). It suggests that this molecule is not produced in very hot and compact regions, but rather in warm envelopes compared to other species. Previous observations show that H$_{2}$CCO has been detected in massive star-forming regions such as Sgr B2 (e.g., \citealt{Turner77,Nummelin00}) and Orion KL \citep{Johansson84,Crockett14}, cold dark cloud TMC-1 \citep{Ohishi91}, and several deeply embedded protostars like NGC 6334I and NGC 7538 I1 \citep{Ruiterkamp07}. Furthermore, these observations reveal that H$_{2}$CCO has abundances of $\sim 10^{-10}$ relative to H$_{2}$, which is slightly lower than the values of the four cores (MM7, MM8, MM4, and MM11) of G9.62+0.19. Previous observations also shown that H$_{2}$CCO is more abundant in the cooler and extended region \citep{Maity14}.  
 
 \emph{Acetaldehyde (CH$_{3}$CHO)}: Under the assumption of fixed temperature, relatively lower column densities ((3.4$-$5.3) $\times$ 10$^{15}$ cm$^{-2}$) and abundances ((1.3$-$1.6) $\times$ 10$^{-8}$) of CH$_{3}$CHO are obtained toward four cores than that for other species. Moreover, the spatial distribution of CH$_{3}$CHO is not well coincident with the positions of MM4, MM8, and MM11. It might imply that CH$_{3}$CHO tend to trace more externally extended gas. 
 
 \emph{Methanol (CH$_{3}$OH)}: CH$_{3}$OH~v=0 has high abundances between 7.1$\times$10$^{-7}$ and 2.3$\times$10$^{-6}$ toward four cores. This species is efficiently formed in molecular clouds from early to evolved stage. The gas distributions cover a more extended region. The much higher temperature (204 K) and abundance (2.8$\times$10$^{-6}$) for CH$_{3}$OH in the first excited torsional state (v$_{t}$=1) derived toward MM4. CH$_{3}$OH~v$_{t}$=1 with very high excitation energy should originate from a very warm and compact region.
 
 \emph{Ethanol (C$_{2}$H$_{5}$OH)}: The abundances toward MM7 ((4.8$\pm$1.0)$\times$10$^{-8}$), MM8 ((5.5$\pm$1.9)$\times$10$^{-8}$), and MM11 ( (3.2$\pm$0.2)$\times$10$^{-8}$) do not present obvious differences. The abundance increases to (1.4$\pm$0.2)$\times$10$^{-7}$ toward MM4. The derived temperature are 149$\pm$32~K for MM8 and 131$\pm$4~K for MM4, which is consistent with the understanding that C$_{2}$H$_{5}$OH usually has rotational temperature higher than 100~K in in massive hot cores (e.g., \citealt{Bisschop07}). Column density ratios between C$_{2}$H$_{5}$OH and CH$_{3}$OH in four dense cores range from 0.037 to 0.061 (see Table~\ref{tab:column density ratio}), which consistent with the mean value 0.064$^{+0.066}_{-0.041}$ measured from several HMSFRs \citep{Taquet15}.   
 
 \emph{Dimethyl ether (CH$_{3}$OCH$_{3}$)}: CH$_{3}$OCH$_{3}$ is an isomer of C$_{2}$H$_{5}$OH. It has a lower temperature (MM8: 111$\pm$30~K, MM4: 106$\pm$13~K) than C$_{2}$H$_{5}$OH. The CH$_{3}$OCH$_{3}$/C$_{2}$H$_{5}$OH abundance ratios are $\sim$1.6, 2.3, 1.4, and 3.2 toward MM7, MM8, MM4, and MM11, respectively, indicating that CH$_{3}$OCH$_{3}$ is more abundant than C$_{2}$H$_{5}$OH in dense cores. Meanwhile, it can be seen from Figure~\ref{fig:abundances_H2} that abundances of CH$_{3}$OCH$_{3}$ are generally higher than those of C$_{2}$H$_{5}$OH in HMSFRs. Also, the averaged abundance ratio between CH$_{3}$OCH$_{3}$ and C$_{2}$H$_{5}$OH for those regions is around 2.5. The abundances of two species are nearly same for three LMSFRs IRAS 16293-2422B, NGC 1333-IRAS 2A, and NGC 1333-IRAS 4A. The formation efficiency of these two isomers in various environments may be different.   
 
 \emph{Methyl Formate (CH$_{3}$OCHO)}: For the G9.62+0.19, its temperatures are between 108~K and 152~K. The abundance increases by an order of magnitude from MM7 (4.8$\times$10$^{-8}$) to MM4 (3.9$\times$10$^{-7}$). The rotational temperatures of ground state and first torsional excited state are similar considering the uncertainties. This is consistent with the findings of \citet{Favre11}, who obtained similar temperatures between CH$_{3}$OCHO~v=0 and CH$_{3}$OCHO~v$_{t}$=1 in Orion KL region.  \citet{Taquet15} gives an averaged CH$_{3}$OCHO abundance with respect to CH$_{3}$OH of 0.14$^{+0.18}_{-0.06}$ for HMSFRs. The abundances in our three cores MM4, MM8, and MM11 are 0.17, 0.15, and 0.16, respectively, which are consistent with 0.14$^{+0.18}_{-0.06}$. However, the abundance of 0.037 toward MM7 are lower than the mean value derived in other HMSFRs.
 
 \emph{Acetone (CH$_{3}$COCH$_{3}$)}: CH$_{3}$COCH$_{3}$ shows low temperature (100$-$113~K) and abundance (2.6$\times10^{-8}-3.3\times10^{-8}$) in MM8 and MM4. The abundances are larger than those derived from Orion KL hot core and compact ridge regions ($\sim$ 10$^{-9}$ for Orion KL; see Table~\ref{tab:column density ratio}).  
 
 \emph{Ethyl Cyanide (C$_{2}$H$_{5}$CN)}: C$_{2}$H$_{5}$CN is usually used as a hot core tracer because of its high rotational temperature. We found that the temperatures are 115$\pm$26~K, 140$\pm$21~K and 124$\pm$21~K for the MM7, MM8 and MM4, respectively. It may be destroyed by UV radiation in the MM11 region. Its abundance is larger by nearly one order of magnitude toward MM8 than that toward MM7 and MM4. Furthermore, C$_{2}$H$_{5}$CN abundances toward HMSFRs (average value of $\sim$10$^{-7}$) are obviously higher than those toward LMSFRs and GCMCs ($\sim$10$^{-10}$) (see Table~\ref{tab:column density ratio}).
 
 \emph{Cyanobutadiyne (HC$_{5}$N)}: Because fewer than three lines were detected, $T_{\rm rot}$ and $N_{\rm T}$ could not be obtained from spectral line fits. We derived column densities and abundances of HC$_{5}$N v=0 toward three cores with the assumption of  $T_{\rm rot}$ = 99 K, 134 K, and 137 K for MM7, MM8, and MM4, respectively. The derived HC$_{5}$N abundance in the range of (7.0$-$18) $\times$ 10$^{-10}$ are slightly lower than that in other four hot cores G10.30-0.15, G12.89+0.49, G16.86-2.16 and G28.28-0.36 (6.3 $\times$ 10$^{-10}$ $-$ 4.2 $\times$ 10$^{-9}$) \citep{Taniguchi17} and in Galactic Center molecular cloud G+0.693-0.027 (1.9 $\times$ 10$^{-9}$) \citep{Zeng18}, and slightly higher than the value measured from L1527 (WCCC) ((1.2$\pm$0.3) $\times$ 10$^{-10}$) \citep{Jorgensen02,Sakai13}. A very low HC$_{5}$N abundance, 1.1 $\times$ 10$^{-11}$ has been measured toward the outer cold envelope of IRAS 16293 \citep{Jaber17}. The peaks of the gas emission are not well associated with the continuum peaks MM4, MM7 and MM8. Such results seem to imply that HC$_{5}$N is not excited in the densest part of three dense cores, especially for MM4 and MM7. \citet{Fontani17} also found that the emission of HC$_{5}$N in the protocluster OMC-2 FIR4 is not associated with continuum emission.

 \emph{Methyl Mercaptan (CH$_{3}$SH)}: Few S-bearing COMs has been detected in the ISM so far. As one of the simplest S-bearing COMs, CH$_{3}$SH has been detected in a few high-mass star-forming regions like Sgr B2(N) \citep{Muller16}, Orion KL \citep{Kolesnikova14}, G327.3-6 \citep{Gibb2000} and G31.41+0.31 \citep{Gorai2021}, low-mass protostar IRAS 16293-2422 \citep{Majumdar16}, and Galactic Center molecular cloud G+0.693-0.027 \citep{Rodriguez-Almeida2021}. In these sources, CH$_{3}$SH has the temperature range $\sim$ 68$-$200~K and abundance from 1.0$\times$10$^{-9}$ to 9.4$\times$10$^{-8}$ for above hot cores, but lower temperature (32~K) and abundance (5.0$\times$10$^{-10}$) for solar-type protostar IRAS 16293-2422, a much lower temperatures of 8.5$-$14.9~K and abundance of 4.8$\times$10$^{-9}$ for G+0.693-0.027. Our observations give the rotational temperature of 156$\pm$21~K and abundance of (9.7$\pm$2.5)$\times10^{-8}$ toward MM8, which is consistent with the upper limit of the parameter range derived from other hot cores.

\setlength{\tabcolsep}{5pt}
 \begin{landscape}
\begin{table}
\caption{\label{tab:column density ratio}\textbf{Molecular abundances relative to CH$_{3}$OH and H$_{2}$ obtained in G9.62+0.19 and other regions.}}
\scriptsize
\begin{threeparttable}
\begin{tabular}{lllllllllllllllllllllllllllllll}
\hline\hline
Source & \multicolumn{2}{c}{[H$_{2}$CCO]} & \multicolumn{2}{c}{[CH$_{3}$OH]} & \multicolumn{2}{c}{[CH$_{3}$CHO]} & \multicolumn{2}{c}{[C$_{2}$H$_{5}$OH]} & \multicolumn{2}{c}{[CH$_{3}$OCH$_{3}$]} & \multicolumn{2}{c}{[CH$_{3}$OCHO]}
 & \multicolumn{2}{c}{[CH$_{3}$COCH$_{3}$]} & \multicolumn{2}{c}{[C$_{2}$H$_{5}$CN]} & \multicolumn{2}{c}{[CH$_{3}$SH]} & Reference \\ 
& /[CH$_{3}$OH] & /[H$_{2}$] & /[CH$_{3}$OH] & /[H$_{2}$] & /[CH$_{3}$OH] & /[H$_{2}$] & /[CH$_{3}$OH] & /[H$_{2}$] & /[CH$_{3}$OH] & /[H$_{2}$] & /[CH$_{3}$OH] & /[H$_{2}$] & /[CH$_{3}$OH] & /[H$_{2}$] & /[CH$_{3}$OH] & /[H$_{2}$] & /[CH$_{3}$OH] & /[H$_{2}$]\\
    \hline
& \multicolumn{18}{c}{High-mass star forming regions} \\
G9.62+0.19 (MM4) & 6.7(-3) & 1.5(-8) & 1 & 2.3(-6) & 7.1(-3) & 1.6(-8) & 6.1(-2) & 1.4(-7) & 8.8(-2) & 2.0(-7) & 1.7(-1) & 3.9(-7) & 1.5(-2) & 3.3(-8) & 5.7(-3) & 1.3(-8)& $-$ & $-$ & this work\\ 
G9.62+0.19 (MM7) & 5.3(-3) & 7.0(-9) & 1 & 1.3(-6) & 1.1(-2) & 1.5(-8) & 3.7(-2) & 4.8(-8) & 6.0(-2) & 7.8(-8) & 3.7(-2) & 4.8(-8) & $-$ & $-$ & 1.5(-2) & 1.9(-8) & $-$ & $-$ & this work\\
G9.62+0.19 (MM8) & 1.4(-2) & 1.5(-8) & 1 & 1.1(-6) & 1.3(-2) & 1.3(-8) & 5.3(-2) & 5.5(-8) & 1.2(-1) & 1.3(-7) & 1.5(-1) & 1.6(-7) & 2.5(-2) & 2.6(-8) & 8.5(-2) & 9.0(-8) & 9.3(-2) & 9.7(-8) & this work\\
G9.62+0.19 (MM11) & 8.0(-3) & 5.7(-9) & 1 & 7.1(-7) & 1.9(-2) & 1.4(-8) & 4.5(-2) & 3.2(-8) & 1.5(-1) & 1.0(-7) & 1.6(-1) & 1.1(-7) & $-$ & $-$ & $-$ & $-$ & $-$ & $-$ & this work \\ 
Orion KL (HC)$^{a}$ & $-$ & $-$ & 1 & 1.8(-6) & $-$ & $-$ & 2.0(-2) & 3.6(-8) & 3.4(-2) & 6.1(-8) & 7.4(-2) & 1.3(-7) & 1.5(-3) & 2.8(-9) & $-$ & $-$ & $-$ & $-$ & 1\\
Orion KL (CR)$^{b}$ & $-$ & $-$ & 1 & 1.3(-6) & $-$ & $-$ & 4.6(-3) & 4.9(-9) & 9.3(-2) & 1.2(-7) & 2.2(-1) & 2.9(-7) & 4.6(-4) & 1.0(-10) & $-$ & $-$ & $-$ & $-$ & 1\\
Orion KL (HC) & $-$ & $-$ & 1 & 2.2(-6) & $-$ & $-$ & $-$ & $-$ & 3.1(-2) & 6.8(-8) & $-$ & $-$ & $-$ & $-$ & 4.1(-2) & 8.9(-8) & $-$ & $-$ & 2 \\
Orion KL (CR) & $-$ & $-$ & 1 & 1.2(-6) & $-$ & $-$ & $-$ & $-$ & 1.4(-1) & 1.7(-7) & 2.8(-1) & 3.3(-7)& $-$ & $-$ & $-$ & $-$ & $-$ & $-$ & 2 \\
Orion KL (HC) & 3.3(-3) & 6.9(-10) & 1 & 2.1(-7) & 2.8(-2) & 5.8(-9) & 3.0(-2) & 6.3(-9) & 6.3(-1) & 1.3(-7) & 1.1(-1) & 2.3(-8) & 2.1(-2) & 4.5(-9) & 4.5(-2) & 9.6(-9) & $-$ & $-$ & 3 \\
Orion KL (CR)$^{c}$ & 2.9(-3) & 1.3(-9) & 1 & 4.3(-7) & 4.5(-2) & 1.9(-8) & 2.0(-2) & 8.6(-9) & 6.9(-1) & 2.9(-7) & 3.4(-1) & 1.5(-7) & 2.9(-3) & 1.2(-9) & 7.7(-3) & 3.3(-9) & $-$ & $-$ & 3 \\
Sgr B2(N1) & 3.4(-2) & 1.3(-8) & 1 & 3.9(-7) & 8.1(-3) & 3.2(-9) & 5.2(-2) & 2.0(-8) & 1.1(-1) & 4.3(-8) & 2.5(-2) & 9.6(-9) & $-$ & $-$ & 1.1(-1) & 4.1(-8) & 1.7(-3) & 6.5(-10) & 4 \\
Sgr B2(N2) & 6.8(-2) & 2.1(-8) & 1 & 3.0(-7) & 1.8(-2) & 5.5(-9) & 9.1(-2) & 2.8(-8) & 2.1(-1) & 6.5(-8) & 3.3(-2) & 1.0(-8) & $-$ & $-$ & 4.9(-1) & 1.5(-7) & 4.6(-3) & 1.4(-9) & 4 \\
Sgr B2(M) & 7.0(-2) & 8.0(-9) & 1 & 1.1(-7) & 1.0(-4) & 1.2(-11) & 3.7(-3) & 4.3(-10) & 1.3(-3) & 1.5(-10) & 4.0(-3) & 4.6(-10) & $-$ & $-$ & 1.8(-3) & 2.0(-10) & 1.8(-3) & 2.1(-10) & 4 \\
Sgr B2(N2) & 8.3(-3) & $-$ & 1 & 2.4(-5) & 1.1(-2) & 2.6(-7) & 5.0(-2) & 1.2(-6) & 5.5(-2) & 1.3(-6) & 3.0(-2) & 7.3(-7) & 1.0(-2) & $-$ & 1.6(-1) & 3.8(-6) & 8.5(-3) & 2.1(-7) & 5, 12 \\
Sgr B2(N3) & $-$ & $-$ & 1 & 8.3(-6) & 1.1(-2) & 9.4(-8) & 4.1(-2) & 3.4(-7) & 1.3(-1) & 1.1(-6) & 2.3(-1) & 1.9(-6) & $-$ & $-$ & 4.3(-2) & 3.6(-7) & 1.1(-2) & 9.4(-8) & 5 \\
Sgr B2(N4) & $-$ & $-$ & 1 & 9.8(-8) & 4.0(-2) & 3.9(-9) & 1.0(-1) & 9.8(-9) & 3.6(-1) & 3.5(-8) & 4.4(-1) & 4.3(-8) & $-$ & $-$ & 5.6(-2) & 5.5(-9)& 3.8(-2) & 3.7(-9) & 5 \\
Sgr B2(N5) & $-$ & $-$ & 1 & 2.2(-6) & 1.3(-2) & 2.8(-8) & 5.0(-2) & 1.1(-7) & 2.3(-1) & 5.0(-7) & 1.2(-1) & 2.7(-7) & $-$ & $-$ & 3.4(-2) & 7.6(-8) & 1.8(-2) & 3.9(-8) & 5\\
AFGL 4176 & 1.2(-3) & 1.6(-8) & 1 & 1.4(-5) & 2.7(-3) & 3.8(-8) & 1.4(-2) & 1.9(-7) & 2.4(-2) & 2.5(-8) & 3.1(-2) & 4.3(-7) & 4.2(-3) & 5.8(-8) & 1.0(-3) & 1.4(-8) & $-$ & $-$ & 6 \\
NGC 6334I & 4.3(-4) & 2.0(-9) & 1 & 4.7(-6) & $-$ & $-$ & 1.0(-2) & 4.7(-8) & 2.1(-1) & 1.0(-6) & 5.1(-2) & 2.4(-7) & $-$ & $-$ & $-$ & $-$ & $-$ & $-$ & 7 \\
G31.41+0.31 & $-$ & $-$ & $-$ & $-$ & $-$ & $-$ & $-$ & 8.4(-9) & $-$ & 8.4(-8) & $-$ & 4.2(-8) & $-$ & $-$ & $-$ & $-$ & 2.3(-2) & 2.7(-8) & 8, 9 \\
7 high-mass YSOs$^{d}$ & 1.7(-4) & 3.2(-10) & 1 & 1.9(-6) & 2.9(-5) & 5.6(-11) & 1.9(-2) & 3.6(-8) & 4.1(-1) & 7.7(-7) & 8.9(-2) & 1.7(-7) & $-$ & $-$ & 8.9(-3) & 1.7(-8) & $-$ & $-$ & 10 \\
\textbf{Average} & $-$ & 8.8(-9) & $-$ & 3.4(-6) & $-$ & 3.4(-8) & $-$ & 1.2(-7) & $-$ & 3.0(-7) & $-$ & 2.7(-7) & $-$ & 1.8(-8) & $-$ & 3.1(-7) & $-$ & 5.3(-8) & $-$ \\
& \multicolumn{18}{c}{Low-mass star forming regions} \\
IRAS 16293-2422B$^{e}$ & 4.8(-3) & 4.0(-9) & 1 & 8.3(-7) & 1.2(-2) & 1.0(-8) & 2.3(-2) & 1.9(-8) & 2.4(-2) & 2.0(-8) & 2.6(-2) & 2.2(-8) & 1.7(-3) & $-$ & 3.6(-4) & 3.0(-10) & 4.8(-4) & 4.0(-10) & 11, 12, 13 \\
NGC 1333-IRAS 2A & 1.4(-3) & 1.4(-9) & 1 & 1.0(-6) & $-$ & $-$ & 1.6(-2) & 1.6(-8) & 1.0(-2) & 1.0(-8) & 1.6(-2) & 1.6(-8) & $-$ & $-$ & 3.0(-4) & 3.0(-10) & $-$ & $-$ & 14 \\
NGC 1333-IRAS 4A & 2.1(-3) & 9.2(-10) & 1 & 4.3(-7) & $-$ & $-$ & 1.0(-2) & 4.3(-9) & 1.0(-2) & 4.3(-9) & 3.1(-2) & 1.4(-8) & $-$ & $-$ & 4.0(-4) & 1.7(-10) & $-$ & $-$ & 14 \\
& \multicolumn{18}{c}{Galactic Center molecular clouds} \\
G+0.693$^{f}$ & 1.6(-2) & 7.0(-9) & 1 & 4.5(-7) & 5.7(-2) & 3.6(-8) & 6.9(-2) & 3.1(-8) & 6.0(-2) & 2.7(-8) & 1.0(-1) & 4.7(-8) & $-$ & $-$ & $-$ & $-$ & $-$ & $-$ & 15, 16\\
G+0.693$^{f}$ & $-$ & $-$ & 1 & 1.1(-7) & $-$ & $-$ & 4.2(-2) & 4.6(-9) & $-$ & $-$ & $-$ & $-$ & $-$ & $-$ & $-$ & 3.0(-10) & 4.3(-2) & 4.8(-9) & 17, 18\\
G-0.02  & 6.4(-3) & 2.0(-9) & 1 & 3.0(-7) & 3.5(-2) & 1.0(-8) & $-$ & $-$ & $-$ & $-$ & 3.4(-2) & 1.0(-8) & $-$ & $-$ & $-$ & $-$ & $-$ & $-$ & 16\\
G-0.11  & 1.5(-2) & 1.6(-8) & 1 & 1.1(-6) & 2.7(-2) & 3.0(-8) & $-$ & $-$ & $-$ & $-$ & 7.1(-2) & 7.8(-8) & $-$ & $-$ & $-$ & $-$ & $-$ & $-$ & 16\\
\hline
\end{tabular}
\begin{tablenotes}
        \footnotesize
         \item[ ] Notes. X(Y) means X $\times$ 10$^{Y}$. $^{a}$ Correspond to "ET peak" in Fig. 1 of \citet{Tercero18}. $^{b}$ Correspond to "MF peak" in Fig. 1 of \citet{Tercero18}. $^{c}$ Correspond to "mm3b" in \citet{Feng15}$. ^{d}$ The abundance for each molecule is the average value from 7 high-mass YSOs including AFGL 2591, G24.78, G75.78, NGC 6334 IRS1, NGC 7538 IRS1, W 3(H$_{2}$O), and W 33A. $^{e}$ The abundances relative to H$_{2}$ are derived by using the lower limit of H$_{2}$ column density (1.2 $\times$ 10$^{25}$ cm$^{-2}$) from \citet{Jorgensen16}. Thus the abundances relative to H$_{2}$ from this source should be upper limits. $^{f}$ \citet{Requena-Torres06} and \citet{Requena-Torres08} used H$_{2}$ column density of 4.1 $\times$ 10$^{22}$ cm$^{-2}$ for abundance calculations, while \citet{Rodriguez-Almeida2021} and \citet{Zeng18} used the value of 1.35 $\times$ 10$^{23}$ cm$^{-2}$. 
\item[] References. (1) \citet{Tercero18}; (2) \citet{Crockett14}; (3) \citet{Feng15}; (4) \citet{Belloche13}; (5) \citet{Bonfand19}; (6) \citet{Bogelund19}; (7) \citet{Zernickel12}; (8) \citet{Rivilla17b}; (9) \citet{Gorai2021}; (10) \citet{Bisschop07}; (11) \citet{Jorgensen18}; (12) \citet{Jorgensen20}; (13) \citet{Drozdovskaya2019}; (14) \citet{Taquet15}; (15)  \citet{Requena-Torres06}; (16) \citet{Requena-Torres08}; (17) \citet{Rodriguez-Almeida2021}; (18) \citet{Zeng18}. 
      \end{tablenotes}
\end{threeparttable}
\end{table}
\end{landscape}

 \subsection{Evolution of dense cores}
 In order to study the variations of temperatures and molecular abundance with the evolutionary sequence of the dense cores, the stage of each core need to be evaluated. According to the detected molecular species, molecular abundance, and line kinematic properties (e.g., outflow) of each core, the stage of the cores are classified as listed in Table~\ref{tab:evolution}. The general criteria for determining the relative evolutionary stage of the cores are summarized as follows. 
 
 \begin{table*}
\caption{\label{tab:evolution}\textbf{Evolutionary stage for each core.}}
\centering
\begin{threeparttable}
\begingroup
\setlength{\tabcolsep}{9pt} 
\renewcommand{\arraystretch}{2.0} 
\begin{tabular}{lcccc}
\hline\hline
Position & Molecular species or continuum emission & Line kinematic property & Stage \\
\hline
MM6  & \parbox[c]{8cm}{simple molecules;  COMs (only one CH$_{3}$OH transition)} & collimated outflow & HMPOs \\
MM7  & \parbox[c]{8cm}{rich COMs; no vibrationally or torsionally excited states of molecules} & molecular outflows & early HMCs    \\
MM8  & \parbox[c]{8cm}{rich COMs; vibrationally excited state of H$_{3}$CN, torsionally excited states of CH$_{3}$OH and CH$_{3}$OCHO} & molecular outflows  & HMCs \\
MM4  & \parbox[c]{8cm}{rich COMs; CH$_{3}$OH~v=0 and CH$_{3}$OH~v$_{t}$=1 with higher upper energy level; centimeter continuum emission $^{a}$}  & no molecular outflows & evolved HMCs \\
MM11  & H$_{40\alpha}$, H$_{50\beta}$; a few COMs; centimeter continuum emission $^{a}$ & no molecular outflows & HMCs/UC H{\sc ii} region \\
MM1  & line-poor; H$_{40\alpha}$ & no molecular outflows & evolved H{\sc ii} region \\
\hline
\end{tabular}
\begin{tablenotes}
        \footnotesize
         \item[ ] Notes. $^{a}$ \citet{Testi2000}
      \end{tablenotes}
\endgroup
\end{threeparttable}
\end{table*}

 Starless core/prestellar core should not drive an outflow. Spectra of this kind of core appears line-poor. MM3 and MM9 were not observed by centimeter continuum \citep{Testi2000} and do not have outflow activities. They also do not excite abundant molecular lines (see Figure~\ref{fig:B1-1}). Those lines (e.g., CS, SO, $^{34}$SO, H$_{2}$CO, CH$_{3}$OH~v=0, HC$_{3}$N~v=0, etc.) detected in MM3 and MM9 are probably remnant emission from MM4 and MM11, respectively. MM3 and MM9 are considered as massive starless core candidates. 
 
\begin{figure}
   \centering
   \includegraphics[width=10cm]{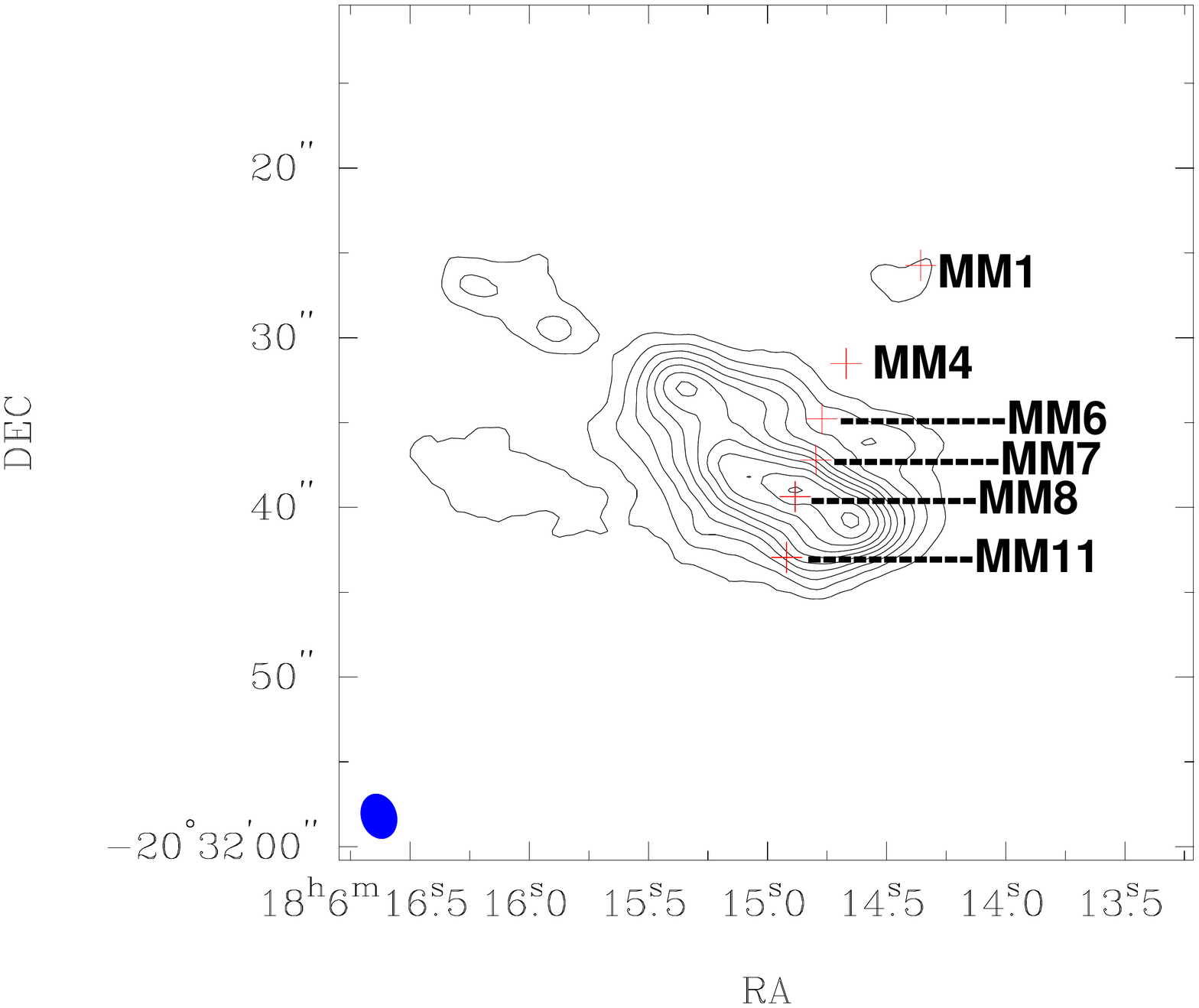}
   \includegraphics[width=10cm]{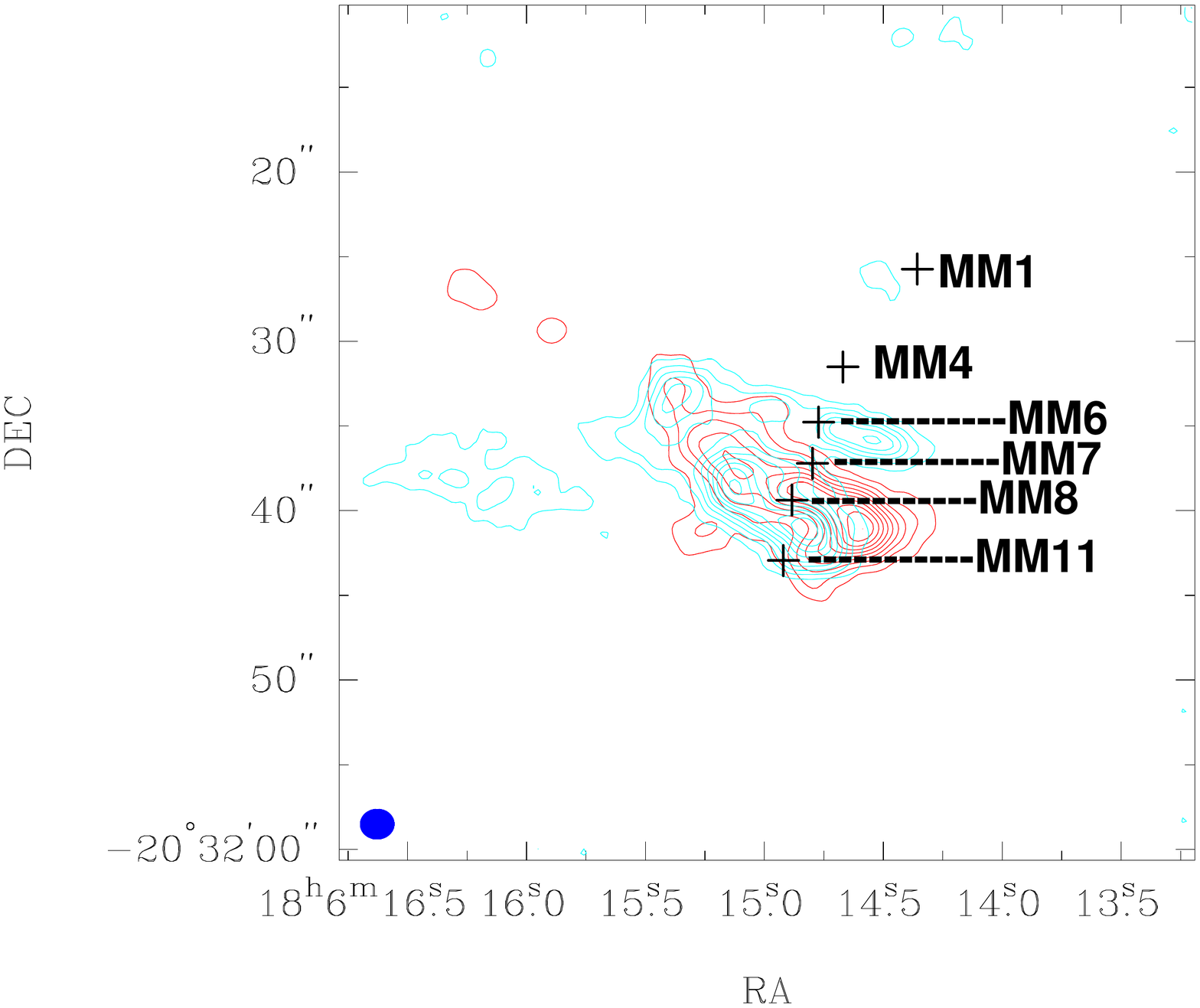}
      \caption{Upper panel: Integrated intensity (-15 to 45 km~s$^{-1}$) of SiO (2$-$1) observed by ALMA 12-m+ACA combined data. The contours are (10\%, 20\%, 30\%, 40\%, 50\%, 60\%, 70\%, 80\%, 90\%, 98\%) $\times$ 5.3 Jy/beam km~s$^{-1}$. Lower panel: SiO outflow observed by 12-m array. The cyan contours represent blueshifted gas with LSR velocities ranging from -15 to -2 km~s$^{-1}$. The red contours represent redshifted emission with velocities 12 to 43 km~s$^{-1}$. The synthesized beam is shown in the bottom left corner. The contours are (10\%, 20\%, 30\%, 40\%, 50\%, 60\%, 70\%, 80\%, 90\%, 100\%) $\times$ 3.6 Jy/beam km~s$^{-1}$. The names of the six densest cores are labeled.
              }
      \label{fig:sio}
   \end{figure}

 MM6 is more evolved, in which a collimated outflow is seen through CS (2--1), SO (3$_{2}$--2$_{1}$) and SiO (2--1) (see Figure \ref{fig:fig-S} and Figure \ref{fig:sio}) emission. SiO (5--4) emission with ALMA 1.3 mm data presented by \citet{Liu17} also showed this bar-like structure associated with MM6, which is very collimated and dominated by blueshifted emission. A small population of molecules such as CH$_{3}$OH are excited in this stage.
 
\begin{figure}
   \centering
   \includegraphics[width=9cm]{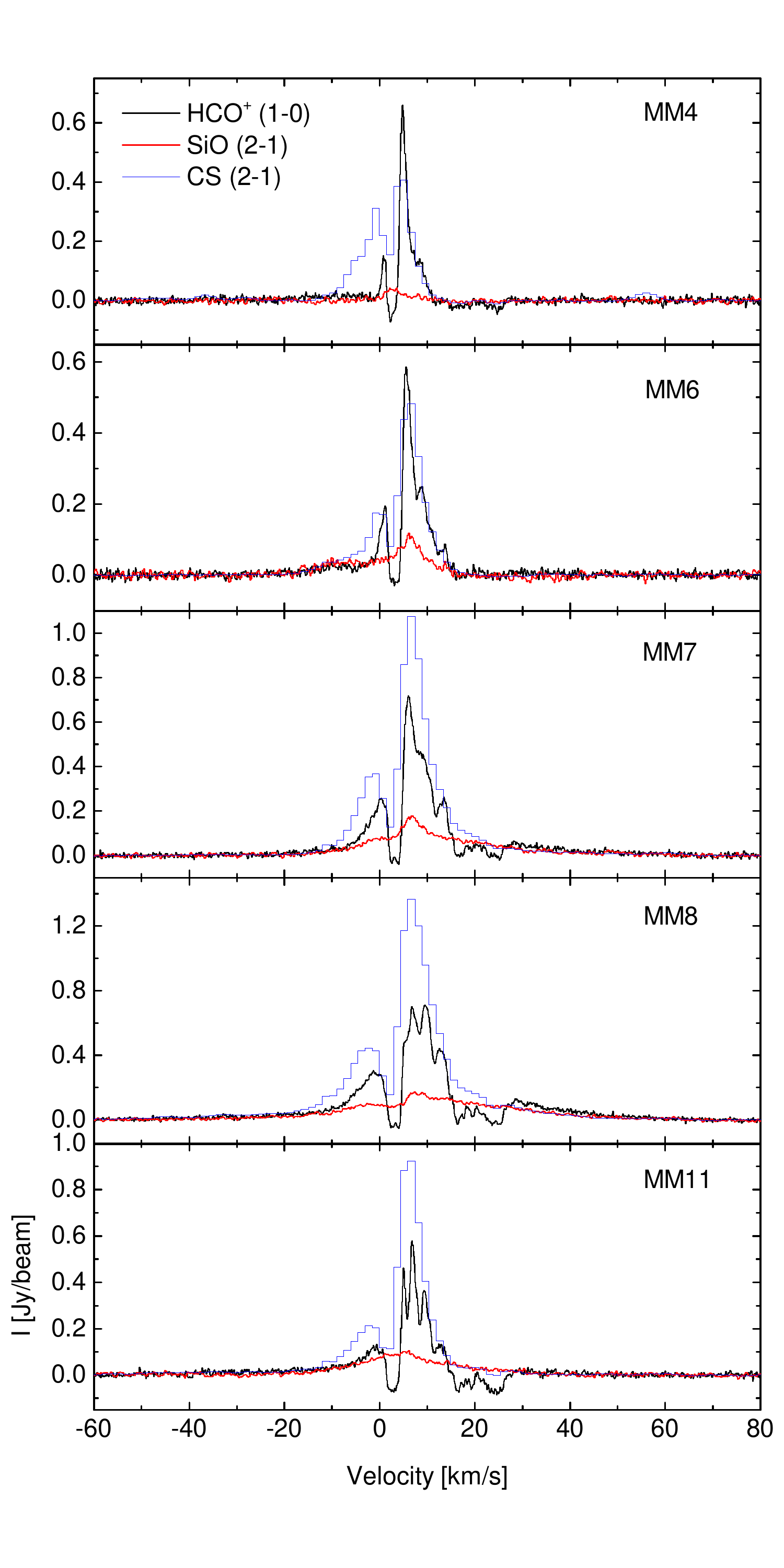}
      \caption{Beam-averaged spectra of HCO$^{+}$ (1-0), SiO (2-1), and CS (2-1) at five protostellar cores, observed by 12-m and ACA combined data. 
              }
      \label{fig:linewidth}
   \end{figure}

 MM7, MM8, and MM4 may evolved into HMC phase. HMCs should have outflows, but this kind of outflows are not as highly collimated as from HMPOs. The significant feature of HMCs is that rich complex organic molecules and even in vibrationally and torsionally excited states (e.g., HC$_{3}$N~v$_{7}$=1, HC$_{3}$N~v$_{6}$=1, CH$_{3}$OH~v$_{t}$=1, and CH$_{3}$OCHO~v$_{t}$=1) exist in this phase. The vibrationally and torsionally excited states of molecules have higher upper energy levels and should be excited at a higher kinetic temperature. It suggests that molecules with higher $E_{\rm u}$ are good indicators for a more evolved region \citep{Gieser21}. CH$_{3}$OCHO~v$_{t}$=1 is not detected toward MM7, suggesting MM7 is less evolved than MM8 and MM4. For MM8 and MM4, we note that CH$_{3}$OH~v=0 with 889~K and CH$_{3}$OH~v$_{t}$=1 with 771~K are excited at MM4 but absent at MM8. Gas emission of outflow tracers (CS, SO, and SiO, see Figure \ref{fig:fig-S} and \ref{fig:sio}), along with the spectra of HCO$^{+}$, SiO, and CS lines (Figure \ref{fig:linewidth}) which show broader wing structure in the MM8 compared to MM4. All tracers do not show any signature of molecular outflows in MM4. So maybe MM4 is older than MM8. 
 
 H$_{40\alpha}$ recombination line toward MM1 and MM11 indicating that they evolved into an H/UC H{\sc ii} stage. In this phase, molecules are hard to be detected since they are destroyed by UV radiation. Meanwhile hydrogen recombination lines emerge. MM1 is line-poor with only a few simple molecules like CS, SO, and HC$_{3}$N~v=0 detected and is not associated with outflows. However, ionized gas tracer H$_{40\alpha}$ is observed in this core. MM1 lies $\sim$0.8$\arcsec$ to the south-east of the H{\sc ii} region C, it is likely that H$_{40\alpha}$ line is from the envelope of the H{\sc ii} region C. A few COMs emission still exist in the MM11 region, meanwhile H$_{40\alpha}$ and H$_{50\beta}$ lines are excited. 
 
 In spite of different phases presenting different physical conditions, it should be noted that during the star-forming process, different evolutionary stages show a vague boundary, and they are not fully separable.

\begin{figure}
   \centering
   \includegraphics[width=10cm]{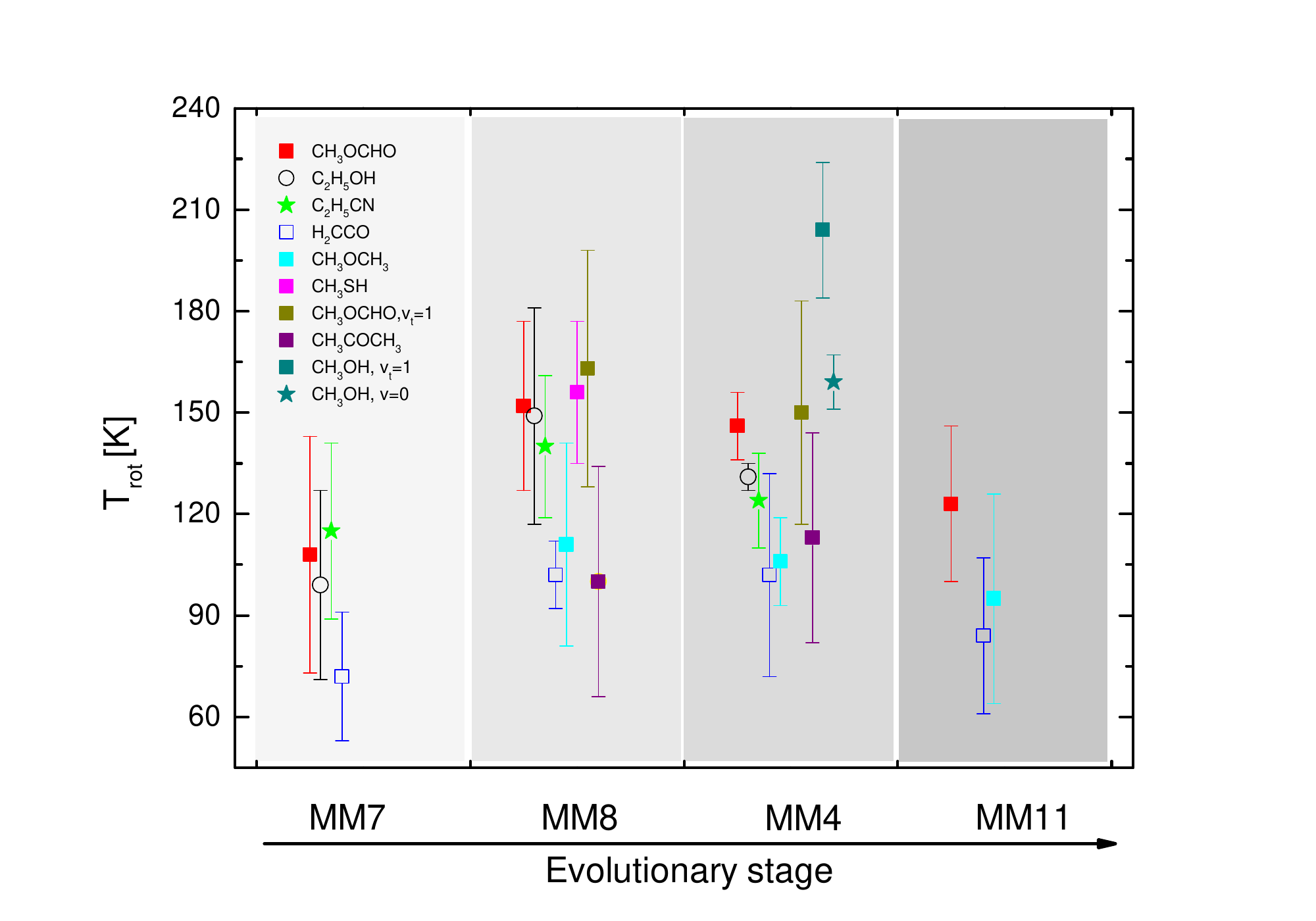}
   \includegraphics[width=10cm]{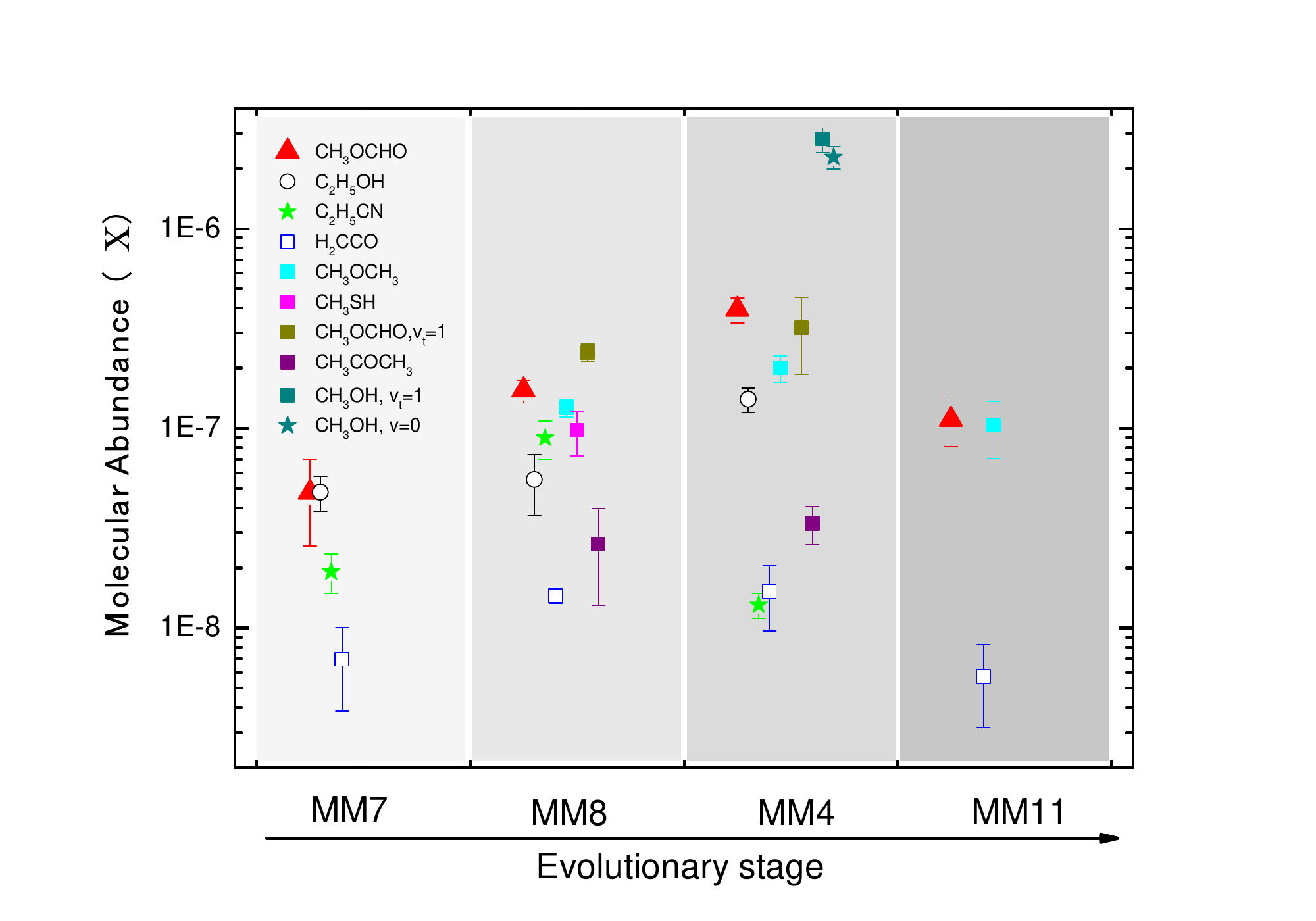}
      \caption{The rotational temperatures (upper) and molecular abundances (lower) of eight species (CH$_{3}$OH (in ground state (v=0) and torsionally excited state (v$_{t}$=1)), CH$_{3}$OCHO (in v=0 and v$_{t}$=1), C$_{2}$H$_{5}$OH, C$_{2}$H$_{5}$CN, H$_{2}$CCO, CH$_{3}$OCH$_{3}$, CH$_{3}$COCH$_{3}$, and CH$_{3}$SH) from four dense cores. 
              }
      \label{fig:trot+abundance}
   \end{figure}

  Figure \ref{fig:trot+abundance} shows that the rotational temperatures $T_{\rm rot}$ and molecular abundances (CH$_{3}$OH, CH$_{3}$OCHO, C$_{2}$H$_{5}$OH, C$_{2}$H$_{5}$CN, CH$_{2}$CO, CH$_{3}$OCH$_{3}$, CH$_{3}$COCH$_{3}$, and CH$_{3}$SH) vary with evolutionary stage. This figure only present those parameters that have been derived via LTE fitting to uncontaminated molecules. The mean values of the two parameters are: $T_{\rm rot}$ = 99$\pm$28 K, $\chi$= (3.0$\pm$1.2)$\times$10$^{-8}$ for MM7, $T_{\rm rot}$ = 134$\pm$27 K, $\chi$= (1.0$\pm$0.2)$\times$10$^{-7}$ for MM8, $T_{\rm rot}$ = 137$\pm$21 K, $\chi$= (6.9$\pm$1.7)$\times$10$^{-7}$ for MM4, and $T_{\rm rot}$ = 100$\pm$18 K, $\chi$= (7.4$\pm$2.6)$\times$10$^{-8}$ for MM11. The data points of molecules (e.g., CH$_{3}$CHO and CH$_{3}$OCH$_{3}$ in MM7) for which column density is derived under the fixed rotational temperature are excluded from the figure. The molecular excitation temperature was adopted to be the gas temperature for each core. Among the four regions, the temperature of MM8/hot-core is similar to MM4/evolved-hot-core, which is higher than those of MM7/early-hot-core and MM11/UC H{\sc ii}. MM7 should not have a comparatively hotter environment because it has not evolved into typical hot core phase. Methanol in torsional state was observed toward MM4 containing three transition lines with $E_{\rm u}$ up to 340$-$902 K, which trace very hot and dense region. So the highest temperature appears toward the MM4. MM11/UC H H{\sc ii} is slightly colder than MM8 and MM4, which is due to the colder and outer part of the gas components traced by CH$_{3}$OCHO, CH$_{3}$OCH$_{3}$ and H$_{2}$CCO. From the Figure \ref{fig:trot+abundance} and Table \ref{tab:parameterstable}, one can see that MM4/evolved-hot-core has higher abundances of O-bearing species than other regions, while the highest abundance of C$_{2}$H$_{5}$CN is seen in MM8. Methanol in ground and torsional states has high abundance $\sim$ 2.5$\times$10$^{-6}$, making the mean value of abundance of MM4 to be seven times higher than those of MM8. In brief, molecular abundance increases with increasing age, except in the last phase MM11 which shows a decrease in abundance probably due to the destruction of molecules by the UV radiation. The molecular abundances increasing with evolution were also found by \citet{Gerner14} and \citet{Coletta2020}.

\subsection{Chemical differentiation}
\subsubsection{Chemical Differentiation between Nitrogen-bearing and Oxygen-bearing species }
  Chemical differentiation between N-bearing and O-bearing species was found both in high-mass star-forming regions such as Orion KL \citep{Friedel08,Wang2009,Crockett15}, W3(OH) \citep{Wyrowski99,Qin15}, W75N \citep{Remijan04b}, and G19.61-0.23 \citep{Qin10}, and in low-mass star-forming regions such as IRAS 16293$-$2422 \citep{Kuan04}. These results imply that N-bearing and O-bearing molecules mainly exist in different environments. Our spectral line survey further revealed that the chemical difference between N-bearing and O-bearing molecules also exists in G9.62+0.19. While O-bearing molecules such as CH$_{3}$OH, CH$_{3}$OCHO, C$_{2}$H$_{5}$OH, and CH$_{3}$OCH$_{3}$ mostly associated with  MM4 (see Figure~\ref{fig:fig-O}), the strongest emission peaks of N-bearing species C$_{2}$H$_{5}$CN and HC$_{3}$N are located toward the MM8 (see Figure~\ref{fig:fig-N}). The derived abundances shows that MM4 is an O-rich source, while MM8 is rich in N-bearing species. This result is consistent with the previous interferometric observations that C$_{2}$H$_{5}$CN emission is the strongest toward the region F (MM8) whereas CH$_{3}$OCHO and CH$_{3}$OCH$_{3}$ are the brightest toward the region E (MM4) \citep{Su05}. 
   
  The N- and O-bearing difference may arise because of different physical properties (initial temperature) during the gravitational collapse phase \citep{Caselli93,Yamamoto17}. As predicted by chemical models, the relatively low initial temperature (lower than 20 K) at molecular accretion phase could keep CO for a long duration on the grain surface and thus may produce O-bearing molecules by the hydrogenation of CO. In contrast, if the temperature at this phase is as high as 40 K \citep{Caselli93}, grain surface CO could be easily desorbed to the gas phase, leading to the formation of more N-bearing COMs compared to the O-bearing species. Furthermore, apart from the desorption of CO, a more important process is that atomic hydrogen is quickly desorbed from the dust grain surface. This hinders the formation of O-bearing species. In contrast, atoms C and N easily migrate on dust grain surface, forming C$_{n}$N and thus initiate the N-bearing COM chemistry. A consistent findings from recent work (Wang et al., in preparation) also presented that the initial temperature of 15 K and 23 K at collapse phase can produce abundant O-bearing and N-bearing species, respectively. Considering that the MM8 region could be externally heated at the gravitationally collapse phase due to its proximity to H {\small II} "B", its initial temperature at accretion phase is likely to be higher than that of MM4. The N/O variation seen in MM8 and MM4 is very likely to be caused by the different initial environments of these two cores.

\subsubsection{Chemical relationships of Oxygen-bearing molecules }
\begin{figure*}
   \centering
   \includegraphics[width=\hsize]{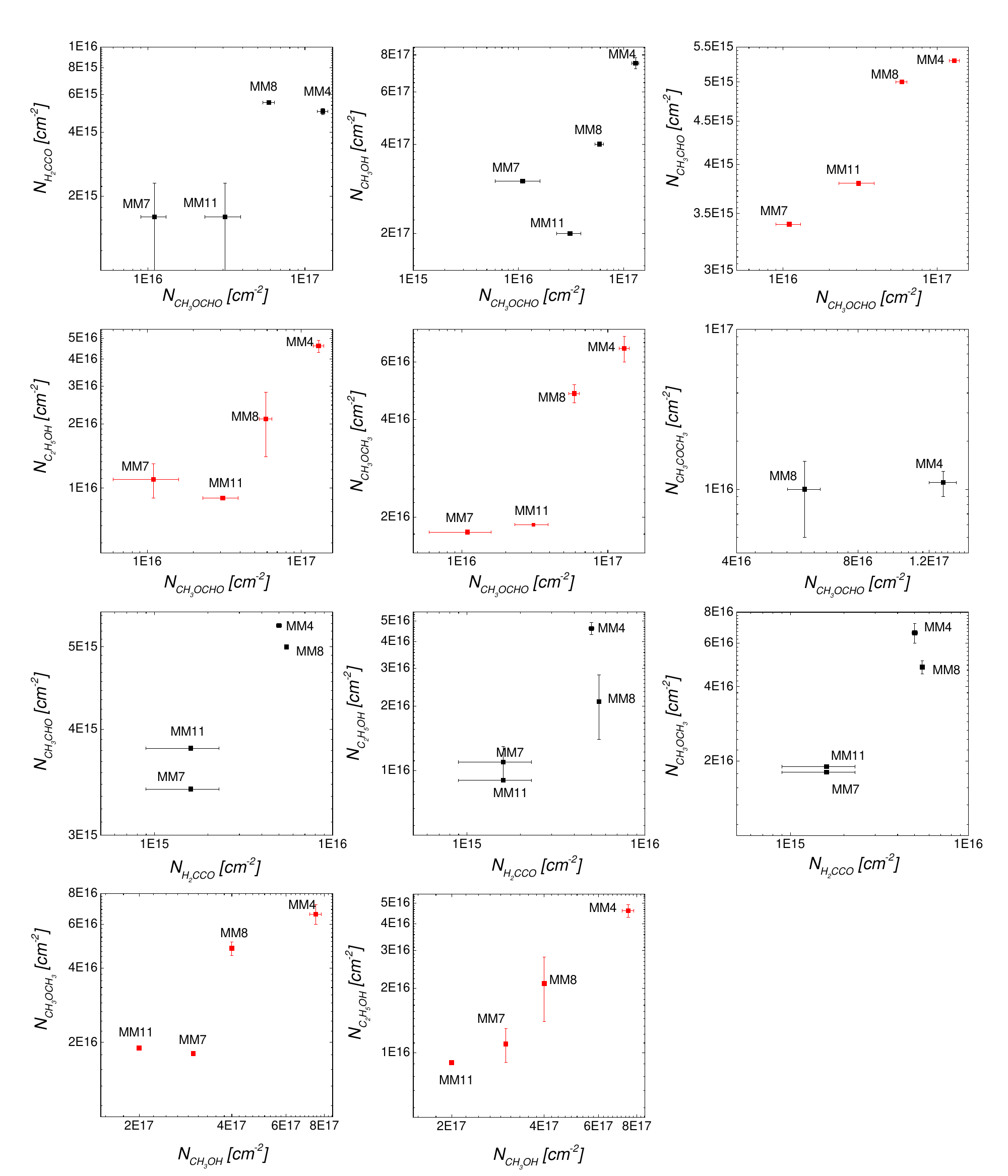}
      \caption{Comparison between the column densities of O-bearing species toward four cores MM7, MM8, MM4, and MM11. The point with error bar means that this value is obtained from model fitting, while one without error bar represents that the value derived by assuming a fixed rotational temperature and source size. The data points with red denote the possible chemical links between CH$_{3}$OCHO/CH$_{3}$CHO, CH$_{3}$OCHO/C$_{2}$H$_{5}$OH, CH$_{3}$OCHO/CH$_{3}$OCH$_{3}$, CH$_{3}$OH/CH$_{3}$OCH$_{3}$, and CH$_{3}$OH/C$_{2}$H$_{5}$OH. 
              }
         \label{fig:chemical-link}
   \end{figure*}

 \citet{Chuang21} suggest that COMs described by the formula C$_{2}$H$_{\rm n}$O$_{2}$ (n=4 and 6) may have similar formation routes. COMs with formula C$_{2}$H$_{\rm n}$O (n=2, 4 and 6) are also chemical links through hydrogenation reaction on interstellar grains: H$_{2}$CCO$\stackrel{2\rm H}{\longrightarrow}$CH$_{3}$CHO$\stackrel{2 \rm H}{\longrightarrow}$C$_{2}$H$_{5}$OH \citep{Charnley05}. In our work, C$_{2}$H$_{\rm n}$O family such as H$_{2}$CCO, CH$_{3}$CHO, C$_{2}$H$_{5}$OH, and CH$_{3}$OCH$_{3}$, C$_{2}$H$_{\rm n}$O$_{2}$ family like CH$_{3}$OCHO, and CH$_{3}$OH which is considered as the precursor of some larger O-bearing COMs (e.g., C$_{2}$H$_{5}$OH, CH$_{3}$OCH$_{3}$, and CH$_{3}$OCHO) \citep{Blake1987,Oberg2009,Maity15,Bergantini17} are simultaneously detected. In order to explore possible chemical links among O-bearing molecules, we compare the column density of several molecules (see Figure \ref{fig:chemical-link}), as the formation process of COMs can be tested by investigating how their abundances/column densities are related. 
 
 Moreover, comparisons of the very limited sample can not testify the potential chemical link among species well. We included more observing results from HMSFRs, to LMSFRs and GCMCs as listed in Table~\ref{tab:column density ratio}. Since column densities of these O-bearing COMs have been derived from various observational equipments with different spatial resolution and spectral setup, comparison of column density is not very helpful to derive sound conclusions for chemical links among O-bearing COMs. It is more reliable to compare the abundance of COMs relative to CH$_{3}$OH. The column density ratios between O-bearing species and CH$_{3}$OH from other types of sources in Table~\ref{tab:column density ratio} are plotted in Figure~\ref{fig:relationship}. By the analyses of the molecular relationship in our four dense cores (see Figure~\ref{fig:chemical-link}), together with molecules from literature (see Figure~\ref{fig:relationship}), the main points for possible chemical correlations between specific molecules are discussed as follows:
 
 \textbf{(1) CH$_{3}$OH/C$_{2}$H$_{5}$OH, CH$_{3}$OH/CH$_{3}$OCH$_{3}$:} The preliminary results from Figure \ref{fig:chemical-link} reveal that there is a tight positive correlation between CH$_{3}$OH and C$_{2}$H$_{5}$OH, and also the positive correlation between CH$_{3}$OH and CH$_{3}$OCH$_{3}$, is suggestive of similar formation route for both the species. Here, the synthesis of C$_{2}$H$_{6}$O isomers ethanol (C$_{2}$H$_{5}$OH) and dimethyl ether (CH$_{3}$OCH$_{3}$) has great potential via radical-radical recombination that the hydroxymethyl radical (CH$_{2}$OH) and methoxy radical (CH$_{3}$O) react with methyl radical (CH$_{3}$) \citep{Maity15}. 
 \begin{equation}\label{reaction1}
 {\rm CH_{3} + CH_{2}OH \longrightarrow C_{2}H_{5}OH~~~~(grain~surface)}
 \end{equation}

 \begin{equation}\label{reaction2}
 {\rm CH_{3} + CH_{3}O \longrightarrow CH_{3}OCH_{3}~~~~(grain~surface)}
 \end{equation}
 The productions of hydroxymethyl radical (CH$_{2}$OH) and methoxy radical (CH$_{3}$O) through unimolecular decomposition of methanol have been identified in \citet{Bennett07a, Bennett07b}. Then the main path to produce the radical CH$_{3}$O and CH$_{2}$OH would likely be the photodissociation process as follows:
 \begin{equation}\label{reaction3}
 {\rm CH_{3}OH +h\nu \longrightarrow CH_{2}OH + H~~~~(grain~surface)}
 \end{equation}
 
 \begin{equation}\label{reaction4}
 {\rm CH_{3}OH +h\nu \longrightarrow CH_{3}O + H~~~~(grain~surface)}
 \end{equation}
 The spatial distributions of CH$_{3}$OH, CH$_{3}$OCH$_{3}$, and C$_{2}$H$_{5}$OH that are mainly shown toward the MM4, MM7, MM8, and MM11, suggest that C$_{2}$H$_{5}$OH and CH$_{3}$OCH$_{3}$ might have a common precursor, CH$_{3}$OH. Correlation coefficient of 0.89 for CH$_{3}$OH/CH$_{3}$OCH$_{3}$ toward G10.6-0.4 \citep{Law2021} and 0.86 for CH$_{3}$OH/CH$_{3}$OCH$_{3}$ and CH$_{3}$OH/C$_{2}$H$_{5}$OH toward seven high-mass hot cores \citep{Bisschop07} also support that C$_{2}$H$_{5}$OH and CH$_{3}$OCH$_{3}$ are tightly correlated with CH$_{3}$OH. 
 
  \textbf{(2) CH$_{3}$OCHO and CH$_{3}$OCH$_{3}$:} Column densities between CH$_{3}$OCHO and CH$_{3}$OCH$_{3}$, from each type of core, are also correlated (see Figure~\ref{fig:chemical-link}). Moreover, abundances of CH$_{3}$OCHO and CH$_{3}$OCH$_{3}$ relative to CH$_{3}$OH (see Figure~\ref{fig:relationship}) from three kinds of objects present the positive correlation. The tight correlation of CH$_{3}$OCHO and CH$_{3}$OCH$_{3}$ was presented from \citet{Bisschop07,Brouillet2013,Coletta2020} as well. The molecular abundance ratio between CH$_{3}$OCHO and CH$_{3}$OCH$_{3}$ is found to be $\sim$1 in different type of source including high-mass star-forming regions (e.g., \citealt{Bisschop07,Brouillet2013,Rivilla17b,Law2021}), intermediate-mass star-forming regions \citep{Ospina-Zamudio18}, hot corinos \citep{Taquet15}, Galactic centre clouds \citep{Requena-Torres06} and protostellar shock region \citep{Lefloch17}. However, the ratio is reported to be $\sim$2 for prestellar cores \citep{Jimenez16} and Comets \citep{Biver19}. In our work, ratios are estimated to be 0.6$\pm$0.3, 1.3$\pm$0.1, 2.0$\pm$0.3, and 1.1$\pm$0.5 toward the MM7, MM8, MM4, and MM11, respectively. In addition, our 3~mm observations showed that CH$_{3}$OCHO and CH$_{3}$OCH$_{3}$ have similar gas distributions. These observation results may imply a potential link between these two species, which shares common precursor methoxy radical (CH$_{3}$O) through reactions (\ref{reaction5}) and (\ref{reaction6}) \citep{Garrod06,Garrod08,Oberg2010}, protonated methanol CH$_{3}$OH$^{+}_{2}$ with reactions (\ref{reaction7}) and (\ref{reaction8}) \citep{Blake1987}, or production of CH$_{3}$OCHO from CH$_{3}$OCH$_{3}$ via the sequence (\ref{reaction9}) \citep{Balucani15}:
  \begin{equation}\label{reaction5}
  {\rm CH_{3}O + HCO \longrightarrow CH_{3}OCHO~~~~(grain~surface)}
  \end{equation}
  
   \begin{equation}\label{reaction6}
   {\rm CH_{3}O + CH_{3} \longrightarrow CH_{3}OCH_{3}~~~~(grain~surface)}
   \end{equation}
   
   \begin{equation}\label{reaction7}
   {\rm CH_{3}OH^{+}_{2} \stackrel{H_{2}CO}{\longrightarrow} CH_{3}OCHO~~~~(gas~phase)}
   \end{equation}
   
   \begin{equation}\label{reaction8}
   {\rm CH_{3}OH^{+}_{2} \stackrel{CH_{3}OH}{\longrightarrow} CH_{3}OCH_{3}~~~~(gas~phase)}
   \end{equation}
   
   \begin{equation}\label{reaction9}
   {\rm CH_{3}OCH_{3} \stackrel{Cl/F}{\longrightarrow} CH_{3}OCH_{2} \stackrel{O}{\longrightarrow} CH_{3}OCHO~~~~(gas~phase)}
   \end{equation}
   
 \textbf{(3) C$_{2}$H$_{\rm n}$O$_{2}$ family:} The correlation plots (see Figure~\ref{fig:chemical-link}) for each pair of H$_{2}$CCO/CH$_{3}$CHO, H$_{2}$CCO/C$_{2}$H$_{5}$OH, and H$_{2}$CCO/CH$_{3}$OCH$_{3}$ showed that there is no clear relationship between H$_{2}$CCO and other C$_{2}$H$_{\rm n}$O group molecules. The gas distribution of CH$_{3}$CHO extend to east, which shows that the emission of CH$_{3}$CHO is not very similar to H$_{2}$CCO and C$_{2}$H$_{5}$OH. These results challenge the prediction that COMs in the formula of C$_{2}$H$_{\rm n}$O have chemical correlation. Figure~\ref{fig:relationship} shows a positive correlations between N(H$_{2}$CCO)/N(CH$_{3}$OH) and N(C$_{2}$H$_{5}$OH)/N(CH$_{3}$OH), and weak correlation between N(H$_{2}$CCO)/N(CH$_{3}$OH) and  N(CH$_{3}$OCH$_{3}$)/N(CH$_{3}$OH). The chemical links among molecules in C$_{2}$H$_{\rm n}$O formula cannot be determined at the moment.
 
 \textbf{(4) CH$_{3}$OCHO and CH$_{3}$CHO:} Although Figure \ref{fig:chemical-link} and Figure~\ref{fig:relationship} showed another possible chemical link between CH$_{3}$OCHO and CH$_{3}$CHO, gas distributions of CH$_{3}$OCHO are considerably different from those of CH$_{3}$CHO. The chemical correlation between them should be further investigated.  
 
 To better characterize the chemical properties, more samples in various astrochemical environments and future chemical models are needed. In the near future, we are on track to study the chemistry of the COMs by considering more sources observed in the ATOMS project. In addition, these sources belong to different evolutionary stages so that we can draw a general picture on the properties of the COMs revealed from these sources. This will help us to set strong constraints on the chemical networks.

 \begin{figure*}
  \centering
   \includegraphics[width=\hsize]{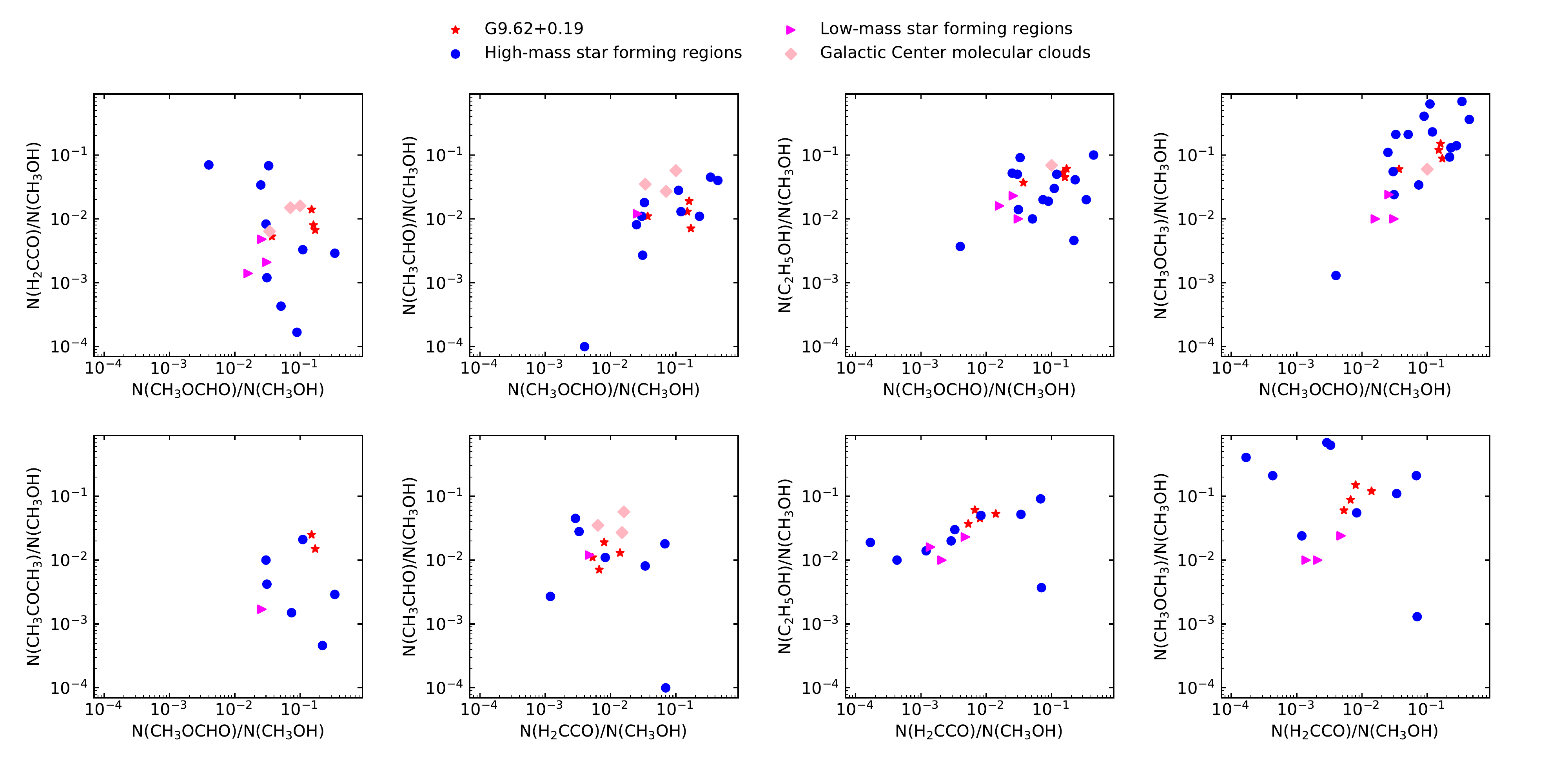}
\caption{ Comparison of the column density ratios with respect to CH$_{3}$OH between O-bearing species for the sources listed in Table~\ref{tab:column density ratio}. 
}
\label{fig:relationship}
\end{figure*}

 \subsubsection{Chemical relationships of Nitrogen-bearing molecules }
 Hot core tracers (HC$_{3}$N and its vibrationally excited lines) found to be abundant with the evolution of the center protostar \citep{Bergin1996,Taniguchi19,Zhang21}. Earlier, longer chain cyanopolyynes HC$_{\rm 2n+1}$N (n>2) have been thought to be detected in the early stage of star formation and absent in protostellar cores (e.g., \citealt{Suzuki92,Hirota09}). Recent observations found that longer cyanopolyynes have been detected in hot cores (e.g., \citealt{Belloche13,Esplugues13,Green14,Taniguchi17,Taniguchi18}). Previous papers proposed that cyanopolyynes are efficiently formed via the neutral–neutral reaction "C$_{2n}$H$_{2}$ + CN $\longrightarrow$ HC$_{2n+1}$N + H" in the warm region (e.g., \citealt{Chapman09,Taniguchi19}). HC$_{5}$N also can be formed via the reaction "CCH + HC$_{3}$N $\longrightarrow$ HC$_{5}$N + H" in the gas phase \citep{Taniguchi19}. At low temperature, the route "N + C$_{n+1}$H $\longrightarrow$ HC$_{n}$N + C", where n = 3, 5, 7, is thought to significantly contribute to the formation of cyanopolyynes. Recently, it has been suggested that HC$_{5}$N/HC$_{3}$N abundance ratios in the cold objects are higher than those in massive hot cores \citep{Jaber17}. So, there exists contradiction between HC$_{5}$N and HC$_{3}$N in production rates. The formation mechanism of cyanopolyynes is not fully understood by now.  
 
 In our work, Figure \ref{fig:fig-N} shows the integrated intensity maps of HC$_{3}$N~v=0, and its isotopes, and HC$_{3}$N in vibrationally excited state v$_{7}$=1 all peak at the position of the continuum peak. However, the distribution of HC$_{5}$N has its own unique characteristics which differ from the distributions of HC$_{3}$N. Therefore, it seems that HC$_{5}$N is not generated from HC$_{3}$N. \citet{Fontani17} studied the intermediate mass protocluster OMC-2 FIR4, and found that the spatial distribution of HC$_{3}$N is not coincident with that of HC$_{5}$N, namely that HC$_{3}$N emission is associated with the continuum peak FIR4, HC$_{5}$N distributes over the eastern part of the continuum peak. They proposed that enhanced cosmic-ray ionisation rate will promote the production of HC$_{5}$N. For HC$_{5}$N in G9.62+0.19, the line intensity ratio between MM3 and MM4 is about 1.3, while for HC$_{3}$N the ratio is around 0.5. Additionally, MM3 and MM4 are supposed to be at starless core and evolved HMCs stage, respectively. This imply that the synthesis of HC$_{5}$N is different from HC$_{3}$N in the early stage. The neutral–neutral reaction between C$_{4}$H$_{2}$ and CN may not be the formation pathway of HC$_{5}$N in G9.62+0.19, which support the idea that formation mechanisms of HC$_{3}$N and HC$_{5}$N may differ in various interstellar environments \citep{Taniguchi19}. 
 
 In addition, the distribution of C$_{2}$H$_{5}$CN (see Figure \ref{fig:fig-N}) is almost identical to that of HC$_{3}$N, predicting that C$_{2}$H$_{5}$CN is correlated with HC$_{3}$N in formation mechanism (e.g., hydrogenation of HC$_{3}$N on grain surface to form C$_{2}$H$_{5}$CN).
 
\section{Conclusions}
 We have preformed line survey using ALMA 3 mm observations toward the complex massive star-forming region. The detailed line modeling and analysis show that there exhibit chemical differentiation in this region. The main results are summarised as follows:
 
 1. Six dense cores are detected in 3 mm continuum emission. The molecular richness, kinematics, and physical and chemical properties are different in these dense cores. MM1 is considered as an H{\sc ii} region because it shows H$_{40\alpha}$ emission. MM6 is at HMPOs with collimated outflow. MM7, MM8, and MM4 are line-rich regions, which are considered to be at HMCs phase. Among them, MM7 is at the earlier stage because of fewer detected molecules and MM4 is at a more evolved stage due to the absence of outflows. MM11 shows a combination of HMCs and UC H{\sc ii} region, as it has the detection of COMs (such as CH$_{3}$CHO, CH$_{3}$OCHO, CH$_{3}$OCH$_{3}$, and C$_{2}$H$_{5}$OH), and hydrogen recombination lines such as H$_{40\alpha}$ and H$_{50\beta}$. In short, their evolutionary sequence is MM6$\rightarrow$MM7$\rightarrow$MM8$\rightarrow$MM4$\rightarrow$MM11$\rightarrow$MM1.
 
 2. Modeling of the detected spectral lines, we have derived rotational temperatures from 72 to 115 K toward MM7, 100 to 163 K toward MM8, 102 to 204 K toward MM4, and 84 to 123 K toward MM11. Molecular column densities are on the orders of 10$^{15}$ $-$ 10$^{16}$~cm$^{-2}$ toward MM7, MM8, and MM11, 10$^{15}$ $-$ 10$^{17}$~cm$^{-2}$ toward MM4. 
 
 3. The emission of N-bearing and S-bearing molecules is strongest toward MM8, while emission of O-bearings is dominated by core MM4. Besides, the largest abundances of N-bearing and O-bearing molecules appear in MM8 and MM4, respectively. These all indicate clear chemical differentiation of N/O between MM8 and MM4. Such chemical differentiation might be affected by different initial temperature toward the MM8 and MM4 at their accretion phase.  
 
 4. We found tight correlations between the column density of molecular pairs  CH$_{3}$OH/C$_{2}$H$_{5}$OH, CH$_{3}$OH/CH$_{3}$OCH$_{3}$, and CH$_{3}$OCHO/CH$_{3}$OCH$_{3}$. Furthermore, the similar spatial distribution between them is seen, indicating that each pair is likely chemically linked, and may share a common precursor. The gas distribution of CH$_{3}$CHO is less correlated to H$_{2}$CCO and C$_{2}$H$_{5}$OH. Their column density/abundance correlations are not very clear. It seems likely that the chemical link of O-bearing COMs in the formula C$_{2}$H$_{\rm n}$O is uncertain at present work. 
 
 5. The comparison of HC$_{3}$N and HC$_{5}$N suggests that they trace different gas structures, and the formation mechanisms of HC$_{3}$N and HC$_{5}$N may differ in hot and cold regions.

\section*{Acknowledgements}

This paper makes use of the following ALMA data: ADS/JAO.ALMA\#2019.1.00685.S. ALMA is a partnership of ESO (representing its member states), NSF (USA), and NINS (Japan), together with NRC (Canada), MOST and ASIAA (Taiwan), and KASI (Republic of Korea), in cooperation with the Republic of Chile. The Joint ALMA Observatory is operated by ESO, AUI/NRAO, and NAOJ. This work has been supported by the National Key R\&D
Program of China (No. 2017YFA0402701), by the National Natural Science Foundation of China (NSFC) under grant No.12033005 and No.11947064, and Yunnan Applied Basic Research Projects under grant No.202001AU070123. Tie Liu acknowledges the supports by National Natural Science Foundation of China (NSFC) through grants No.12073061 and No.12122307, the international partnership program of Chinese academy of sciences through grant No.114231KYSB20200009, and Shanghai Pujiang Program 20PJ1415500. H.-L. Liu is supported by NSFC through the grant No.12103045. G.G. and L.B. gratefully acknowledge support from ANID BASAL project FB210003. C.W.L is supported by the Basic Science Research Program (2019R1A2C1010851) through the National Research Foundation of Korea. D.L. Li acknowledges support from NSFC through grants No.12173075 and support from Youth Innovation Promotion Association CAS. K.T. was supported by JSPS KAKENHI Grant Number 20H05645. Mengyao Tang is supported by the scientific research fund project of Yunnan Provincial Education Department (No. 2022J0855). YZ thanks the National Science Foundation of China (NSFC, Grant No. 11973099) and the science research grants from the China Manned Space Project (NO. CMS-CSST-2021-A09 and CMS-CSST-2021-A10) for financial supports.

Best

\textbf{Data availability}

The data underlying this article are available in the ALMA archive.




\begin{thebibliography}{}
 \bibitem[Abplanalp et al.(2016)]{Abplanalp16} Abplanalp, M. J., Gozem, S., \& Krylov, A. I. 2016, PNAS, 113, 7727
 \bibitem[Ag\'undez et al.(2021)]{Agundez21} Ag\'undez, M., Marcelino, N., Tercero, B., Cabezas, C., de Vicente, P., Cernicharo, J. 2021, A\&A, 649, 4
 \bibitem[Bacmann et al.(2012)]{Bacmann12} Bacmann, A., Taquet, V., Faure, A., Kahane, C., \& Ceccarelli, C. 2012, A\&A, 541, L12
 \bibitem[Bally et al.(2005)]{Bally05} Bally, J., Moeckel, N., \& Throop, H. Evolution of UV-Irradiated Protoplanetary Disks. Chondrites and the Protoplanetary Disk, ed. Alexander N. Krot, Edward R. D. Scott, \& Bo Reipurth. ASP Conference. 2005, Ser. 341: 81
 \bibitem[Balucani et al.(2015)]{Balucani15} Balucani, N., Ceccarelli, C., \& Taquet, V. 2015, MNRAS, 449, L16
 \bibitem[Belloche et al.(2013)]{Belloche13} Belloche, A., M\"uller, H. S. P., Menten, K. M., Schilke, P., \& Comito, C. 2013, A\&A, 559, A47
 \bibitem[Belloche et al.(2016)]{Belloche16} Belloche, A., M\"uller, H. S. P., Garrod, R. T., Menten, K. M. 2016, A\&A, 587, A91
 \bibitem[Belloche et al.(2017)]{Belloche17} Belloche, A., Meshcheryakov, A. A., Garrod, R. T., Ilyushin, V. V., Alekseev, E. A., Motiyenko, R. A., Margul\'es, L., M\"uller, H. S. P., et al. 2017, A\&A, 601, A49
 \bibitem[Beltr\'an et al.(2009)]{Beltran09} Beltr\'an, M. T., Massi, F., L\'opez, R., Girart, J. M., \& Estalella, R. 2009, A\&A, 504, 97
 \bibitem[Bennett \& Kaiser (2007)]{Bennett07a} Bennett, Chris J., \& Kaiser, Ralf I. 2007, ApJ, 661, 899
 \bibitem[Bennett et al. (2007)]{Bennett07b} Bennett, Chris J., Chen, Shih-Hua., Sun, Bing-Jian., et al. 2007, ApJ, 660, 1588
 \bibitem[Bergantini, Maksyutenko \&  Kaiser. (2017)]{Bergantini17} Bergantini, Alexandre., Maksyutenko, Pavlo., Kaiser, Ralf I. 2017, ApJ, 841, 96
 \bibitem[Bergner et al.(2017)]{Bergner17} Bergner, J. B., Oberg, K. I., \& Rajappan, M. 2017, ApJ,
  845, 29
 \bibitem[Bergin et al.(1996)]{Bergin1996} Bergin, E. A., Snell, R. L., \& Goldsmith, P. F. 1996, ApJ, 460, 343
 \bibitem[Bisschop et al.(2007)]{Bisschop07} Bisschop, S. E., J{\o}rgensen, J. K., van Dishoeck, E. F., \& de Wachter, E. B. M. 2007, A\&A, 465, 913
 \bibitem[Biver \& Bockel\'{e}e-Morvan (2019)]{Biver19} Biver, N. \& Bockel\'ee-Morvan, D. 2019, ACS Earth and Space Chemistry, 3, 1550
 \bibitem[Blake et al.(1987)]{Blake1987} Blake, Geoffrey A., Sutton, E. C., Masson, C. R., Phillips, T. G. 1987, ApJ, 315, 621
 \bibitem[B{\o}gelund et al.(2018)]{Bogelund18} B{\o}gelund, Eva G., McGuire, Brett A., Ligterink, Niels F. W. et al. 2018, A\&A, 615, A88B
 \bibitem[B{\o}gelund et al.(2019)]{Bogelund19} B{\o}gelund, E. G., Barr, A. G., Taquet, V., et al. 2019, A\&A, 628, A2
 \bibitem[Bonfand et al.(2019)]{Bonfand19} Bonfand, M., Belloche, A., Garrod, R. T., et al. 2019, A\&A, 628, 27
 \bibitem[Brouillet et al.(2013)]{Brouillet2013} Brouillet, N., Despois, D., Baudry, A., et al. 2013, A\&A, 550, 46
 \bibitem[Caselli et al.(1993)]{Caselli93} Caselli, P., Hasegawa, T. I., \& Herbst, E. 1993, ApJ, 408, 548
 \bibitem[Cernicharo et al.(2016)]{Cernicharo16} Cernicharo, J., Kisiel, Z., Tercero, B., et al. 2016, A\&A, 587, 4
 \bibitem[Charnley \& Rodgers (2005)]{Charnley05} Charnley, S. B., \& Rodgers, S. D. 2005, IAU Symp., 231, 237
 \bibitem[Chapman et al.(2009)]{Chapman09} Chapman, J. F., Millar, T. J., Wardle, M., et al. 2009, MNRAS, 394, 221
 \bibitem[Chuang et al.(2021)]{Chuang21} Chuang, K. -J., Fedoseev, G., Scirè, C., et al. 2021, A\&A, 650, A85
 \bibitem[Coletta et al.(2020)]{Coletta2020} Coletta, A., Fontani, F., Rivilla, V. M., et al. 2020, A\&A, 641, 54 
 \bibitem[Colzi et al.(2021)]{Colzi21} Colzi, L., Rivilla, V. M., Beltrán, M. T., et al. 2021, A\&A, 653, A129
 \bibitem[Crockett et al.(2014)]{Crockett14} Crockett, N. R., Bergin, E. A., Neill, J. L., et al. 2014, ApJ, 787, 112
 \bibitem[Crockett et al.(2015)]{Crockett15} Crockett, N. R., Bergin, E. A., Neill, J. L., Favre, C., Blake, G. A., et al. 2015, ApJ, 806, 239 
 \bibitem[Dall$^{'}$Olio et al.(2019)]{Dallolio19} Dall$^{'}$Olio, D., Vlemmings, W. H. T., Persson, M.V., et al. 2019, A\&A, 626, 36
 \bibitem[Dickens et al.(2000)]{Dickens2000} Dickens, J. E., Irvine, W. M., Snell, R. L., et al. 2000, ApJ, 542, 870
 \bibitem[Drozdovskaya et al.(2019)]{Drozdovskaya2019} Drozdovskaya, M. N., van Dishoeck, E. F., Rubin, M., J{\o}rgensen, J. K., Altwegg, K. 2019, MNRAS, 490, 50
 \bibitem[Duronea et al.(2019)]{Duronea2019} Duronea, N. U., Bronfman, L., Mendoza, E., et al. 2019, MNRAS, 489, 1519
 \bibitem[Esplugues et al.(2013)]{Esplugues13} Esplugues, G. B., Cernicharo, J., Viti, S., et al. 2013, A\&A, 559, A51
 \bibitem[Favre et al.(2011)]{Favre11} Favre, C., Despois, D., Brouillet, N., et al. 2011, A\&A, 532, A32
 \bibitem[Feng et al.(2015)]{Feng15} Feng, S., Beuther, H., Henning, T., et al. 2015, A\&A, 581, A71
 \bibitem[Fontani et al.(2017)]{Fontani17} Fontani, F., Ceccarelli, C., Favre, C., et al. 2017, A\&A, 605, A57
 \bibitem[Frau et al.(2010)]{Frau10} Frau, P., Girart, J. M., Beltrán, M. T., et al. 2010, ApJ, 723, 1665
 \bibitem[Friedel \& Snyder (2008)]{Friedel08} Friedel, D. N., \& Snyder, L. E. 2008, ApJ, 672, 962
 \bibitem[Garay et al.(1993)]{Garay93} Garay, G., Rodriguez, L. F., Moran, J. M., \& Churchwell, E. 1993, ApJ, 418,368
 \bibitem[Garrod \& Herbst. (2006)]{Garrod06} Garrod, R. T., \& Herbst, E. 2006, A\&A, 457, 927
 \bibitem[Garrod et al.(2008)]{Garrod08} Garrod, R. T., Weaver, S. L. W., \& Herbst, E. 2008, ApJ, 682, 283
 \bibitem[Gerner et al.(2014)]{Gerner14} Gerner, T., Beuther, H., Semenov, D., et al. 2014, A\&A, 563, 97
 \bibitem[Gibb et al.(2000)]{Gibb2000} Gibb, E., Nummelin, A., Irvine, W. M., et al. 2000, ApJ, 545, 309
 \bibitem[Gieser et al.(2021)]{Gieser21} Gieser, C., Beuther, H., Semenov, D., et al. 2021, A\&A, 648, 66
 \bibitem[Goldsmith (2001)]{Goldsmith2001} Goldsmith, Paul F. 2001, ApJ, 557, 736
 \bibitem[Gorai et al.(2021)]{Gorai2021} Gorai, Prasanta., Das, Ankan., Shimonishi, Takashi., et al. 2021, ApJ, 907, 108 
 \bibitem[Green et al.(2014)]{Green14} Green, C.-E., Green, J. A., Burton, M. G., et al. 2014, MNRAS, 443, 2252
 \bibitem[Hasegawa et al.(1992)]{Hasegawa1992} Hasegawa, T. I., Herbst, E., Leung, C. M. 1992, ApJS, 82, 167
 \bibitem[Herbst \& van Dishoeck.(2009)]{Herbst09} Herbst, E., \& van Dishoeck, E. F. 2009, ARA\&A, 47, 427
 \bibitem[Herv\'ias-Caimapo et al.(2019)]{Hervias2019} Herv\'ias-Caimapo, Carlos., Merello, Manuel., Bronfman, Leonardo., et al. 2019, ApJ, 872, 200
 \bibitem[Hirota et al.(2009)]{Hirota09} Hirota, T., Ohishi, M., \& Yamamoto, S. 2009, ApJ, 699, 585
 \bibitem[Hoare et al.(2007)]{Hoare07} Hoare, M. G., Kurtz, S. E., Lizano, S., Keto, E., \& Hofner, P. 2007, in Protostars and Planets V, ed. B. Reipurth, D. Jewitt, \& K. Keil, 181
 \bibitem[Hofner et al.(1994)]{Hofner94} Hofner, P., Churchwell, E., Kurtz, S., Cesaroni, R., \& Walmsley, C. M. 1994, ApJ, 429, L85
 \bibitem[Hofner et al.(2001)]{Hofner01} Hofner, P., Wiesemeyer, H., \& Henning, T. 2001, ApJ, 549, 425
 \bibitem[Jaber et al.(2017)]{Jaber17} Jaber, A. A., Ceccarelli, C., Kahane, C., et al. 2017, A\&A, 597, A40
 \bibitem[Jim\'enez-Serra et al.(2016)]{Jimenez16} Jim\'enez-Serra, I., Vasyunin, A. I., Caselli, P., et al. 2016, ApJ, 830, L6
 \bibitem[Johansson et al.(1984)]{Johansson84} Johansson, L. E. B., Andersson, C., Ellder, J., et al. 1984, A\&A, 130, 227
 \bibitem[J{\o}rgensen et al.(2002)]{Jorgensen02} J{\o}rgensen, J. K., Schöier, F. L., \& van Dishoeck, E. F. 2002, A\&A, 389, 908
 \bibitem[J{\o}rgensen et al.(2012)]{Jorgensen12} J{\o}rgensen, J. K., Favre, C., Bisschop, S. E., et al. 2012, ApJ, 757, L4
 \bibitem[J{\o}rgensen et al.(2016)]{Jorgensen16} J{\o}rgensen, J. K., van der Wiel, M. H. D., Coutens, A., et al. 2016, A\&A, 595, A117
 \bibitem[J{\o}rgensen et al.(2018)]{Jorgensen18} J{\o}rgensen, J. K., M\"uller, H.S.P., Calcutt, H., Coutens, A., Drozdovskaya, M. N., et al. 2018. A\&A 620, A170
 \bibitem[J{\o}rgensen et al.(2020)]{Jorgensen20} J{\o}rgensen, J. K., Belloche, A., Garrod, R. T. 2020, ARA\&A, 58, 727
 \bibitem[Kauffmann et al.(2008)]{Kauffmann08} Kauffmann, J., Bertoldi, F., Bourke, T. L., Evans, N. J., I., \& Lee, C. W. 2008, A\&A, 487, 993
 \bibitem[Kennicutt (1998)]{Kennicutt98} Kennicutt, R. C. 1998, ARA\&A, 36, 189
 \bibitem[Kolesnikov\'a (2014)]{Kolesnikova14} Kolesnikov\'a, L., Tercero, B., Cernicharo, J., et al. 2014, ApJ, 784, L7
 \bibitem[Krumholz et al.(2014)]{Krumholz14} Krumholz, M. R., Bate, M. R., Arce, H. G., et al., 2014, Protostars and Planets VI, Henrik Beuther, Ralf S. Klessen, Cornelis P. Dullemond, and Thomas Henning (eds.), University of
 Arizona Press, Tucson, 914pp., p.243-266
 \bibitem[Kuan et al.(2004)]{Kuan04} Kuan, Yi-Jehng., Huang, Hui-Chun., Charnley., Steven B., et al. 2004, ApJ, 616, L27
 \bibitem[Kurtz et al.(2000)]{Kurtz2000} Kurtz, S., Cesaroni, R., Churchwell, E., Hofner, P., \& Walmsley, C. M. 2000, in Protostars and Planets IV, ed. V. Mannings, A. P. Boss, \& S. S. Russell, 299–326
 \bibitem[Kurtz et al.(1994)]{Kurtz94} Kurtz, S., Churchwell, E., \& Wood, D. O. S. 1994, ApJS, 91, 659
 \bibitem[Laas et al.(2011)]{Laas11} Laas, J. C., Garrod, R. T., Herbst, E., \& Widicus Weaver, S. L. 2011, ApJ, 728, 71
 \bibitem[Langer \& Penzias (1990)]{Langer1990} Langer, W. D., \& Penzias, A. A. 1990, ApJ, 357, 477
 \bibitem[Langer \& Penzias (1993)]{Langer1993} Langer, W. D., \& Penzias, A. A. 1993, ApJ, 408, 539
 \bibitem[Law et al.(2021)]{Law2021} Law, C. J., Zhang, Q., \"Oberg, K. I., et al. 2021, ApJ, 909, 214
 \bibitem[Lee et al.(2019)]{Lee2019} Lee, Jeong-Eun., Lee, Seokho., Baek, Giseon., et al, 2019, Nature Astronomy, 3, 314
 \bibitem[Lefloch et al.(2017)]{Lefloch17} Lefloch, B., Ceccarelli, C., Codella, C., et al. 2017, MNRAS, 469, L73
 \bibitem[Ligterink et al.(2020)]{Ligterink20} Ligterink, Niels F. W., El-Abd, Samer J., Brogan, Crystal L., Hunter, Todd R., Remijan, Anthony J., Garrod, Robin T., McGuire, Brett M. 2020, ApJ, 901, 37
 \bibitem[Lis et al.(1991)]{Lis91} Lis, D. C., Carlstrom, J. E., \& Keene, J. 1991, ApJ, 380, 429
 \bibitem[Liu et al.(2011)]{Liu11} Liu, T., Wu, Y., Liu, S.-Y., et al. 2011, ApJ, 730, 102
 \bibitem[Liu et al.(2017)]{Liu17} Liu, T., Lacy, J., Li, P. k., Wang, K., et al, 2017, ApJ, 849, 25
 \bibitem[Liu et al.(2020a)]{Liu20a} Liu, Tie., Evans, Neal J., Kim, Kee-Tae., et al, 2020a, MNRAS, 496, 2790
 \bibitem[Liu et al.(2020b)]{Liu20b} Liu, Tie., Evans, Neal J., Kim, Kee-Tae., et al, 2020b, MNRAS, 496, 2821
 \bibitem[Liu et al.(2021)]{Liu2021} Liu, Hong-Li., Liu, Tie., Evans, Neal J., et al, 2021, MNRAS, 505, 2801 
 \bibitem[Lykke et al.(2015)]{Lykke15} Lykke, J. M., Favre, C., Bergin, E. A., et al. 2015, A\&A, 582, A64
 \bibitem[Maity et al.(2014)]{Maity14} Maity, S., Kaiser, R. I., \& Jones, B. M. 2014, ApJ, 789, 36
 \bibitem[Maity et al.(2015)]{Maity15} Maity, S., Kaiser, R. I., \& Jones, B. M. 2015, PCCP, 17, 3081
 \bibitem[Majumdar et al.(2016)]{Majumdar16} Majumdar, L., Gratier, P., Vidal, T., et al. 2016, MNRAS, 458, 1859
 \bibitem[McGuire et al.(2018)]{McGuire18} McGuire, Brett A., Burkhardt, Andrew M., Kalenskii, Sergei., Shingledecker, Christopher N., Remijan, Anthony J., Herbst, Eric., McCarthy, Michael C. 2018, Science, 359, 202
 \bibitem[Mininni et al.(2020)]{Mininni20} Mininni, C., Beltr\'an, M. T., Rivilla, V. M., et al. 2020, A\&A, 644, A84
 \bibitem[Molet et al.(2019)]{Molet19} Molet, J., Brouillet, N., Nony, T., Gusdorf, A., Motte, F., et al. 2019. A\&A, 626, A132
 \bibitem[M\"oller et al.(2013)]{Moller13} M\"oller, T., Bernst, I., Panoglou, D., et al. 2013, A\&A, 549, 21
 \bibitem[M\"oller et al.(2017)]{Moller17} M\"oller, T., Endres, C., Schilke, P. 2017, A\&A, 598, 7
 \bibitem[M\"uller et al.(2001)]{Muller01} M\"uller, H. S. P., Thorwirth, S., Roth, D. A.,\& Winnewisser, G. 2001, A\&A, 370, 49
 \bibitem[M\"uller et al.(2005)]{Muller05} M\"uller, H. S. P., Schl\"{o}der, F., Stutzki, J., \& Winnewisser, G. 2005, Journal of Molecular Structure, 742, 215
 \bibitem[M\"uller et al.(2016)]{Muller16} M\"uller, H.S.P., Belloche, A., Xu, L. et al. 2016, A\&A, 587, A92
 \bibitem[Nummelin et al.(2000)]{Nummelin00} Nummelin, A., Bergman, P., Hjalmarson, A., et al. 2000, ApJS, 128, 213
 \bibitem[Ohishi et al.(1991)]{Ohishi91} Ohishi, M., Kawaguchi, K., Kaifu, N., et al. 1991, in ASP Conf. Ser. 16, Atoms, Ions and Molecules: New Results in Spectral Line Astrophysics (San Francisco, CA: ASP), 387
\bibitem[\"Oberg et al.(2009)]{Oberg2009} \"Oberg, K. I., Garrod, R. T.; van Dishoeck, E. F., Linnartz, H. 2009, A\&A, 504, 891
\bibitem[\"Oberg et al.(2010)]{Oberg2010} \"Oberg, K. I., Bottinelli, S., Jørgensen, J. K., \& van Dishoeck, E. F. 2010, ApJ, 716, 825
\bibitem[Ospina-Zamudio et al.(2018)]{Ospina-Zamudio18} Ospina-Zamudio, J., Lefloch, B., Ceccarelli, C., et al. 2018, A\&A, 618, A145
\bibitem[Ossenkopf \& Henning (1994)]{Ossenkopf1994} Ossenkopf, V. \& Henning, T. 1994, A\&A, 291, 943
 \bibitem[Peng et al.(2017)]{Peng17} Peng, Y. P., Qin, S. L., Schilke, P., et al. 2017, ApJ, 837, 49
 \bibitem[Pickett et al.(1998)]{Pickett98} Pickett, H. M., Poynter, R. L., Cohen, E. A., et al. 1998, Quant. Spectrosc. \& Rad. Transfer, 60, 883
 \bibitem[Qin et al.(2010)]{Qin10} Qin, Sheng-Li., Wu, Yuefang., Huang, Maohai., et al. 2010, ApJ, 711, 399
 \bibitem[Qin et al.(2015)]{Qin15} Qin, Sheng-Li., Schilke, Peter., Wu,Jingwen., et al. 2015, ApJ, 803, 39
  \bibitem[Requena-Torres et al.(2006)]{Requena-Torres06} Requena-Torres, M. A., Mart\'in-Pintado, J., Rodr\'iguez-Franco, A., et al. 2006, A\&A, 455, 971
  \bibitem[Requena-Torres et al.(2008)]{Requena-Torres08} Requena-Torres, M. A., Mart\'n-Pintado, J., Mart\'n, S., \& Morris, M. R. 2008, ApJ, 672, 352
  \bibitem[Remijan et al.(2004a)]{Remijan04a} Remijan, A., Sutton, E. C., Snyder, L. E., Friedel, D. N., Liu, S. -Y., Pei, C. -C. 2004a, ApJ, 606, 917 
  \bibitem[Remijan et al.(2004b)]{Remijan04b} Remijan, A., Shiao, Y. S., Friedel, D. N., Meier, D. S., Snyder, L. E, 2004b, ApJ, 617, 384
  \bibitem[Richard et al.(2013)]{Richard13} Richard, C., Margul\`es, L., Caux, E., et al. 2013, A\&A, 552, A117
  \bibitem[Rivilla et al.(2017a)]{Rivilla17a} Rivilla, V. M., Beltr\'an, M. T., Mart\'in-Pintado, J., Fontani, F., Caselli, P., Cesaroni, R. 2017a, A\&A, 599, A26 
  \bibitem[Rivilla et al.(2017b)]{Rivilla17b} Rivilla, V. M., Beltr\'an, M. T., Cesaroni, R., et al. 2017b, A\&A, 598, A59
  \bibitem[Rodr\'iguez-Almeida et al.(2021)]{Rodriguez-Almeida2021} Rodr\'iguez-Almeida, Lucas F., Jim\'enez-Serra, I., Rivilla, V. M., et al. 2021, ApJ, 912, 11
  \bibitem[Ruaud et al.(2015)]{Ruaud15} Ruaud, M., Loison, J. C., Hickson, K. M. et al. 2015, MNRAS, 447, 4004
 \bibitem[Ruiterkamp et al.(2007)]{Ruiterkamp07} Ruiterkamp, R., Charnley, S. B., Butner, H. M., et al. 2007, Ap\&SS, 310, 181
 \bibitem[Sakai \& Yamamoto (2013)]{Sakai13} Sakai, N., \& Yamamoto, S. 2013, ChRv, 113, 8981
 \bibitem[Sanna et al.(2009)]{Sanna09} Sanna, A., Reid, M. J., Moscadelli, L., et al. 2009, ApJ, 706, 464
 \bibitem[Shingledecker et al.(2019)]{Shingledecker19} Shingledecker, C. N., Vasyunin, A., Herbst, E., \& Caselli, P. 2019, ApJ, 876, 140
 \bibitem[Skouteris et al.(2018)]{Skouteris18} Skouteris, D., Balucani, N., Ceccarelli, C., et al. 2018, ApJ, 854, 135
 \bibitem[Suzuki et al.(1992)]{Suzuki92} Suzuki, H., Yamamoto, S., Ohishi, M., et al. 1992, ApJ, 392, 551
 \bibitem[Su et al.(2005)]{Su05} Su, Y.-N., Liu, S.-Y., Lim, J., \& Chen, H.-R. 2005, in Conf. Proc. Protostars and Planets V, Submillimeter Observations of the High-mass Star Forming Complex G9.62+0.19 (HI: Waikoloa Village),1286, 8336
 \bibitem[Tan et al.(2013)]{Tan13} Tan, J. C., Kong, S., Butler, M. J., Caselli, P., \& Fontani, F. 2013, ApJ, 779, 96
 \bibitem[Taniguchi et al.(2017)]{Taniguchi17} Taniguchi, Kotomi., Saito, Masao., Hirota, Tomoya., et al. 2017, ApJ, 844, 68
 \bibitem[Taniguchi et al.(2018)]{Taniguchi18} Taniguchi, K., Saito, M., Sridharan, T. K., \& Minamidani, T. 2018, ApJ, 854, 133
 \bibitem[Taniguchi et al.(2019)]{Taniguchi19} Taniguchi, K., Herbst, E., Caselli, P., et al. 2019, ApJ, 881, 57
 \bibitem[Taquet et al.(2015)]{Taquet15} Taquet, V., L\'{o}pez-Sepulcre, A., Ceccarelli, C., et al. 2015, ApJ, 804, 81
 \bibitem[Tercero et al.(2018)]{Tercero18} Tercero, B., Cuadrado, S., L\'opez, A., Brouillet, N., Despois, D., Cernicharo, J. 2018, A\&A, 620, 6
 \bibitem[Testi et al.(1998)]{Testi98} Testi, Leonardo., Felli, Marcello., Persi, Paolo., \& Roth, Miguel. 1998, A\&A, 329, 233
  \bibitem[Testi et al.(2000)]{Testi2000} Testi, L., Hofner, P., Kurtz, S., \& Rupen, M. 2000, A\&A, 359, L5
 \bibitem[Turner (1977)]{Turner77} Turner, B. E. 1977, ApJL, 213, L75
 \bibitem[Vastel et al.(2014)]{Vastel14} Vastel, C., Ceccarelli, C., Lefloch, B., \& Bachiller, R. 2014, ApJL, 795, L2
 \bibitem[Vasyunin \& Herbst.(2013)]{Vasyunin13} Vasyunin, A. I., \& Herbst, E. 2013, ApJ, 762, 86
 \bibitem[Wang et al.(2009)]{Wang2009} Wang, K. S., Kuan, Y. J., Liu, S. Y., et al. 2009, ASPC, 420.
  \bibitem[Woods et al.(2013)]{Woods13} Woods, P. M., Slater, B., Raza, Z., et al. 2013, ApJ, 777, 90
 \bibitem[Wyrowski et al.(1999)]{Wyrowski99} Wyrowski, F., Schilke, P., Walmsley, C. M., Menten, K. M. 1999, ApJ, 514, 43
 \bibitem[Yamamoto (2017)]{Yamamoto17} Yamamoto, S. 2017, Introduction to Astrochemistry: Chemical Evolution from Interstellar Clouds to Star and Planet Formation (Berlin: Springer)
 \bibitem[Yan et al.(2019)]{Yan19} Yan, Y. T., Zhang, J. S., Henkel, C., et al. 2019, ApJ, 877, 154
 \bibitem[Zeng et al.(2018)]{Zeng18} Zeng, S., Jim\'enez-Serra, I., Rivilla, V. M., et al. 2018, MNRAS, 478, 2962
 \bibitem[Zernickel et al.(2012)]{Zernickel12} Zernickel, A., Schilke, P., Schmiedeke, A., Lis, D. C., Brogan, C. L., et al. 2012. A\&A, 546, A87
 \bibitem[Zhang et al.(2021)]{Zhang21} Zhang, C., Wu, Y., Liu, X. -C., et al. 2021, A\&A, 648, 83



\end{thebibliography}



\appendix
\section{Spectra from 12-m array and 12-m + ACA combined data.}
Interferometric observations can filter out large scale structure. The combination data from interferometric and single-dish observations sample small-scale components and extended structures simultaneously. In this work, the spectra we used is from 12-m observations alone. We claim that this is safe for our data modeling, because the dense cores of G9.62+0.19 are very compact (see Table~\ref{tab:continuum}). The information from these cores are not lost. Figure~\ref{fig:A1_12mvscombine} compare the spectrum of MM8 from 12-m with 12-m + ACA combined data. It exhibited that 2\arcsec-averaged spectrum from two observations are very similar in line intensity and line profile (see panel (a) and (b) in Figure~\ref{fig:A1_12mvscombine}). Spectrum extracted from 10\arcsec area showed that strong lines such as CS and HC$_{3}$N appeared minor difference in line intensity(see panel (c) and (d)). This implies that the structures smaller than 10\arcsec are almost unaffected by missing flux, while larger scale structures may be affected. Therefore, 12-m array and combined data have no difference when using them to study the hot core chemistry, but 12-m array data have higher angular resolution for better investigating the compact cores.  
\begin{figure*}
\setcounter{figure}{0}
\centerline{\resizebox{1.0\hsize}{!}{\includegraphics{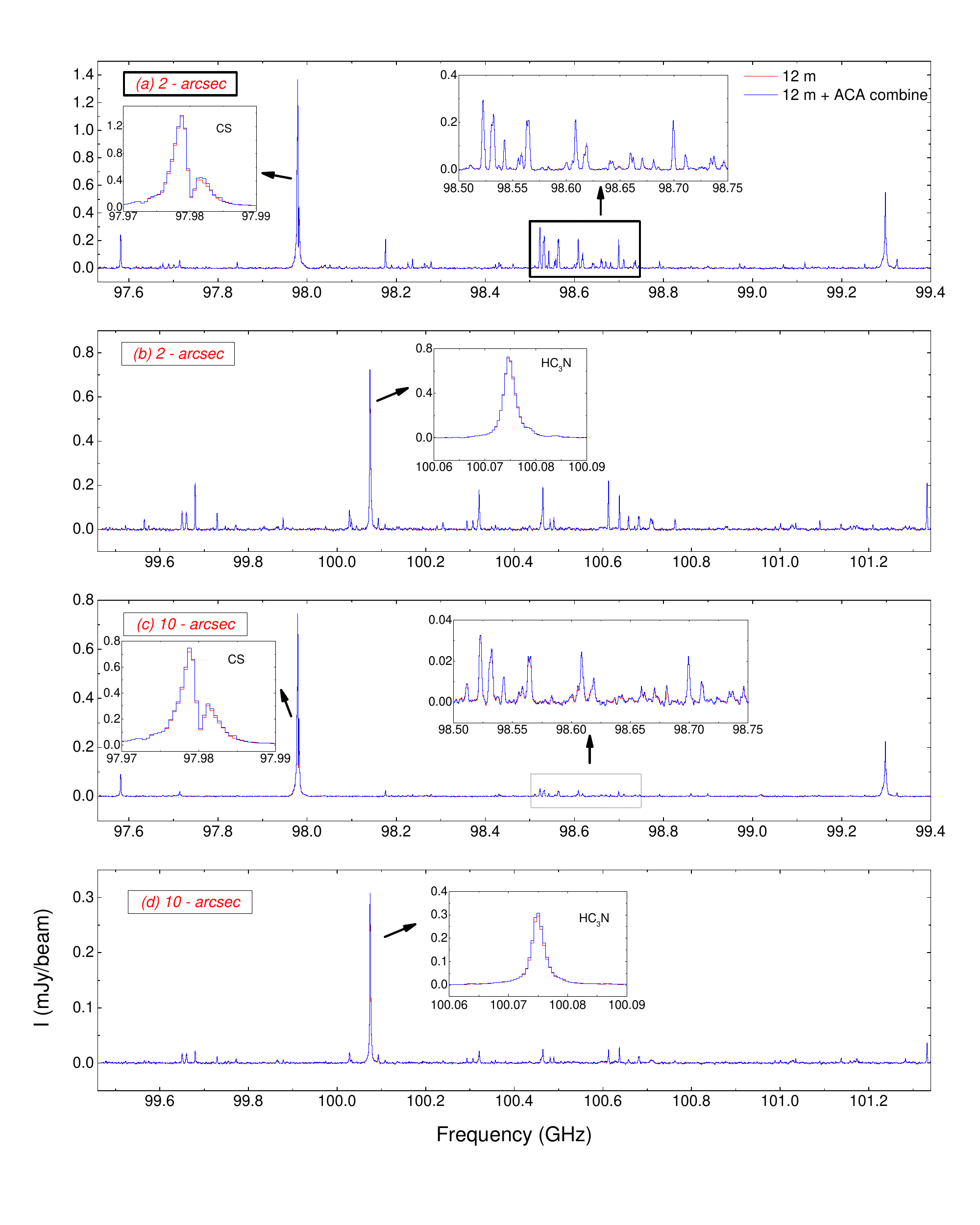}}}
\caption{Spectra at the MM8 peak from 12-m array (red) and 12-m + ACA combined data (blue). Spectra in panel (a) and (b) are averaged over 2\arcsec area. Panel (c) and (d) show the averaged spectra over 10\arcsec. The frequency range is 97.53 $-$ 99.40 GHz for (a) and (C), while 99.46 $-$ 101.34 GHz for (b) and (d). The zoom-in spectra of CS, HC$_{3}$N, and a part of relatively weak lines from 98.50 to 98.75 GHz are also showed.  
}
\label{fig:A1_12mvscombine}
\end{figure*}

\section{Detected molecular lines from six dense cores.}
\setlength{\tabcolsep}{5pt}
\onecolumn
\begin{longtable}{llcccccccccccccc}
\caption{Transitions of detected molecules toward G9.62+0.19 }
\label{tab:Transitions} \\
\hline\hline
Rest Frequency & Quantum Numbers$^{a}$ & $S_{ij}\mu^{2}$ & E$_{\rm u }$ & \multicolumn{6}{c}{Line Optical Depth($\tau^{\rm line}$)}  \\
\cline{5-10}\\
(GHz) & & (Debye$^{2}$) & (K) &  MM1 & MM6 & MM7/G & MM8/F & MM4/E & MM11/D  \\
\hline
\endfirsthead
\caption{continued.}\\
\hline\hline
Rest Frequency & Quantum Numbers$^{a}$ & $S_{ij}\mu^{2}$ & E$_{\rm u }$ & \multicolumn{6}{c}{Line Optical Depth($\tau^{\rm line}$)}  \\
\cline{5-10}\\
(GHz) & & (Debye$^{2}$) & (K) & MM1 & MM6 & MM7/G & MM8/F & MM4/E & MM11/D  \\
\hline
\endhead
\hline
\endfoot
\multicolumn{8}{c}{\textbf{Hydrogen Recombination Line} }\\
99.02295 & H(40)$\alpha$ & $\cdots$ & $\cdots$ & $\surd^b$ & $\cdots^c$ & $\cdots$ & $\cdots$ & $\cdots$ & $\surd$ \\
99.22520 &  H(50)$\beta$ & $\cdots$ & $\cdots$ & $\cdots$ & $\cdots$ & $\cdots$ & $\cdots$& $\cdots$ & $\surd$ \\
\hline
\multicolumn{8}{c}{\textbf{Formaldehyde} (H$_{2}$CO)} \\
101.33299  & $J(K_{a},K_{c})=6( 1, 5)- 6( 1, 6)$ & 5.06 & 88 & $\cdots$ & 7.46(-1) & 1.01(-1) & 1.83(-1) & 3.25(-1) & 4.59(-2) \\
\hline
\multicolumn{8}{c}{\textbf{Ketene} (H$_{2}$CCO)} \\
100.09451  & $J(K_{a},K_{c})=5(1,5)-4(1,4)$ & 28.79 & 27 & $\cdots$ & $\cdots$ & 7.10(-3) & 4.23(-2) & 7.02(-2) & 3.56(-2) \\
101.00245  & $5(3,3)-4(3,2)$ & 19.20 & 133 & $\cdots$  & $\cdots$  & 5.92(-3) & 1.98(-2) & 3.33(-2) & 1.33(-2) \\
101.02447  & $5(2,4)-4(2,3)$ & 8.40  & 67 & $\cdots$ & $\cdots$  & 6.42(-3) & b & 1.40(-2) & $\cdots$\\
101.03223  & $5(2,3)-4(2,2)$ & 8.40  & 67 & $\cdots$ &$\cdots$ & 6.76(-3) & 8.26(-3) & 1.33(-2) & $\cdots$\\
101.03671  & $5(0,5)-4(0,4)$ & 10  & 15 & $\cdots$ & $\cdots$ & 1.81(-2) & 1.66(-2) & 2.64(-2) & 1.54(-2) \\
\hline
\multicolumn{8}{c}{\textbf{Methanol} (CH$_{3}$OH v$_{t}$ = 0)} \\
97.58279    & $J(K_{a},K_{c})=2(1, 1)-1(1, 0)A$  & 4.85 & 22 & $\cdots$ & 2.46(-2) & 2.33(-1) & 2.09(-1) & 3.05(-1) & 2.35(-1)\\
97.67768  & $21( 6,16)-22( 5,17)A$ & 22.90 & 729 & $\cdots$ & $\cdots$ & $\cdots$ & b$^d$ & 2.42(-2) & $\cdots$\\
97.67880  & $21( 6,15)-22( 5,18)A$ & 22.90 & 729 &$\cdots$ &$\cdots$ & $\cdots$ & b & 2.42(-2) & $\cdots$\\
98.03064  & $24( 6,19)-23( 7,16) A$ & 6.4 & 889 &$\cdots$ & $\cdots$ & $\cdots$ & $\cdots$ & 1.33(-2) & $\cdots$\\
98.03068  & $24( 6,18)-23( 7,17) A$ & 6.4 & 889 &$\cdots$ & $\cdots$ & $\cdots$ & $\cdots$ & 1.33(-2) & $\cdots$\\
100.63887   & $13(-2, 12)-12(-3, 10)E$ & 15 & 234 & $\cdots$ & $\cdots$ & 8.87(-2) & 1.45(-1) & 2.63(-1) & 9.21(-2)\\
101.18545   & $6(2, 4)-6(-1, 5)E$ & 0.09 & 75 & $\cdots$ & $\cdots$ & $\cdots$& $\cdots$ & b & $\cdots$ \\
101.29341   & $7(2, 5)-7(-1, 6)E$ & 0.19 & 91 & $\cdots$ & $\cdots$ & $\cdots$ & b & b & $\cdots$ \\
\multicolumn{8}{c}{(CH$_{3}$OH v$_{t}$ = 1)} \\
99.37434    & $J(K_{a},K_{c})=15( 6,10)-14( 7, 8) E$   & 11.11 & 771  & $\cdots$ & $\cdots$ & $\cdots$ & $\cdots$ & 7.13(-3) & $\cdots$ \\
99.73094    & $6(- 1, 6)- 5(- 0, 5) E$  & 12.56 & 340 & $\cdots$ & $\cdots$ & 1.60(-2) & 3.62(-2) & 6.82(-2) & 1.94(-2)\\
99.77283    & $20(- 3,17)-21(- 4,18) E$ & 70.53 & 902 & $\cdots$ & $\cdots$ & $\cdots$ & 8.65(-3) & 2.40(-2) & b\\
\hline
\multicolumn{8}{c}{\textbf{Acetaldehyde} (CH$_{3}$CHO)} \\
98.86331   & $J(K_{a},K_{c})=5(1,4) - 4(1,3) E$ & 60.69 & 17 & $\cdots$ & $\cdots$ & 1.58(-2) & 8.09(-3) & 1.22(-2) & 1.72(-2)\\
98.90094   & $5(1,4) - 4(1,3) A--$ & 60.70 & 16 & $\cdots$ & $\cdots$ & 1.58(-2) & 8.30(-3) & 1.21(-2) & 1.71(-2)\\
\hline
\multicolumn{8}{c}{\textbf{Methyl Formate} (CH$_{3}$OCHO,v=0)} \\
97.65127    & $J(K_{a},K_{c})=10(4, 7)-10(3, 8)E$ & 2.43 & 43 & $\cdots$ & $\cdots$ & $\cdots$ & $\cdots$ & b & $\cdots$\\
97.69426    & $10(4, 7)-10(3, 8)A$ & 2.45 & 43 & $\cdots$ & $\cdots$ & $\cdots$ & $\cdots$ & 6.52(-3) & $\cdots$\\
98.18233    & $8(7, 1)-7(7, 0)E$ & 5 & 54 & $\cdots$ & $\cdots$ & $\cdots$ & $\cdots$ & 1.24(-2) & $\cdots$\\
98.19065    & $8(7, 2)-7(7, 1)A$ & 5 & 54 & $\cdots$ & $\cdots$ & $\cdots$ & 1.22(-2) & 3.25(-2) & 1.38(-2)\\
98.19065    & $8(7, 1)-7(7, 0)A$ & 5 & 54 & $\cdots$ & $\cdots$ & $\cdots$ & 1.22(-2) & 3.25(-2) & 1.38(-2)\\
98.19146    & $8(7, 2)-7(7, 1)E$ & 5 & 54 & $\cdots$ & $\cdots$ & $\cdots$ & 1.22(-2) & 3.25(-2) & 1.38(-2)\\
98.27050    & $8(6, 2)-7(6, 1)E$ & 9.3 & 45 & $\cdots$ & $\cdots$ & $\cdots$ & 8.87(-3) & 2.42(-2) & 1.09(-2)\\
98.27892    & $8(6, 3)-7(6, 2)E$ & 9.3 & 45 & $\cdots$ & $\cdots$ & 4.49(-3) & 2.34(-2) & 6.21(-2) & 2.57(-2)\\
98.27976    & $8(6, 3)-7(6, 2)A$ & 9.3 & 45 & $\cdots$ & $\cdots$ & 4.49(-3) & 2.34(-2) & 6.21(-2) & 2.57(-2)\\
98.27976    & $8(6, 2)-7(6, 1)A$ & 9.3 & 45 & $\cdots$ & $\cdots$ & 4.49(-3) & 2.34(-2) & 6.21(-2) & 2.57(-2)\\
98.42420    & $8(5, 3)-7(5, 2)E$ & 13  & 38 & $\cdots$ & $\cdots$ & 6.01(-3) & 1.30(-2) & 3.54(-2) & 1.63(-2)\\
98.43180    & $8(5, 4)-7(5, 3)E$ & 13  & 38 & $\cdots$ & $\cdots$ & 9.69(-3) & 2.18(-2) & 5.71(-2) & 2.33(-2)\\
98.43276    & $8(5, 4)-7(5, 3)A$ & 13  & 38 & $\cdots$ & $\cdots$ & 9.69(-3) & 2.18(-2) & 5.71(-2) & 2.33(-2)\\
98.43580    & $8(5, 3)-7(5, 2)A$ & 13  & 38 & $\cdots$ & $\cdots$ & 7.43(-3) & 1.32(-2) & 3.61(-2) & 1.68(-2)\\
98.60685    & $8(3, 6)-7(3, 5)E$ & 18.24 & 27 & $\cdots$ & $\cdots$ & 1.05(-2) & 2.00(-2) & 5.50(-2) & 2.59(-2)\\
98.61116    & $8(3, 6)-7(3, 5)A$ & 18.27 & 27 & $\cdots$ & $\cdots$ & b & b & b & 2.46(-2)\\
98.68261    & $8(4, 5)-7(4, 4)A$ & 15.96 & 32 & $\cdots$ & $\cdots$ & 9.80(-3) & 1.62(-2) & 4.39(-2) & 2.00(-2)\\
98.71200    & $8(4, 5)-7(4, 4)E$ & 15.43 & 32 & $\cdots$ & $\cdots$ & 9.22(-3) & b & 4.42(-2) & 2.03(-2)\\
98.74790    & $8(4, 4)-7(4, 3)E$ & 15.44 & 32 & $\cdots$ & $\cdots$ & 9.05(-3) & 1.64(-2) & 4.50(-2) & 2.10(-2)\\
98.79228    & $8(4, 4)-7(4, 3)A$ & 15.96 & 32 & $\cdots$ & $\cdots$ & 9.74(-3) & 1.70(-2) & 4.64(-2) & 2.17(-2)\\
98.83952    & $11(4, 8)-11(3, 9)E$ & 2.74 & 50 & $\cdots$ & $\cdots$ & $\cdots$ & $\cdots$ & 6.92(-3) & $\cdots$\\
98.87522    & $11(4, 8)-11(3, 9)A$ & 2.75 & 50 & $\cdots$ & $\cdots$ & $\cdots$ & $\cdots$ & 7.01(-3) & $\cdots$\\
99.13327    & $9(0, 9)-8(1, 8)E$ & 3.35 & 25 & $\cdots$ & $\cdots$ & $\cdots$ & 3.78(-3) & 1.02(-2) & $\cdots$\\
99.13576    & $9(0, 9)-8(1, 8)A$ & 3.35 & 25 & $\cdots$ & $\cdots$ & $\cdots$ & 3.76(-3) & 1.02(-2) & $\cdots$\\
100.07860   & $9(1, 9)-8(1, 8)E$ & 23.50 & 25 & $\cdots$ & $\cdots$ & b & b & b & b\\
100.08054   & $9(1, 9)-8(1, 8)A$ & 23.51 & 25 & $\cdots$ & $\cdots$ & b & b & b & 3.32(-2) \\
100.29460   & $8(3, 5)-7(3, 4)E$ & 18.25 & 27 & $\cdots$ & $\cdots$ & $\cdots$ & 1.95(-2) & 5.29(-2) & 2.42(-2) \\
100.30817   & $8(3, 5)-7(3, 4)A$ & 18.28 & 27 & $\cdots$ & $\cdots$ & $\cdots$ & 1.98(-2) & 5.40(-2) & 2.49(-2) \\
100.48224   & $8(1, 7)-7(1, 6)E$ & 20.63 & 23 & $\cdots$ & $\cdots$ & 1.40(-2) & 2.34(-2) & 6.43(-2) & 3.01(-2) \\
100.49068   & $8(1, 7)-7(1, 6)A$ & 20.63 & 23 & $\cdots$ & $\cdots$ & 1.41(-2) & 2.31(-2) & 6.31(-2) & 2.92(-2) \\
100.68154   & $9(0, 9)-8(0, 8)E$ & 23.54 & 25 & $\cdots$ & $\cdots$ & 1.43(-2) & 2.89(-2) & 7.44(-2) & 3.34(-2) \\
100.68336   & $9(0, 9)- 8(0, 8)A$ & 23.54 & 25 & $\cdots$ & $\cdots$ & 1.44(-2) & 2.86(-2) & 7.64(-2) & 3.50(-2) \\
100.69466   & $5(3, 3)-5(1, 4)E$ & 0.04 & 15 & $\cdots$ & $\cdots$ & $\cdots$ & $\cdots$ & 7.20(-3) & $\cdots$\\
\multicolumn{8}{c}{ (CH$_{3}$OCHO,v=1)} \\
97.57730   & $J(K_{a},K_{c})=8( 5, 3)- 7( 5, 2) E$  & 13.00 & 225 & $\cdots$ & $\cdots$ & $\cdots$ & 6.87(-3) & 1.01(-2) & $\cdots$\\
97.59716   & $8( 3, 6)- 7( 3, 5) A$  & 18.21 & 215 & $\cdots$ & $\cdots$ & $\cdots$ & 9.90(-3) & 1.47(-2) & $\cdots$\\
97.66140   & $8( 4, 5)- 7( 4, 4) A$  & 15.91 & 220 & $\cdots$ & $\cdots$ & $\cdots$ & 8.66(-3) & 1.29(-2) & $\cdots$\\
97.75288   & $8( 4, 4)- 7( 4, 3) A$  & 15.91 & 220 & $\cdots$ & $\cdots$ & $\cdots$ & 8.70(-3) & b & $\cdots$\\
97.88566   & $8( 5, 4)- 7( 5, 3) E$  & 12.99 & 225 & $\cdots$ & $\cdots$ & $\cdots$ & $\cdots$ & 9.90(-3) & $\cdots$\\
97.89711   & $8( 4, 4)- 7( 4, 3) E$  & 16.02 & 219 & $\cdots$ & $\cdots$ & $\cdots$ & $\cdots$ & b & $\cdots$\\
98.17629   & $8( 4, 5)- 7( 4, 4) E$  & 15.97 & 219 & $\cdots$ & $\cdots$ & $\cdots$ & b & b & $\cdots$\\
98.68242   & $8( 3, 6)- 7( 3, 5) E$  & 18.08 & 215 & $\cdots$ & $\cdots$ & $\cdots$ & b & b & $\cdots$\\
98.81529   & $8( 3, 5)- 7( 3, 4) E$  & 18.22 & 215 & $\cdots$ & $\cdots$ & $\cdots$ & $\cdots$ & 1.55(-2) & $\cdots$\\
99.08951   & $8( 3, 5)- 7( 3, 4) A$  & 18.22 & 215 & $\cdots$ & $\cdots$ & $\cdots$ & $\cdots$ & 1.51(-2) & $\cdots$\\
99.48905   & $9( 1, 9)- 8( 1, 8) A$  & 23.42 & 213 & $\cdots$ & $\cdots$ & $\cdots$ & 1.30(-2) & 1.96(-2) & 9.33(-3)\\
99.57554   & $8( 1, 7)- 7( 1, 6) A$  & 20.58 & 210 & $\cdots$ & $\cdots$ & $\cdots$ & 1.17(-2) & 1.79(-2) & 8.46(-3)\\
99.57741   & $9( 1, 9)- 8( 1, 8) E$  & 23.56 & 212 & $\cdots$ & $\cdots$ & $\cdots$ & 1.43(-2) & 1.73(-2) & 8.68(-3)\\
99.86910   & $8( 1, 7)- 7( 1, 6) E$  & 20.67 & 210 & $\cdots$ & $\cdots$ & $\cdots$ & 1.15(-2) & 1.74(-2) & b \\
100.13691  & $9( 0, 9)- 8( 0, 8) A$  & 23.46 & 213 & $\cdots$ & $\cdots$ & $\cdots$ & $\cdots$ & 2.05(-2) & $\cdots$\\
100.22668  & $9( 0, 9)- 8( 0, 8) E$  & 23.60 & 212 & $\cdots$ & $\cdots$ & $\cdots$ & 1.33(-2) & 2.01(-2) & 9.28(-3)\\
\hline
\multicolumn{8}{c}{\textbf{Dimethyl ether} (CH$_{3}$OCH$_{3}$)} \\
97.99062   & $J(K_{a},K_{c})=16( 3,14)-15( 4,11) AA$ & 38.07 & 137 & $\cdots$ & $\cdots$ & $\cdots$ & $\cdots$ & 7.30(-3) & $\cdots$\\
97.99338   & $16( 3,14)-15( 4,11) EE$ & 60.92 & 137 & $\cdots$ & $\cdots$ & $\cdots$ & $\cdots$ & 1.22(-2) & $\cdots$\\
97.99617   & $16( 3,14)-15( 4,11) AE$ & 22.84 & 137 & $\cdots$ & $\cdots$ & $\cdots$ & $\cdots$ & b & $\cdots$\\
99.18316   & $25( 6,19)-24( 7,18) EE$ & 90.22 & 347 & $\cdots$ & $\cdots$ & $\cdots$ & $\cdots$ & 4.89(-3) & $\cdots$\\
99.32436   & $4( 1, 4)- 3( 0, 3) AE$ & 26.18 & 10 & $\cdots$ & $\cdots$ & 1.48(-2) & 4.79(-2) & 7.45(-2) & 4.19(-2)\\
99.32521   & $4( 1, 4)- 3( 0, 3) EE$ & 69.82 & 10 & $\cdots$ & $\cdots$ & 1.48(-2) & 4.79(-2) & 7.45(-2) & 4.19(-2)\\
99.32607   & $4( 1, 4)- 3( 0, 3) AA$ & 43.63 & 10 & $\cdots$ & $\cdots$ & 1.48(-2) & 4.79(-2) & 7.45(-2) & 4.19(-2)\\
99.83366   & $14( 2,13)- 13( 3,10) AA$ & 28.85 & 101 & $\cdots$ & $\cdots$ & $\cdots$ & $\cdots$ & 8.03(-3) & $\cdots$\\
99.83645   & $14( 2,13)- 13( 3,10) EE$ & 46.17 & 101 & $\cdots$ & $\cdots$ & $\cdots$ & 8.93(-3) & 1.32(-2) & 6.81(-3)\\
99.83924   & $14( 2,13)- 13( 3,10) AE$ & 17.31 & 101 & $\cdots$ & $\cdots$ & $\cdots$ & $\cdots$ & 8.20(-3) & $\cdots$\\
100.43416  & $22( 5,18)- 21( 6,15) AA$ & 50.94 & 266 &$\cdots$ & $\cdots$ & $\cdots$ & $\cdots$ & b& $\cdots$\\
100.43546  & $22( 5,18)- 21( 6,15) EE$ & 81.42 & 266 & $\cdots$ & $\cdots$ & $\cdots$ & $\cdots$ & b & $\cdots$\\
100.46043  & $6(2, 5)-6(1, 6)AE$  & 28.84 & 25 & $\cdots$ & $\cdots$ & b & 1.61(-2) & 2.85(-2) & 1.58(-2)\\
100.46307  & $6( 2, 5)- 6( 1, 6)EE$  & 76.92 & 25 & $\cdots$ & $\cdots$ & b & b & b & 2.55(-2)\\
100.46572  & $6(2, 5)- 6(1, 6) AA$  & 48.07 & 25 & $\cdots$ & $\cdots$ & b & b & b & 1.54(-2)\\
100.94683  & $19( 4,16)- 18( 5,13) AA$ & 26.95 & 196 & $\cdots$ & $\cdots$ & $\cdots$ & $\cdots$ & $\cdots$ & $\cdots$\\
100.94900  & $19( 4,16)- 18( 5,13) EE$ & 71.88 & 196 & $\cdots$ & $\cdots$ & $\cdots$ & 6.03(-3) & 8.49(-3) & $\cdots$\\
100.95109  & $19( 4,16)- 18( 5,13) EA$ & 17.96 & 196 & $\cdots$ & $\cdots$ & $\cdots$ & $\cdots$ & $\cdots$& $\cdots$\\
\hline
\multicolumn{8}{c}{\textbf{Acetone} (CH$_{3}$COCH$_{3}$)} \\
98.05239   & $J(K_{a},K_{c})=17( 6,11)-17( 5,12) EE$ & 872 & 111 & $\cdots$ & $\cdots$ & $\cdots$ & b & 8.25(-3) & $\cdots$\\
98.05353   & $17( 7,11)-17( 6,12) EE$ & 872 & 111 & $\cdots$ & $\cdots$ & $\cdots$ & b & 8.25(-3) & $\cdots$\\
98.60072   & $16( 5,11)-16( 4,12) EE$ & 736 & 96 & $\cdots$ & $\cdots$ & $\cdots$ & b & 1.30(-2) & $\cdots$ \\ 
98.60097   & $16( 6,11)-16( 5,12) EE$ & 736 & 96 & $\cdots$ & $\cdots$ & $\cdots$ & b & 1.30(-2) & $\cdots$\\ 
98.65151   & $5( 5, 1)- 4( 4, 1) EE$ & 496 & 14 & $\cdots$ & $\cdots$ & $\cdots$ & b & 8.96(-3) & $\cdots$\\ 
98.73857   & $16( 5,11)-16( 4,12) AA$ & 276 & 96 & $\cdots$ & $\cdots$ & $\cdots$ & b & 6.42(-3) & $\cdots$\\
98.73883   & $16( 6,11)-16( 5,12) AA$ & 461 & 96 & $\cdots$ & $\cdots$ & $\cdots$ & b & 6.42(-3) & $\cdots$\\
98.80089   & $5( 5, 0)- 4( 4, 0) EE$ & 498 & 14 & $\cdots$ & $\cdots$ & $\cdots$ & 6.40(-3) & 9.39(-3) & $\cdots$\\
99.05250   & $15( 4,11)-15( 3,12) EE$ & 598 & 81 & $\cdots$ & $\cdots$ & $\cdots$ & 9.17(-3) & 1.18(-2) & $\cdots$\\
99.05255   & $15( 5,11)-15( 4,12) EE$ & 598 & 81 & $\cdots$ & $\cdots$ & $\cdots$ & 9.17(-3) & 1.18(-2) & $\cdots$\\
100.35030  & $8( 2, 6)- 7( 3, 5) EE$ & 694 & 25 & $\cdots$ & $\cdots$ & $\cdots$ & 8.60(-3) & 1.21(-2) &  $\cdots$\\
100.50706  & $8( 3, 6)- 7( 2, 5) AA$ & 434 & 25 & $\cdots$ & $\cdots$ & $\cdots$ & 5.07(-3) & 7.14(-3) & $\cdots$\\
\hline
\multicolumn{8}{c}{\textbf{gauche-Ethanol} (g-C$_{2}$H$_{5}$OH)} \\
97.53590   & $J(K_{a},K_{c})=21(1,21)-21(0,21), vt=1-0$ & 26.06 & 246 & $\cdots$ & $\cdots$ & 9.23(-3) & 1.49(-2) & 3.73(-2) & $\cdots$\\
97.53684   & $23( 1,23)-23( 0,23), vt= 1- 0$ & 28.49 & 282 & $\cdots$ & $\cdots$ & 9.36(-3) & 1.17(-2) & 3.55(-2) & $\cdots$\\
97.54687   & $29(1,28)-29(2,28), vt=1-0$ & 17.84 & 420 & $\cdots$ & $\cdots$ & $\cdots$ & $\cdots$ & 7.12(-3) & $\cdots$\\
97.54969   & $26(0,26)-26(1,26), vt=1-0$ & 32.12 & 341 & $\cdots$ & $\cdots$ & $\cdots$  & 5.99(-3) & 1.54(-2) & $\cdots$\\
97.56281   & $24(1,24)-24(0,24), vt=1-0$ & 29.69 & 301 & $\cdots$ & $\cdots$ & $\cdots$ & $\cdots$ & 1.86(-2) & $\cdots$\\
97.57400   & $20(1,20)-20(0,20), vt=1-0$ & 24.85 & 230 & $\cdots$ & $\cdots$ & 7.92(-3) & 1.05(-2) & 2.66(-2) & $\cdots$\\
97.60039   & $25(1,25)-25(0,25), vt=1-0$ & 30.91 & 320 & $\cdots$ & $\cdots$ & $\cdots$ & $\cdots$ & 1.64(-2) & $\cdots$\\
97.63132   & $27(0,27)-27(1,27), vt=1-0$ & 33.34 & 362 & $\cdots$ & $\cdots$ & $\cdots$ & $\cdots$ & 1.34(-2) & $\cdots$\\
97.64950   & $19(1,19)-19(0,19), vt=1-0$ & 23.64 & 214 & $\cdots$ & $\cdots$ & 8.56(-3) & 1.11(-2) & 2.81(-2) & $\cdots$\\
97.69853   & $27( 1,27)-27( 0,27), vt= 1- 0$ & 33.33 & 362 & $\cdots$ & $\cdots$ & $\cdots$ & $\cdots$ & 1.33(-2) & $\cdots$\\
97.70888   & $28( 0,28)-28( 1,28), vt= 1- 0$ & 34.54 & 384 & $\cdots$ & $\cdots$ & $\cdots$ & $\cdots$ & 1.16(-2) & $\cdots$\\
97.75561   & $28(1,28)-28(0,28), vt=1-0$ & 34.55 & 384 & $\cdots$ & $\cdots$ & $\cdots$ & $\cdots$ & 1.19(-2) & $\cdots$\\
97.77430   & $18(1,18)-18(0,18), vt=1-0$ & 22.42 & 199 & $\cdots$ & $\cdots$ & b & 1.14(-2) & 2.96(-2) & $\cdots$ \\
97.78411   & $29( 0,29)-29( 1,29), vt= 1- 0$ & 35.76 & 407 & $\cdots$ & $\cdots$ & $\cdots$ & $\cdots$ & 1.03(-2) & $\cdots$\\
97.81598   & $29( 1,29)-29( 0,29), vt= 1- 0$ & 35.76 & 407 & $\cdots$ & $\cdots$ & $\cdots$ & $\cdots$ & 9.88(-3) & $\cdots$\\
97.82895   & $30( 1,29)-30( 2,29), vt= 1- 0$ & 18.45 & 444 & $\cdots$ & $\cdots$ & $\cdots$ & $\cdots$ & 6.19(-3) & $\cdots$\\
97.85747   & $30( 0,30)-30( 1,30), vt= 1- 0$ & 36.98 & 430 & $\cdots$ & $\cdots$ & $\cdots$ & $\cdots$ & 8.22(-3) & $\cdots$\\
97.87745   & $51( 4,48)-51( 3,48), vt= 1- 0$ & 36.98 & 430 & $\cdots$ & $\cdots$& $\cdots$ & $\cdots$ & 8.21(-3) & $\cdots$\\
97.96283   & $17(1,17)-17(0,17), vt=1-0$ & 21.21 & 185 & $\cdots$ & $\cdots$ & 1.10(-2) & 1.20(-2) & 3.06(-2) & $\cdots$\\
98.23031   & $16(1,16)-16(0,16), vt=1-0$ & 20    & 171 & $\cdots$ & $\cdots$ & 1.18(-2) & 1.22(-2) & 3.19(-2) & 8.52(-3) \\
98.58509   & $15(1,15)-15(0,15), vt=1-0$ & 18.79 & 158 & $\cdots$ & $\cdots$ & 1.16(-2) & 1.25(-2) & 3.30(-2) & $\cdots$\\
98.82369   & $29( 2,28)-29( 1,28), vt= 1- 0$ & 17.84 & 420 & $\cdots$ & $\cdots$ & $\cdots$ & $\cdots$ & 7.62(-3) & $\cdots$\\
98.87828   & $13(1,13)-13(0,13), vt=1-0$ & 16.37 & 135 & $\cdots$ & $\cdots$ & $\cdots$ & $\cdots$ & 1.87(-2) & $\cdots$\\
98.94626   & $28( 2,27)-28( 1,27), vt= 1- 0$ & 17.23 & 397 & $\cdots$ & $\cdots$ & $\cdots$ & $\cdots$ & 8.86(-3) & $\cdots$\\
98.98354   & $14( 1,14)-14( 0,14), vt= 1- 0$ & 17.58 & 147 & $\cdots$ & $\cdots$ & 1.19(-2) & 1.22(-2) & 3.19(-2) & $\cdots$\\
99.86441   & $25( 2,24)-25( 1,24), vt= 1- 0$ & 15.41 & 332 & $\cdots$ & $\cdots$ & $\cdots$ & $\cdots$ & 1.19(-2) & $\cdots$\\
100.19432  & $6( 1, 6)- 5( 1, 5), vt= 0- 0$ & 9.32  & 75 & $\cdots$ & $\cdots$ & $\cdots$ & 6.83(-3) & 1.93(-2) & $\cdots$\\
100.36505  & $6( 1, 6)- 5( 1, 5), vt= 1- 1$ & 9.32  & 80 & $\cdots$ & $\cdots$ & $\cdots$ & 6.08(-3) & 2.07(-2) & $\cdots$\\
100.45207  & $24(2,23)-24(1,23), vt=1-0$ & 14.80  & 312 & $\cdots$ & $\cdots$ & $\cdots$ & $\cdots$ & 1.31(-2) & $\cdots$\\
101.24363  & $23( 2,22)-23( 1,22), vt= 1- 0$ & 14.19 & 292 & $\cdots$ & $\cdots$ & $\cdots$ & $\cdots$ & 1.46(-2) & $\cdots$\\
\multicolumn{8}{c}{\textbf{trans-Ethanol} (t-C$_{2}$H$_{5}$OH)} \\
99.52409   & $J(K_{a},K_{c})=17(3,14)-17(2,15)$ & 18.34 & 141 & $\cdots$ & $\cdots$ & 1.35(-2) & 1.37(-2) & 3.58(-2) & 1.04(-2) \\
99.97588   & $18(3,15)- 18(2,16)$ & 19.42 & 157 & $\cdots$ & $\cdots$ & 1.30(-2) & 1.28(-2) & 3.41(-2) & $\cdots$\\
100.35895  & $16(3,13)- 16(2,14)$ & 25.70 & 127 & $\cdots$ & $\cdots$ & 1.50(-2) & 1.39(-2) & 3.56(-2) & $\cdots$\\
100.99010  & $8(2, 7)- 8(1, 8)$  & 8.83 & 35 & $\cdots$ & $\cdots$ & 1.00(-2) & 7.55(-3) & 2.09(-2) & $\cdots$\\
\hline
\multicolumn{8}{c}{\textbf{Cyanoacetylene} \quad (HC$_{3}$N, v=0)} \\
100.07638  & $J=11-10$ & 152.56 & 29 & 4.87(-2) & 2.44(-1) & 2.69(-1) & 2.53(-1) & 2.50(-1) & 2.23(-1)\\
\multicolumn{8}{c}{ \quad (HC$^{13}$CCN, v=0)} \\
99.65184  & $J=11-10, F=11-10$ & 153.19 & 29 & $\cdots$ & $\cdots$ & 4.58(-2) & 5.25(-2) & 5.53(-2) & 8.44(-3)\\
\multicolumn{8}{c}{ (HCC$^{13}$CN, v=0)} \\
99.66146  & $J=11-10, F=11-10$ & 153.19 & 29 & $\cdots$ & $\cdots$ & 4.41(-2) & 3.95(-2) & 4.60(-2) & 6.66(-3)\\
\multicolumn{8}{c}{ (HC$_{3}$N,v7=1)} \\
100.32241  & $J=11-10, l=1e$ & 151.18 & 350 & $\cdots$ & $\cdots$ & 4.90(-2) & 1.43(-1) & 1.00(-1) & 5.94(-3)\\
100.46617  & $J=11-10, l=1f$ & 151.15 & 350 & $\cdots$ & $\cdots$ & 4.86(-2) & 1.44(-1) & 9.79(-2) & b\\
\multicolumn{8}{c}{ (HC$_{3}$N,v7=2)} \\
100.70878  & $J=11-10, l=0$  & 151.73 & 671 & $\cdots$ & $\cdots$ & 9.46(-3) & 4.28(-2) & 9.01(-3) & $\cdots$\\
100.71106  & $J=11-10, l=2e$ & 146.72 & 674 & $\cdots$ & $\cdots$ & 9.37(-3) & 4.16(-2) & 8.52(-3) & $\cdots$\\
100.71439  & $J=11-10, l=2f$ & 146.71 & 674 & $\cdots$ & $\cdots$ & 7.65(-3) & 3.19(-2) & 8.28(-3) & $\cdots$\\
\multicolumn{8}{c}{ (HC$_{3}$N,v7=3)} \\
100.88044  & $J=11-10, l=1e$ & 150.48 & 985 & $\cdots$ & $\cdots$ & $\cdots$ & b & $\cdots$ & $\cdots$\\
101.02780  & $J=11-10, l=3$  & 140.45 & 991 & $\cdots$ & $\cdots$ & $\cdots$ & b & $\cdots$ & $\cdots$\\
101.16989  & $J=11-10, l=1f$ & 150.48 & 985 & $\cdots$ & $\cdots$ & $\cdots$ & b & $\cdots$ & $\cdots$\\
\multicolumn{8}{c}{ (HC$_{3}$N,v6=1)} \\
100.24058  & $J=11-10, l=1e$ & 151.48 & 747 & $\cdots$ & $\cdots$ & $\cdots$ & 1.74(-2) & $\cdots$ & $\cdots$\\
100.31938  & $J=11-10, l=1f$ & 151.46 & 747 & $\cdots$ & $\cdots$ & b & b & b & $\cdots$\\
\multicolumn{8}{c}{ (HC$^{13}$CCN,v7=1)} \\
99.88792   & $J=11-10, l=1e$ & 151.17 & 345 & $\cdots$ & $\cdots$ & $\cdots$ & 1.06(-2) & $\cdots$ & $\cdots$\\
100.03251  & $J=11-10, l=1f$ & 151.17 & 345 & $\cdots$ & $\cdots$ & $\cdots$ & b & $\cdots$ & $\cdots$\\
\multicolumn{8}{c}{ (HCC$^{13}$CN,v7=1)} \\
99.90177   & $J=11-10, l=1e$ & 151.17 & 347 & $\cdots$ & $\cdots$ & $\cdots$ & b & $\cdots$ & $\cdots$\\
100.04531  & $J=11-10, l=1f$ & 151.17 & 347 & $\cdots$ & $\cdots$ & $\cdots$ & 1.02(-2) & $\cdots$ & $\cdots$\\
\hline
\multicolumn{8}{c}{\textbf{Cyanodiacetylene} (HC$_{5}$N,v=0)} \\
98.51252   & $J = 37 - 36$ & 2080.91 & 90 & $\cdots$ & $\cdots$ & 3.03(-2) & 1.86(-2) & 3.36(-2) & b\\
101.17467  & $J = 38 - 37$ & 2137.61 & 95 & $\cdots$ & $\cdots$ & 3.10(-2) & 1.85(-2) & 3.52(-2) & $\cdots$\\
\hline
\multicolumn{8}{c}{\textbf{Ethyl Cyanide} (CH$_{3}$CH$_{2}$CN)} \\
97.84469   & $J(K_{a},K_{c})=19( 3,16)-19( 2,17)$ & 18.91 & 92  & $\cdots$ & $\cdots$ & $\cdots$ & b & $\cdots$ & $\cdots$\\
97.87509   & $34( 4,31)-33( 5,28)$ & 8.22  & 274 & $\cdots$ & $\cdots$ & $\cdots$ & $\cdots$ & $\cdots$ & $\cdots$\\
98.17757   & $11( 2,10)-10( 2, 9)$ & 157.60 & 33 & $\cdots$ & $\cdots$ & 4.19(-2) & 2.09(-1) & 4.61(-2) & $\cdots$\\
98.52387   & $11( 6, 5)-10( 6, 4)$ & 114.54 & 68  & $\cdots$ & $\cdots$ & 7.25(-2) & 3.79(-1) & 7.75(-2) & $\cdots$\\
98.52467   & $11( 7, 4)-10( 7, 3)$ & 97.02  & 83  & $\cdots$ & $\cdots$ & 6.57(-2) & 3.71(-1) & 5.80(-2) & $\cdots$\\
98.53208   & $11( 8, 3)-10( 8, 2)$ & 76.81  & 99  & $\cdots$ & $\cdots$ & 6.00(-2) & 3.03(-1) & 5.74(-2) & $\cdots$\\
98.53398   & $11( 5, 6)-10( 5, 5)$ & 129.36 & 56  & $\cdots$ & $\cdots$ & 5.78(-2) & 3.11(-1) & 6.59(-2) & $\cdots$\\
98.54416   & $11( 9, 2)-10( 9, 1)$ & 53.90 & 118  & $\cdots$ & $\cdots$ & 1.37(-2) & b & 1.59(-2) & $\cdots$\\
98.55992   & $11(10, 1)-10(10, 0)$ & 28.30 & 139  & $\cdots$ & $\cdots$ & b & b & $\cdots$ & $\cdots$\\
98.56679   & $11( 4, 7)-10( 4, 6)$ & 141.49 & 46  & $\cdots$ & $\cdots$ & 4.08(-2) & 2.07(-1) & 3.95(-2) & $\cdots$\\
98.61009   & $11( 3, 9)-10( 3, 8)$ & 150.93 & 38  & $\cdots$ & $\cdots$ & 3.84(-2) & 1.93(-1) & b & $\cdots$\\
98.70110   & $11( 3, 8)-10( 3, 7)$ & 150.93 & 38  & $\cdots$ & $\cdots$ & 3.85(-2) & 1.93(-1) & 4.19(-2) & $\cdots$\\
99.07060   & $32( 3,29)-32( 2,30)$ & 37.29 & 240  & $\cdots$ & $\cdots$ & b & b & $\cdots$ & $\cdots$\\
99.25344   & $15( 2,14)-15( 1,15)$ & 8.83 & 56  & $\cdots$ & $\cdots$ & b & b & $\cdots$ & $\cdots$\\
99.68146   & $11( 2, 9)-10( 2, 8)$ & 157.63 & 33  & $\cdots$ & $\cdots$ & b & 2.12(-1) & 4.86(-2) & $\cdots$\\
100.03442  & $18( 3,15)-18( 2,16)$ & 17.34 & 84 & $\cdots$ & $\cdots$ & $\cdots$ & b & $\cdots$ & $\cdots$\\
100.61428  & $11( 1,10)-10( 1, 9)$ & 161.58 & 30  & $\cdots$ & $\cdots$ & 4.13(-2) & 2.18(-1) & 5.16(-2) & $\cdots$\\
101.09167  & $10( 1,10)- 9( 0, 9)$ & 9.77 & 24  & $\cdots$ & $\cdots$ & b & b & $\cdots$ & $\cdots$\\
\multicolumn{8}{c}{(CH$_{3}$CH$_{2}$$^{13}$CN)} \\
97.69154   & $J(K_{a},K_{c})=11( 2,10)- 10( 2, 9)$ & 154.84 & 33 & $\cdots$ & $\cdots$ & $\cdots$ & b & $\cdots$ & $\cdots$ \\
98.03285   & $11( 6, 6)- 10( 6, 5)$ & 112.51 & 68 & $\cdots$ & $\cdots$ & $\cdots$ & 1.69(-2) & $\cdots$ & $\cdots$\\
98.03285   & $11( 6, 5)- 10( 6, 4)$ & 112.51 & 68 & $\cdots$ & $\cdots$ & $\cdots$ & 1.69(-2) & $\cdots$ & $\cdots$\\
98.04150   & $11( 8, 4)- 10( 8, 3)$ & 75 & 99 & $\cdots$ & $\cdots$ & $\cdots$ & 1.84(-2) & $\cdots$ & $\cdots$\\
98.04150   & $11( 8, 3)- 10( 8, 2)$ & 75 & 99 & $\cdots$ & $\cdots$ & $\cdots$ & 1.84(-2) & $\cdots$ & $\cdots$\\
98.05374   & $11( 9, 3)- 10( 9, 2)$ & 52.96 & 118 & & & & b & & \\
98.07271   & $11(4,8)-10(4,7)$ & 138.99 & 46 & $\cdots$ & $\cdots$ & $\cdots$ & 9.25(-3) & $\cdots$ & $\cdots$\\
98.07461   & $11(4,7)-10(4,6)$ & 139.02 & 46 & $\cdots$ & $\cdots$ & $\cdots$ & 9.59(-3) & $\cdots$ & $\cdots$\\
98.11741   & $11(3,9)-10(3,8)$ & 148.26 & 38 & $\cdots$ & $\cdots$ & $\cdots$ & 1.04(-2) & $\cdots$ & $\cdots$\\
98.20621   & $11( 3, 8)- 10( 3, 7)$ & 148.27 & 38 & $\cdots$ & $\cdots$ & $\cdots$ & 1.01(-2) & $\cdots$ & $\cdots$\\
99.17252   & $11(2,9)-10(2,8)$ & 154.88 & 33 & $\cdots$ & $\cdots$ & $\cdots$ & 1.16(-2) & $\cdots$ & $\cdots$\\
100.10973  & $11(1,10)-10(1,9)$ & 	158.73 & 40 & $\cdots$ & $\cdots$ & $\cdots$ & 1.11(-2) & $\cdots$ & $\cdots$\\
\multicolumn{8}{c}{ (CH$_{3}$$^{13}$CH$_{2}$CN)} \\
97.67201   & $J(K_{a},K_{c})=11(2,10)-10(2,9)$ & 154.84 & 33  & $\cdots$ & $\cdots$ & $\cdots$ & 1.13(-2) & $\cdots$ & $\cdots$\\
98.03964   & $11( 7, 5)- 10( 7, 4)$ & 95.31 & 81  & $\cdots$ & $\cdots$ & $\cdots$ & 1.77(-2) & $\cdots$ & $\cdots$\\
98.04058   & $11( 6, 6)- 10( 6, 5)$ & 112.52 & 67 & $\cdots$ & $\cdots$ & $\cdots$ & 1.74(-2) & $\cdots$ & $\cdots$\\
98.04556   & $11( 8, 4)- 10( 8, 3)$ & 75.45 & 98 & $\cdots$ & $\cdots$ & $\cdots$ & $\cdots$ & $\cdots$ & $\cdots$\\
98.05296   & $11( 5, 6)- 10( 5, 5)$ & 127.09 & 55 & $\cdots$ & $\cdots$ & $\cdots$ & 1.58(-2) & $\cdots$ & $\cdots$\\
98.08734   & $11( 4, 8)- 10( 4, 7)$ & 139 & 46 & $\cdots$ & $\cdots$ & $\cdots$ & 9.14(-3) & $\cdots$ & $\cdots$\\
98.08968   & $11( 4, 7)- 10( 4, 6)$ & 138.99 & 46 & $\cdots$ & $\cdots$ & $\cdots$ & 9.06(-3) & $\cdots$ & $\cdots$\\
98.13485   & $11( 3, 9)- 10( 3, 8)$ & 148.26 & 38 & $\cdots$ & $\cdots$ & $\cdots$ & 1.02(-2) & $\cdots$ & $\cdots$\\
98.23779   & $11( 3, 8)- 10( 3, 7)$ & 148.26 & 38 & $\cdots$ & $\cdots$ & $\cdots$ & b & $\cdots$ & $\cdots$\\
99.27963   & $11( 2, 9)- 10( 2, 8)$ & 154.89 & 33 & $\cdots$ & $\cdots$ & $\cdots$ & 1.13(-2) & $\cdots$ & $\cdots$\\
100.15582  & $11( 1,10)- 10( 1, 9)$ & 158.73 & 30 & $\cdots$ & $\cdots$ & $\cdots$ & 1.15(-2) & $\cdots$ & $\cdots$\\
\multicolumn{8}{c}{ ($^{13}$CH$_{3}$CH$_{2}$CN)} \\
98.16534   & $J(K_{a},K_{c})=11( 1,10)- 10( 1, 9)$ & 158.73 & 29 & $\cdots$ & $\cdots$ & $\cdots$ & 1.17(-2) & $\cdots$ & $\cdots$\\
\multicolumn{8}{c}{ (C$_{2}$H$_{5}$C$^{15}$N)} \\
97.72519   & $J(K_{a},K_{c})=11( 1,10)- 10( 1, 9)$ & 161.60 & 29 & $\cdots$ & $\cdots$ & $\cdots$ & $\cdots$ & $\surd$ & $\cdots$ \\
101.17615  & $12( 1,12)- 11( 1,11)$ & 175.64 & 33 & $\cdots$ & $\cdots$ & $\cdots$ & b & b & $\cdots$\\
\hline
\multicolumn{8}{c}{\textbf{Ammonia} (NH2D)} \\
99.11881    & $J(K_{a},K_{c})=5( 2, 4)0a- 5( 1, 4)0s$ & 87 & 261 & $\cdots$ & $\cdots$ & $\cdots$ & 4.24(-2) & 3.24(-2) & 9.76(-3)\\
\hline
\multicolumn{8}{c}{\textbf{Carbon Monosulfide} (CS)} \\
97.98095    & $J=2-1$ & 8 & 7 & 2.33(-1) & 4.62(-1) & 1.32(0) & 1.17(0) & 3.28(-1) & 6.55(-1)\\
\hline
\multicolumn{8}{c}{\textbf{Sulfur Monoxide} (SO)} \\
99.29987   & $J(N)=3( 2)- 2( 1)$ & 6.91 & 9 & 5.06(-2) & 1.05(-1) & 4.12(-1) & 4.51(-1) & 3.55(-1) & 1.94(-1) \\
100.02964  & $4( 5)- 4( 4)$ & 0.84 & 39 & $\cdots$ & $\cdots$ & 3.79(-2) & 4.11(-2) & 2.89(-2) & 1.72(-2)\\
\multicolumn{8}{c}{ ($^{34}$SO)} \\
97.71532   & $J(N)=3( 2)- 2( 1)$ & 6.91 & 9 & $\cdots$ & 1.12(-2) & 3.81(-2) & 6.35(-2) & 4.46(-2) & 1.26(-2)\\
\hline
\multicolumn{8}{c}{\textbf{Sulfur dioxide} (SO$_{2}$, v=0)} \\
97.70233   & $J(K_{a},K_{c})=7( 3, 5)- 8( 2, 6)$ & 2.51 & 48 & $\cdots$ & $\cdots$ & 3.76(-2) & 3.06(-2) & 5.12(-2) & $\cdots$\\
100.87810  & $2( 2, 0)- 3( 1, 3)$  & 0.43 & 13 & $\cdots$ & $\cdots$ & $\cdots$ & b & 1.20(-2) & $\cdots$\\
\multicolumn{8}{c}{ ($^{33}$SO$_{2}$, v=0)} \\
98.25793   & $J(K_{a},K_{c})=2( 2, 0)- 3( 1, 3), F=7/2-7/2$ & 0.07 & 13 & $\cdots$ & $\cdots$ & $\cdots$ & 1.63(-2) & $\cdots$ & $\cdots$\\
98.25810   & $2( 2, 0)- 3( 1, 3), F=5/2-7/2$ & 0.41 & 13 & $\cdots$ & $\cdots$ & $\cdots$ & 1.63(-2) &$\cdots$ & $\cdots$\\
98.26062   & $2( 2, 0)- 3( 1, 3), F=3/2-5/2$ & 0.27 & 13 & $\cdots$ & $\cdots$ & $\cdots$ & $\cdots$ & $\cdots$ & $\cdots$\\
98.26089   & $2( 2, 0)- 3( 1, 3), F=5/2-5/2$ & 0.09 & 13 & $\cdots$ & $\cdots$ & $\cdots$ & $\cdots$ & $\cdots$ & $\cdots$ \\
98.26380   & $2( 2, 0)- 3( 1, 3), F=7/2-9/2$ & 0.60 & 13 & $\cdots$ & $\cdots$ & $\cdots$ & b & $\cdots$ & $\cdots$\\
98.26623   & $2( 2, 0)- 3( 1, 3), F=1/2-3/2$ & 0.17 & 13 & $\cdots$ & $\cdots$ & $\cdots$ & b & $\cdots$ & $\cdots$\\
\multicolumn{8}{c}{ (SO$_{2}$, v2=1)} \\
98.26469   & $J(K_{a},K_{c})=16( 2,14)-15( 3,13)$ & 8.20 & 900 & $\cdots$ & $\cdots$ & $\cdots$ & 2.35(-2) & 5.60(-3) & $\cdots$ \\
99.17739   & $27( 3,25)-26( 4,22)$ & 8.74 & 1136 & $\cdots$ & $\cdots$ & $\cdots$ & $\cdots$ & $\cdots$ & $\cdots$\\
\hline
\multicolumn{8}{c}{\textbf{Methyl Mercaptan} (CH$_{3}$SH)} \\
100.11021  & $J(K_{a},K_{c})=4(1,3) - 3(1,2) A$ & 6.23 & 17 & $\cdots$ & $\cdots$ & $\cdots$ & b & $\cdots$ & $\cdots$\\
101.02974  & $4(-1,4) - 3(-1,3) E$ & 6.22 & 17 & $\cdots$ & $\cdots$ & $\cdots$ & b & $\cdots$ & $\cdots$\\
101.13915  & $4(0,4) - 3(0,3) A$ & 6.64 & 12 & $\cdots$ & $\cdots$ & $\cdots$ & 1.21(-2) & $\cdots$ & $\cdots$\\
101.13965  & $4(0,4) - 3(0,3) E$ & 6.64 & 14 & $\cdots$ & $\cdots$ & $\cdots$ & 1.21(-2) & $\cdots$ & $\cdots$\\
101.15687  & $4(3,1) - 3(3,0) E$ & 2.91 & 51 & $\cdots$ & $\cdots$ & $\cdots$ & $\cdots$ & $\cdots$ & $\cdots$\\
101.15932  & $4(-2,3) - 3(-2,2) A$ & 4.98 & 31 & $\cdots$ & $\cdots$ & $\cdots$ & 8.79(-3) & $\cdots$ & $\cdots$\\
101.15999  & $4(-3,2) - 3(-3,1) E$ & 2.91 & 52 & $\cdots$ & $\cdots$ & $\cdots$ & 8.79(-3) & $\cdots$ & $\cdots$\\
101.16065  & $4(3,1) - 3(3,0) A$  & 2.91 & 53 & $\cdots$ & $\cdots$ & $\cdots$ & 8.27(-3) & $\cdots$ & $\cdots$\\
101.16069  & $4(-3,2) - 3(-3,1) A$ & 2.91 & 53 & $\cdots$ & $\cdots$ & $\cdots$ & 8.27(-3) & $\cdots$ & $\cdots$\\
101.16715  & $4(-2,3) - 3(-2,2) E$ & 4.98 & 30 & $\cdots$ & $\cdots$ & $\cdots$ & b & $\cdots$ & $\cdots$ \\
101.16830  & $4(2,2) - 3(2,1) E$ & 4.98 & 30 & $\cdots$ & $\cdots$ & $\cdots$ & b & $\cdots$ & $\cdots$\\
101.17981  & $4(2,2) - 3(2,1) A$ & 4.98 & 31 & $\cdots$ & $\cdots$ & $\cdots$ & $\cdots$ & $\cdots$ & $\cdots$ \\
101.28436  & $4(1,3) - 3(1,2) E$ & 6.22 & 18 & $\cdots$ & $\cdots$ & $\cdots$ & 5.71(-3) & $\cdots$ & $\cdots$\\
\hline
\multicolumn{10}{l}{\footnotesize For the line optical depth, X(Y) means X $\times$ 10$^{Y}$.}\\
\multicolumn{10}{l}{\footnotesize$^a$ Quantum numbers of the upper and lower levels. For each species, the levels are labelled at the first transition. A and E refer to symmetry state. }  \\
\multicolumn{10}{l}{\footnotesize$^b$ "$\surd$" indicates that the transition of H(40)$_{\alpha}$ is detected. } \\
\multicolumn{10}{l}{\footnotesize$^c$ "$\cdots$" indicates that the transition is not detected, or the line intensity of transition is below 3$\sigma$.}\\
\multicolumn{10}{l}{\footnotesize$^d$ "b" indicates that the transition is blended with other molecules.}
\end{longtable}
\twocolumn

\begin{figure*}
\setcounter{figure}{0}
\centerline{\resizebox{0.9\hsize}{!}{\includegraphics{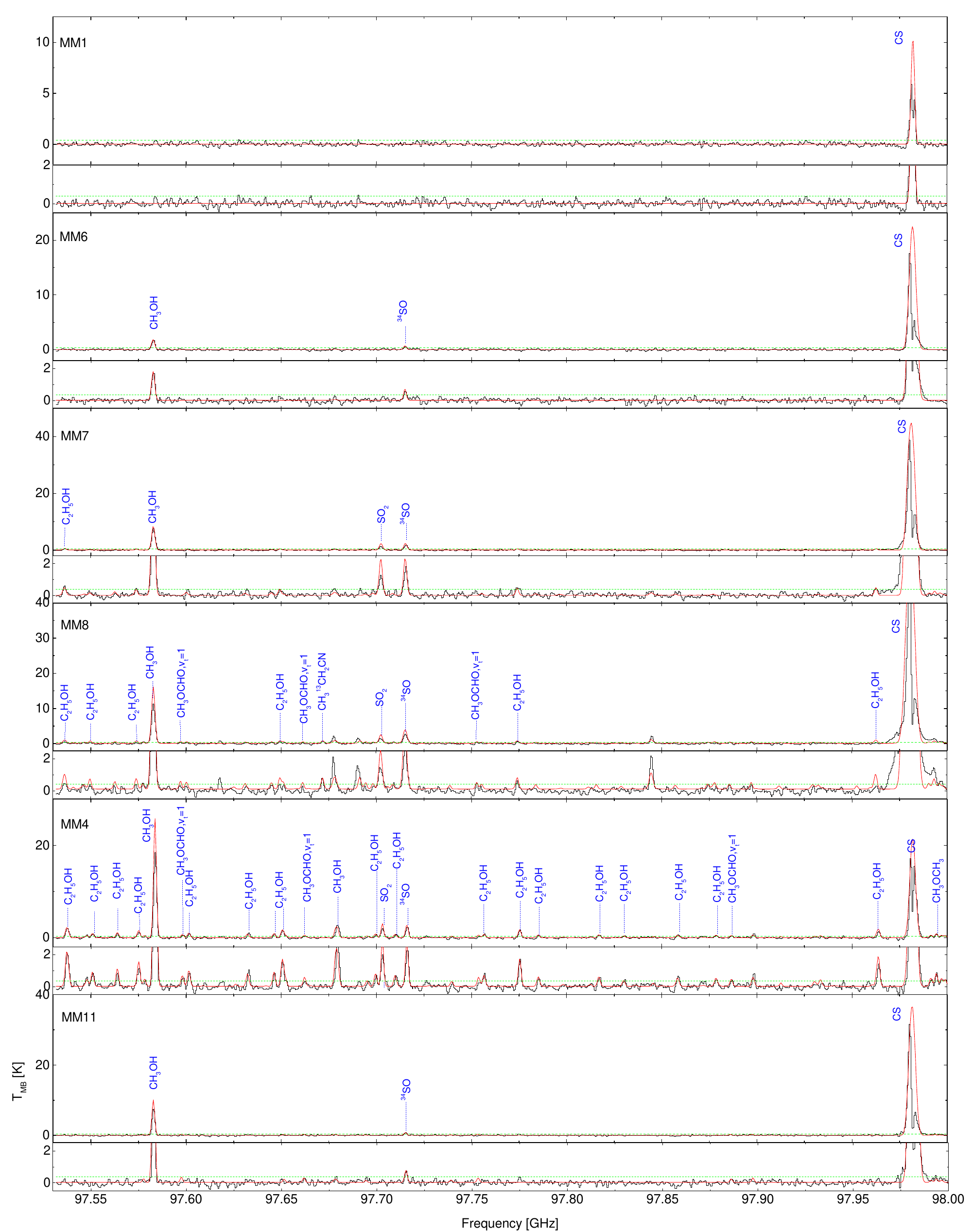}}}
\caption{Same as Figure \ref{fig:survey_1}, but for frequency range from 97.53 to 98.00 GHz. The black curves are the observed spectra, and the red curves indicate the simulated LTE spectra. The horizontal green dashed line indicates the 3 $\sigma$ noise level in each core.
}
\label{fig:spec_2}
\end{figure*}

\begin{figure*}
\setcounter{figure}{0}
\centerline{\resizebox{0.9\hsize}{!}{\includegraphics{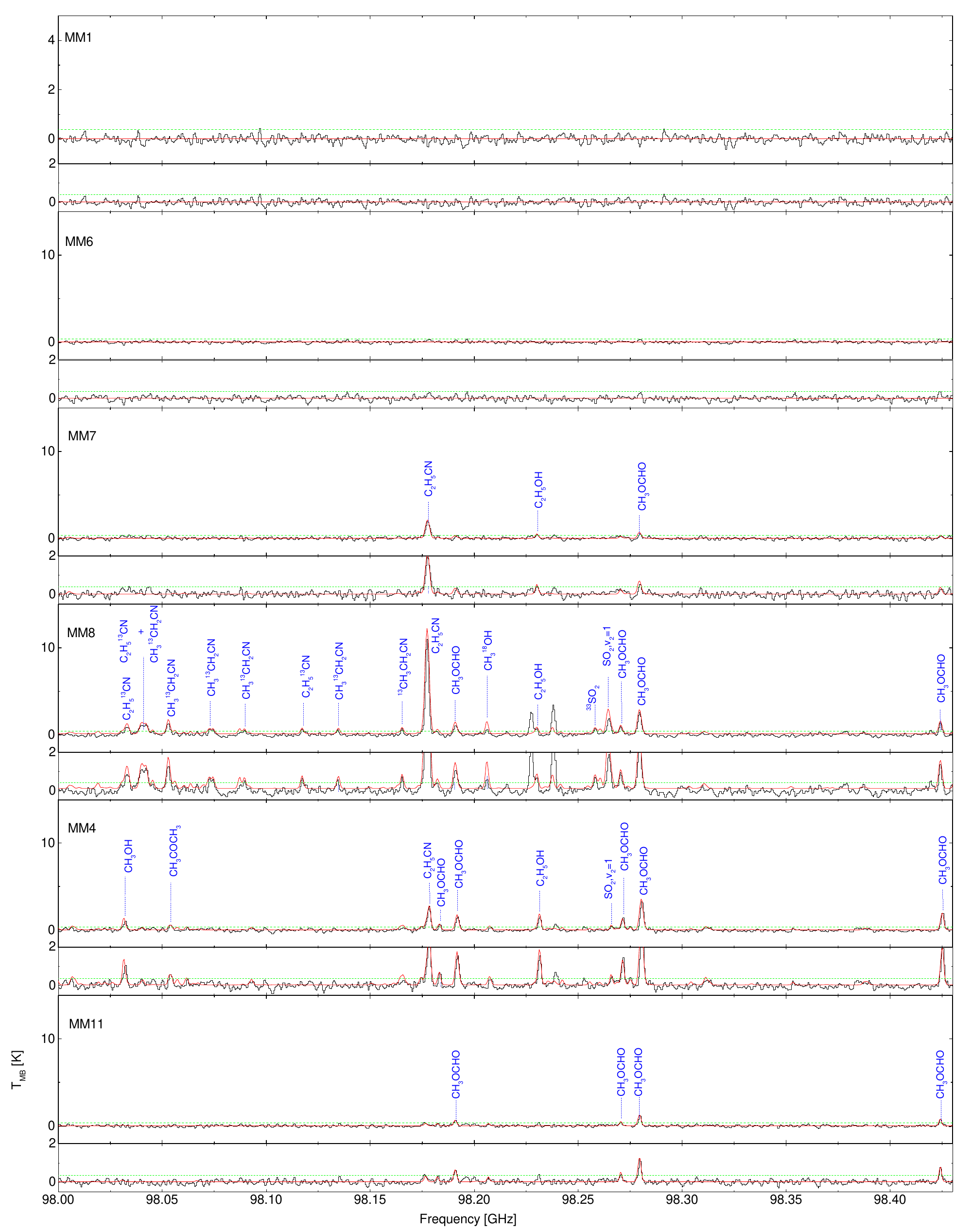}}}
\caption{Continued. Frequency range from 98.00 to 98.43 GHz. 
}
\label{fig:spec_3}
\end{figure*}

\begin{figure*}
\setcounter{figure}{0}
\centerline{\resizebox{0.9\hsize}{!}{\includegraphics{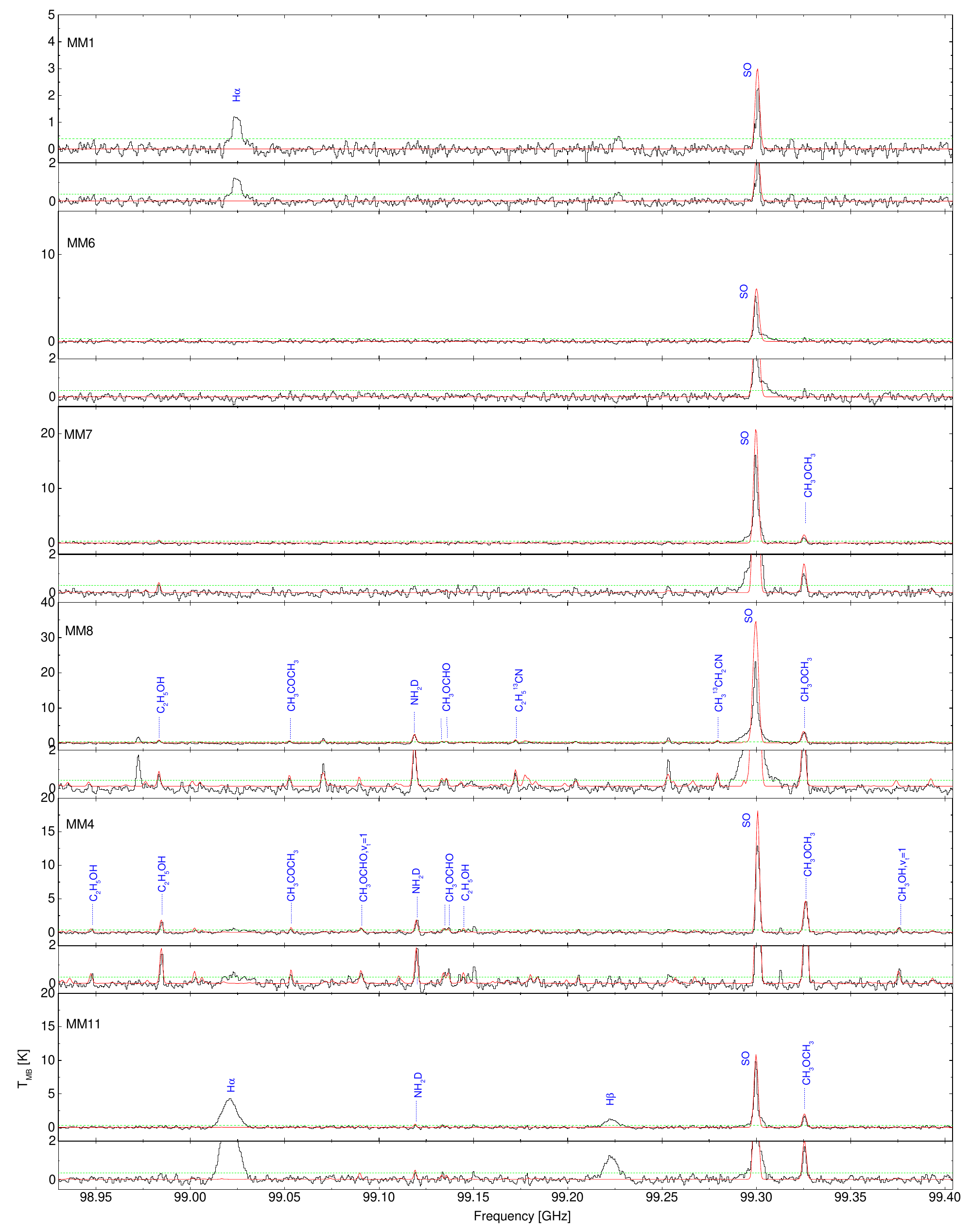}}}
\caption{Continued. Frequency range from 98.93 to 99.404 GHz.
}
\label{fig:spec_4}
\end{figure*}

\begin{figure*}
\setcounter{figure}{0}
\centerline{\resizebox{0.9\hsize}{!}{\includegraphics{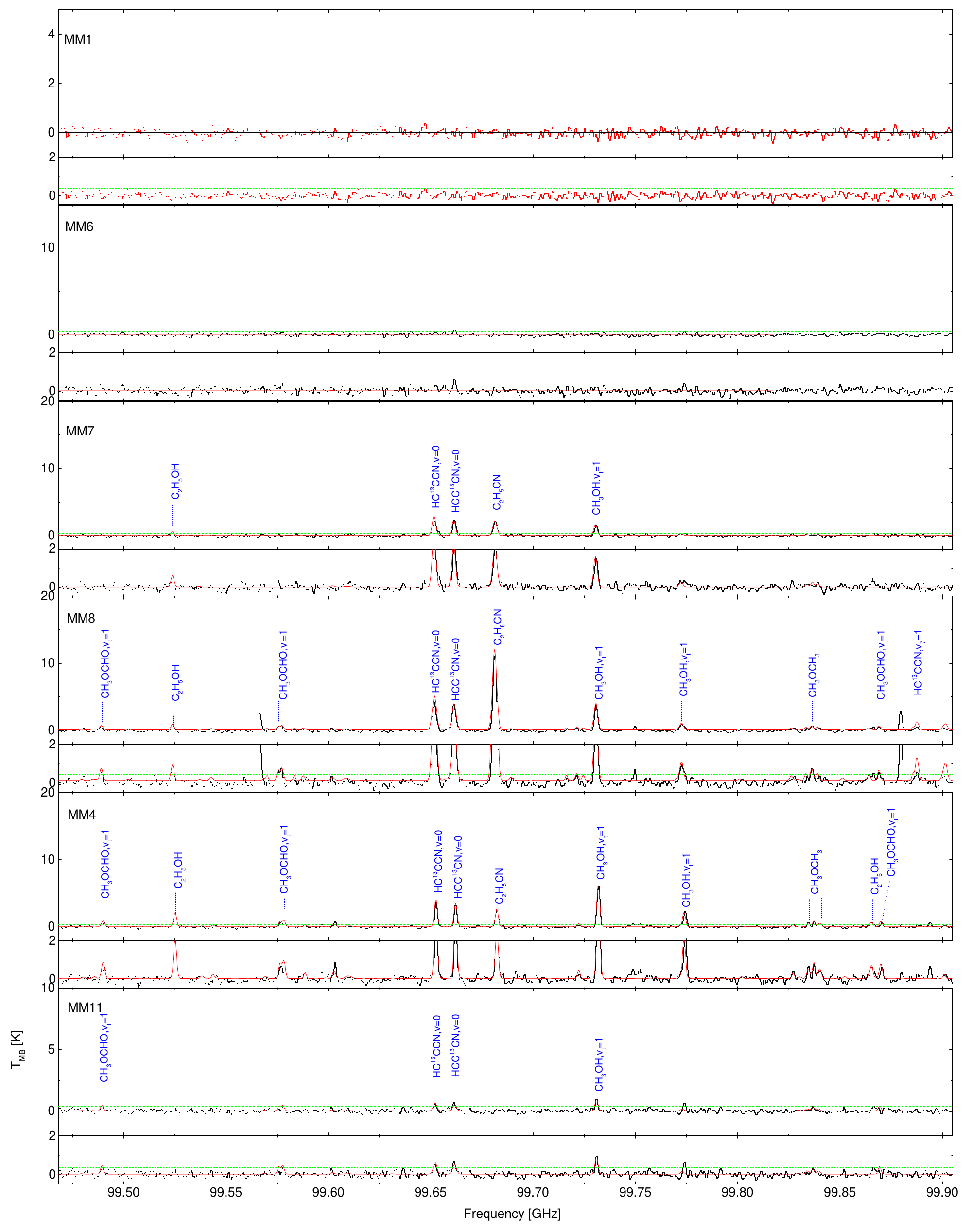}}}
\caption{Continued. Frequency range from 99.458 to 99.905 GHz.
}
\label{fig:spec_5}
\end{figure*}

\begin{figure*}
\setcounter{figure}{0}
\centerline{\resizebox{0.9\hsize}{!}{\includegraphics{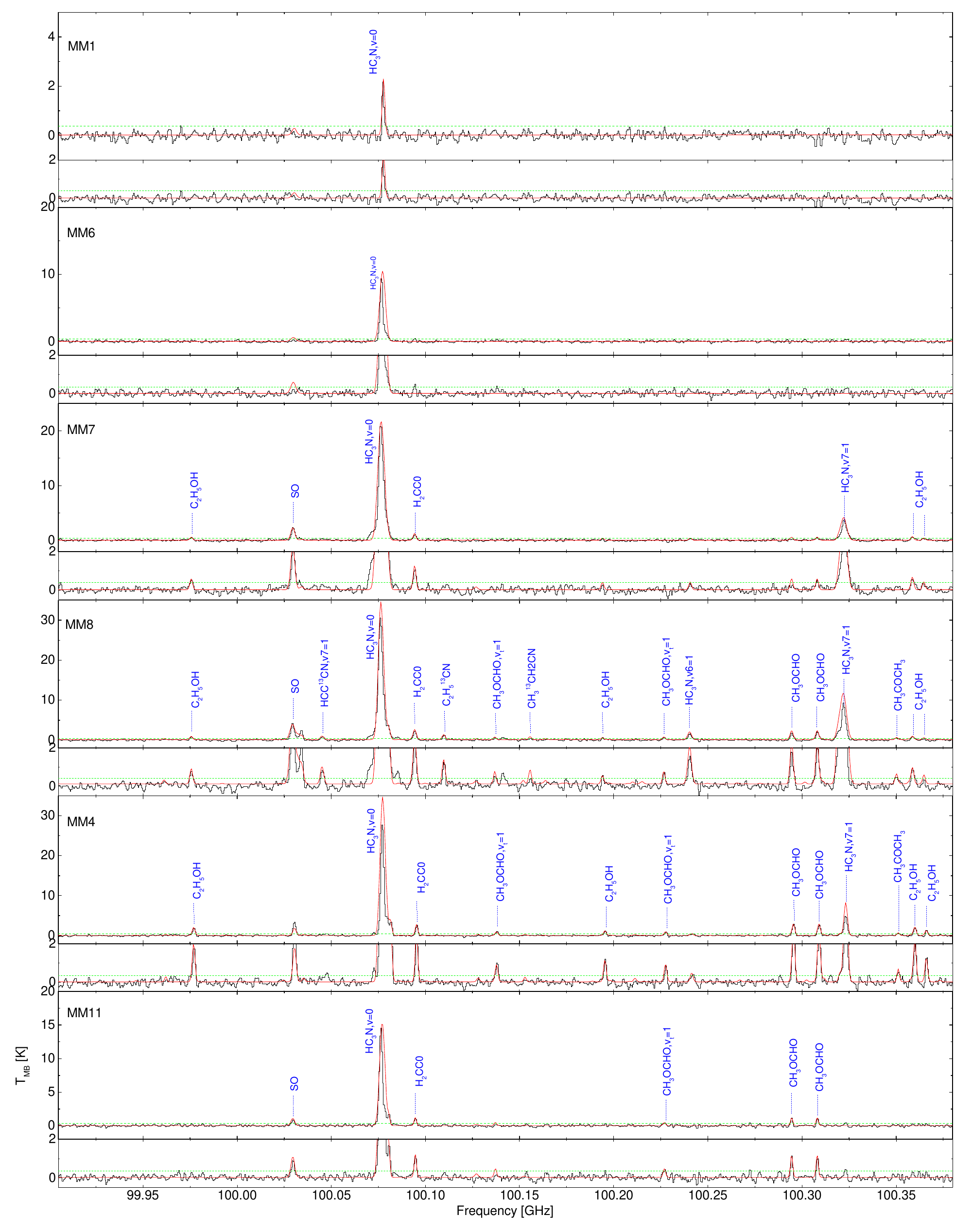}}}
\caption{Continued. Frequency range from 99.905 to 100.38 GHz.
}
\label{fig:spec_6}
\end{figure*}

\begin{figure*}
\setcounter{figure}{0}
\centerline{\resizebox{0.9\hsize}{!}{\includegraphics{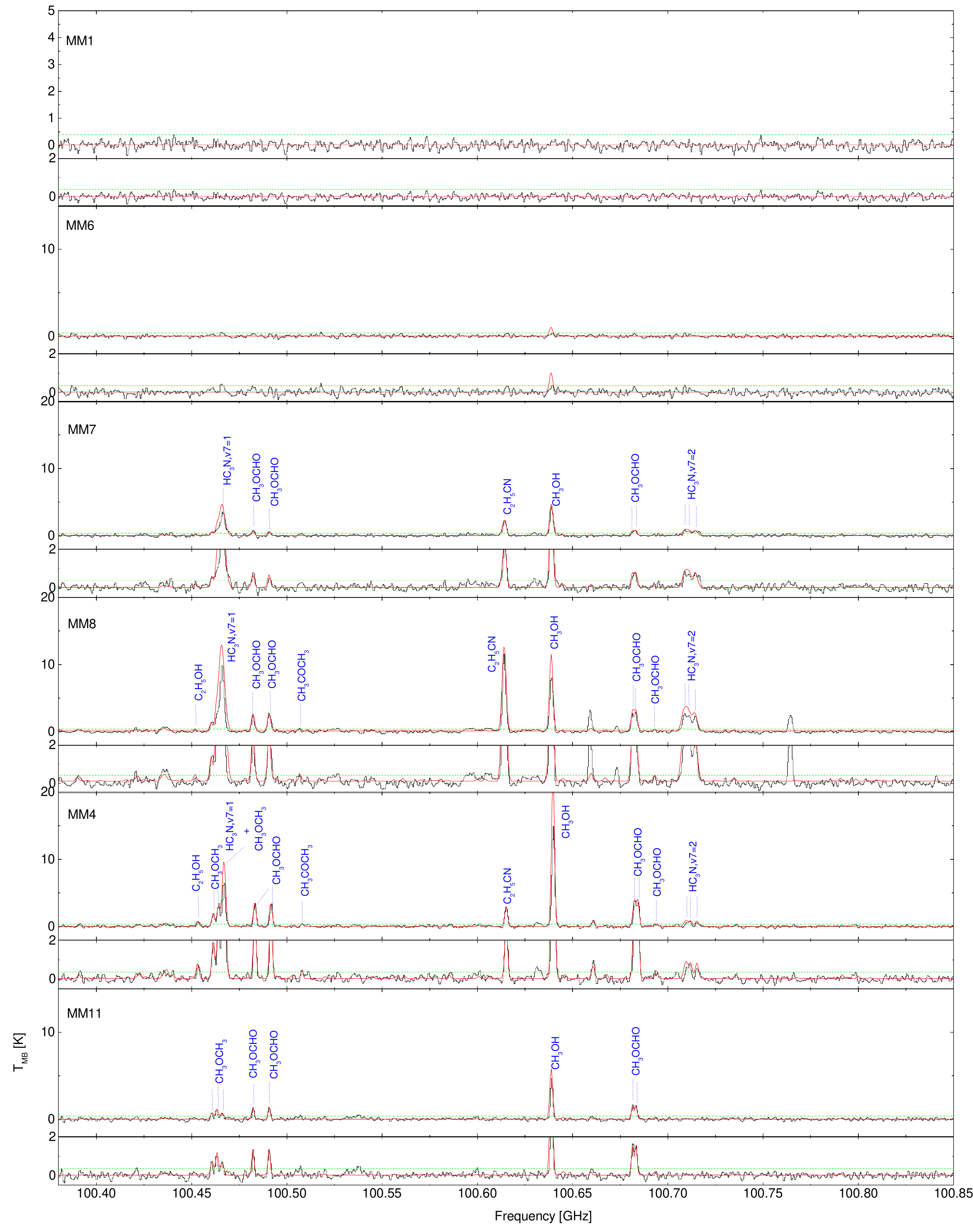}}}
\caption{Continued. Frequency range from 100.38 to 100.85 GHz.
}
\label{fig:spec_7}
\end{figure*}

\begin{figure*}
\setcounter{figure}{0}
\centerline{\resizebox{0.9\hsize}{!}{\includegraphics{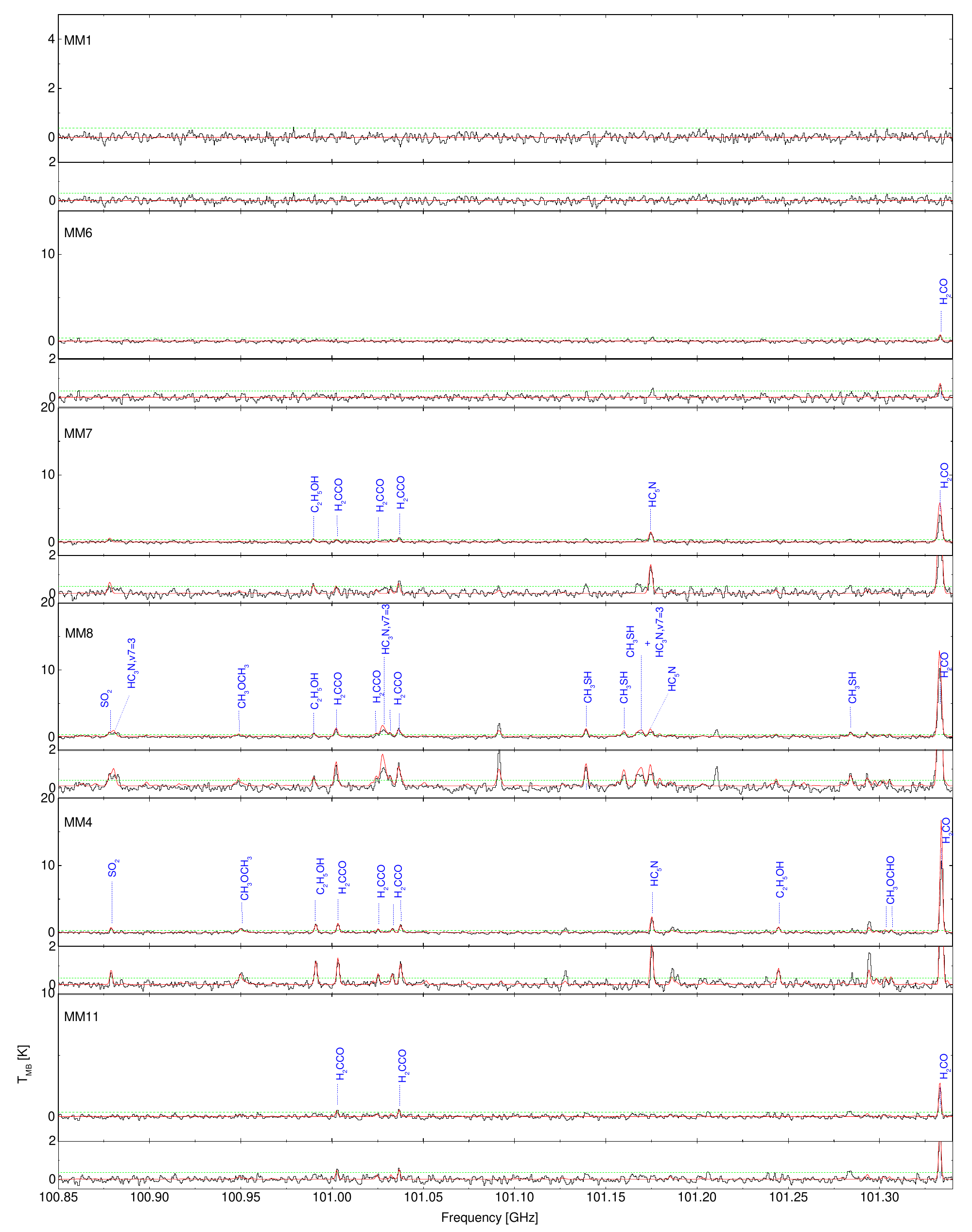}}}
\caption{Continued. Frequency range from 100.85 to 101.34 GHz.
}
\label{fig:spec_8}
\end{figure*}

\section{Line spectra at eight dense cores of G9.62+0.19.}
\begin{figure*}
\setcounter{figure}{0}
\centerline{\resizebox{0.9\hsize}{!}{\includegraphics{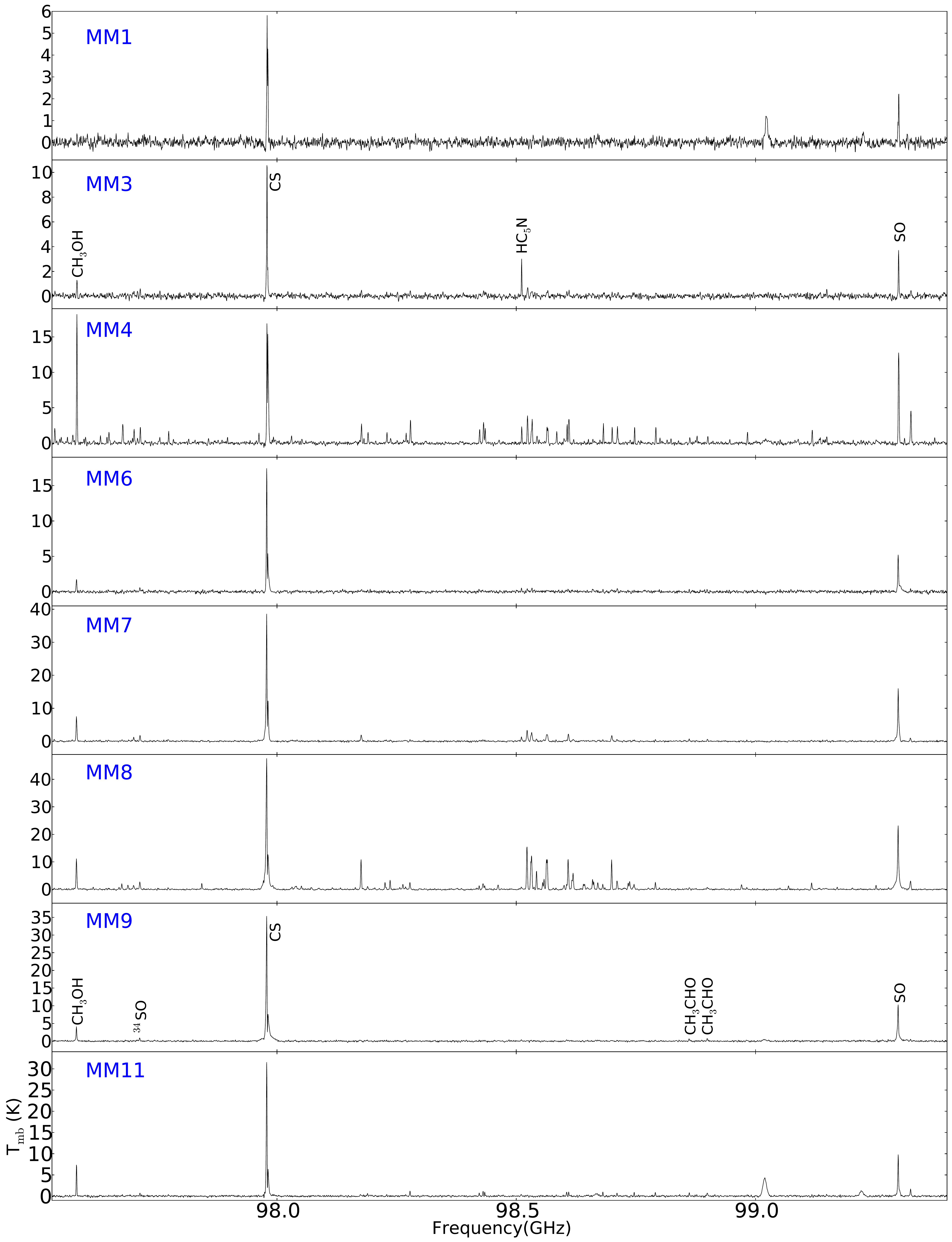}}}
\caption{Beam-averaged spectra toward the MM1, MM3, MM4, MM6, MM7, MM8, MM9, and MM11. This segment covers frequency range from 97.53 to 99.40 GHz.
}
\label{fig:B1-1}
\end{figure*}

\begin{figure*}
\setcounter{figure}{0}
\centerline{\resizebox{0.9\hsize}{!}{\includegraphics{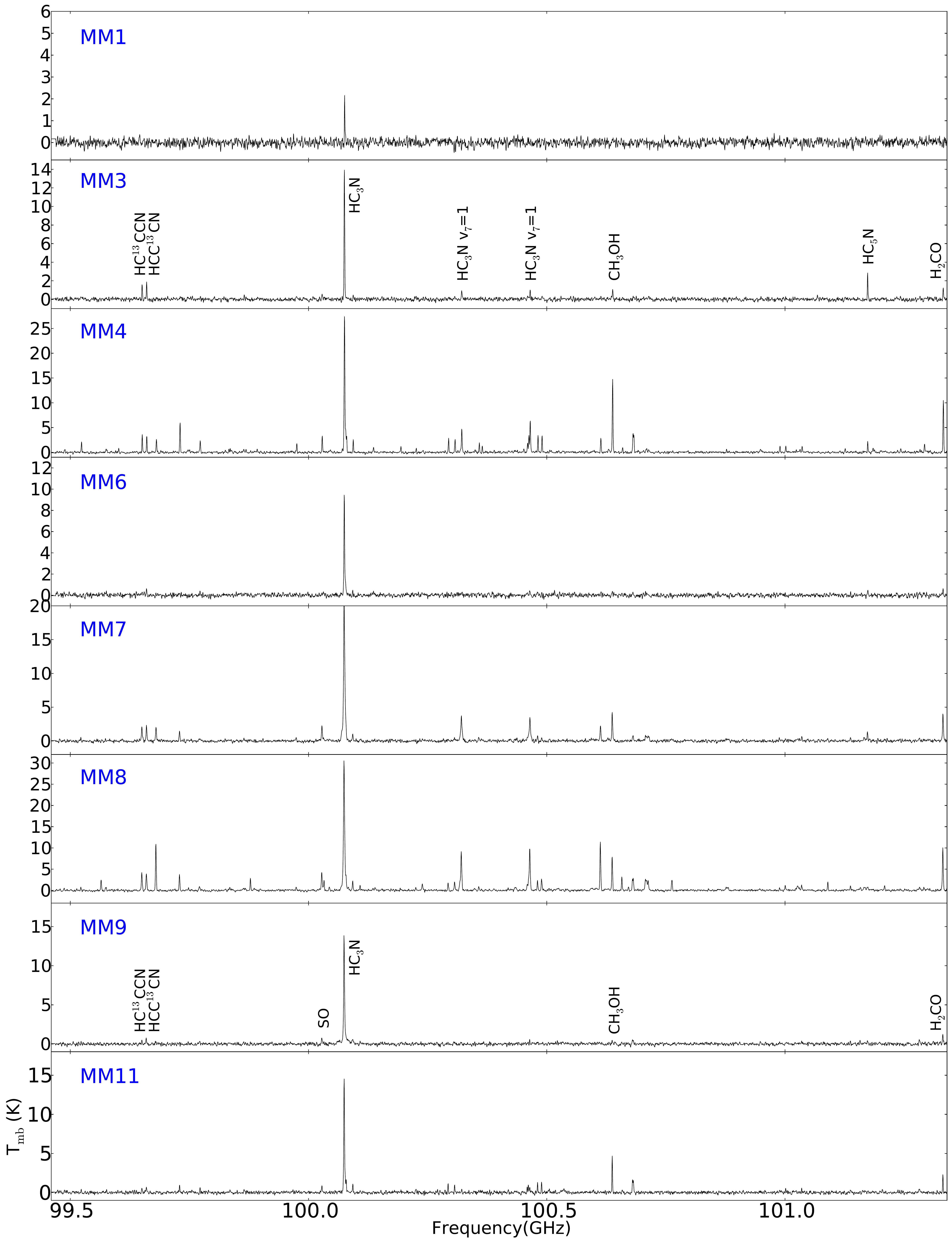}}}
\caption{Continued. This segment covers frequency range from 99.46 to 101.34 GHz.
}
\label{fig:B1-2}
\end{figure*}

\clearpage
\noindent
Author affiliations: \\
$^{1}$ College of Science, Yunnan Agricultural University, Kunming 650201, People’s Republic of China\\
$^{2}$ Department of Astronomy, and Key Laboratory of Astroparticle Physics of
         Yunnan Province, Yunnan University, Kunming 650091, People’s Republic
        of China \\
$^{3}$ Shanghai Astronomical Observatory, Chinese Academy of Sciences, 80 Nandan Road, Shanghai 200030, People's Republic of China \\
$^{4}$ S. N. Bose National Centre for Basic Sciences, Block-JD, Sector-III, Salt Lake, Kolkata 700106, India \\
$^{5}$ Kavli Institute for Astronomy and Astrophysics, Peking University, 5 Yiheyuan Road, Haidian District, Beijing 100871, People’s Republic of China \\
$^{6}$ Departamento de Astronom\'ia, Universidad de Chile, Las Condes,
7591245 Santiago, Chile \\
$^{7}$ Institute of Astronomy and Astrophysics, Anqing Normal University, Anqing 246133, People’s Republic of China \\
$^{8}$ National Astronomical Observatories, Chinese Academy of Sciences, Beijing 100101, China \\
$^{9}$ Korea Astronomy and Space Science Institute, 776 Daedeokdaero,
Yuseong-gu, Daejeon 34055, Republic of Korea \\
$^{10}$ University of Science and Technology, Korea (UST), 217 Gajeong-ro, Yuseong-gu, Daejeon 34113, Republic of Korea \\
$^{11}$ Department of Physics, University of Helsinki, PO Box 64, FI-00014
Helsinki, Finland \\
$^{12}$ Xinjiang Astronomical Observatory, Chinese Academy of Sciences, 830011 Urumqi, People’s Republic of China \\
$^{13}$ Nobeyama Radio Observatory, National Astronomical Observatory of Japan, National Institutes of Natural Sciences, 462-2 Nobeyama, Minamimaki, \\
\quad Minamisaku, Nagano 384-1305, Japan \\
$^{14}$ School of Space Research, Kyung Hee University, Yongin-Si, Gyeonggi-Do 17104, Republic of Korea\\
$^{15}$ School of Astronomy and Space Science, Nanjing University, Nanjing 210093, People’s Republic of China \\
$^{16}$ Physical Research Laboratory, Navrangpura, Ahmedabad380009, India \\
$^{17}$ Department of Astronomy, Peking University, 100871 Beijing, People’s Republic of China \\
$^{18}$ Chinese Academy of Sciences South America Center for Astronomy, National Astronomical Observatories, Chinese Academy of Sciences, Beijing 100012, China\\
$^{19}$ Institute of Astrophysics, School of Physics and Electronical Science, Chuxiong Normal University, Chuxiong 675000, China \\
$^{20}$ School of Physics and Astronomy, Sun Yat-sen University, 2 Daxue Road, Zhuhai, Guangdong 519082, China \\
$^{21}$ CSST Science Center for the Guangdong-Hongkong-Macau Greater Bay Area, Sun Yat-Sen University, Guangdong Province, China \\
$^{22}$ Laboratory for Space Research, The University of Hong Kong, Hong Kong, China \\
$^{23}$ Purple Mountain Observatory and Key Laboratory of Radio Astronomy, Chinese Academy of Sciences, 10 Yuanhua Road, 210023 Nanjing, PR China 


\bsp	
\label{lastpage}
\end{document}